\documentclass[fleqn,usenatbib]{mnras}
\usepackage{newtxtext,newtxmath}
\usepackage[T1]{fontenc}
\DeclareRobustCommand{\VAN}[3]{#2}
\let\VANthebibliography\thebibliography
\def\thebibliography{\DeclareRobustCommand{\VAN}[3]{##3}\VANthebibliography}
\usepackage{graphicx}	
\usepackage{amsmath}	
\usepackage{subfigure}
\usepackage{orcidlink}

\newcommand{\rev}[1]{\textcolor{black}{#1}}

\title[EMU/POSSUM Pilot: Galactic SNRs]{A Catalogue of Radio Supernova Remnants and Candidate Supernova Remnants in the EMU/POSSUM Galactic Pilot Field}

\author[B. D. Ball et al.]{
Brianna D. Ball\orcidlink{0009-0003-2088-9433},$^{1}$\thanks{E-mail: bdball@ualberta.ca} Roland Kothes\orcidlink{0000-0001-5953-0100},$^{2,1}$
Erik Rosolowsky\orcidlink{0000-0002-5204-2259},$^{1}$ Jennifer West\orcidlink{0000-0001-7722-8458},$^{2,3}$ Werner Becker\orcidlink{0000-0003-1173-6964},$^{4,5}$ 
\newauthor
Miroslav D. Filipovi\'c\orcidlink{0000-0002-4990-9288},$^{6}$ B.M. Gaensler\orcidlink{0000-0002-3382-9558},$^{3}$ Andrew M. Hopkins\orcidlink{0000-0002-6097-2747},$^{7}$ B\"arbel Koribalski\orcidlink{0000-0003-4351-993X},$^{8,6}$ 
\newauthor
Tom Landecker\orcidlink{0000-0003-1455-2546},$^{2}$ Denis Leahy\orcidlink{0000-0002-4814-958X},$^{9}$ Joshua Marvil\orcidlink{0000-0003-1111-8066},$^{10}$ Xiaohui Sun\orcidlink{0000-0002-3464-5128},$^{11}$ Filomena Bufano\orcidlink{0000-0002-3429-2481},$^{12}$ 
\newauthor
Ettore Carretti\orcidlink{0000-0002-3973-8403},$^{13}$ Adriano Ingallinera\orcidlink{0000-0002-3137-473X},$^{12}$ Cameron L. Van Eck\orcidlink{0000-0002-7641-9946},$^{14}$ and Tony Willis\orcidlink{0000-0002-2173-6151}$^{2}$
\\
$^{1}$Department of Physics, University of Alberta, Edmonton, Alberta, T6G 2E1, Canada\\
$^{2}$Dominion Radio Astrophysical Observatory, Herzberg Astronomy \& Astrophysics, National Research Council Canada, \\~~~P.O. Box 248, Penticton, BC V2A 6J9, Canada\\
$^{3}$Dunlap Institute for Astronomy and Astrophysics, University of Toronto, 60 St. George Street, Toronto, M5S 3H4, Canada \\
$^{4}$Max-Planck-Institut für Extraterrestrische Physik, Gießenbachstraße 1, 85748 Garching, Germany \\
$^{5}$Max-Planck-Institut für Radioastronomie, Auf dem Hügel 69, 53121 Bonn, Germany \\
$^{6}$Physical Sciences, Western Sydney University, Locked Bag 1797, Penrith South DC, Sydney, NSW 1797, Australia \\
$^{7}$Australian Astronomical Optics, Macquarie University, 105 Delhi Rd, North Ryde, NSW 2113, Australia \\
$^{8}$Australia Telescope National Facility, CSIRO Astronomy and Space Science, P.O. Box 76, NSW 1710, Epping, Australia \\
$^{9}$Department of Physics \& Astronomy, University of Calgary, Calgary, AB T2N 1N4, Canada \\
$^{10}$National Radio Astronomy Observatory, P.O. Box O, Socorro, NM 87801, USA \\
$^{11}$School of Physics and Astronomy, Yunnan University, Kunming 650500, PR China \\
$^{12}$INAF - Osservatorio Astrofisico di Catania, Via Santa Sofia, 78, 95123 Catania, Italy \\
$^{13}$INAF, Istituto di Radioastronomia, Via Gobetti 101, 40129 Bologna, Italy \\
$^{14}$Research School of Astronomy \& Astrophysics, The Australian National University, Canberra, ACT 2611, Australia
}

\date{Accepted 2023 June 27. Received 2023 June 23; in original form 2023 May 10}

\pubyear{2023}

\begin{document}
\label{firstpage}
\pagerange{\pageref{firstpage}--\pageref{lastpage}}
\maketitle

\begin{abstract}
 We use data from the pilot observations of the EMU/POSSUM surveys to study the "missing supernova remnant (SNR) problem", the discrepancy between the number of Galactic SNRs that have been observed and the number that are estimated to exist. The Evolutionary Map of the Universe (EMU) and the Polarization Sky Survey of the Universe's Magnetism (POSSUM) are radio sky surveys that are conducted using the Australian Square Kilometre Array Pathfinder (ASKAP). We report on the properties of 7 known SNRs in the joint Galactic pilot field, with an approximate longitude and latitude of \(323^\circ\leq l \leq 330^\circ\) and \(-4^\circ\leq b \leq 2^\circ\) respectively, and identify 21 SNR candidates. \rev{Of these, 4 have been previously identified as SNR candidates, 3 were previously listed as a single SNR, 13 have not been previously studied, and 1 has been studied in the infrared.} These are the first discoveries of Galactic SNR candidates with EMU/POSSUM and, if confirmed, they will increase the SNR density in this field by a factor of 4. By comparing our SNR candidates to the known Galactic SNR population, we demonstrate that many of these sources were likely missed in previous surveys due to their small angular size and/or low surface brightness. We suspect that there are SNRs in this field that remain undetected due to limitations set by the local background and confusion with other radio sources. The results of this paper demonstrate the potential of the full EMU/POSSUM surveys to uncover more of the missing Galactic SNR population.
\end{abstract}




\section{Introduction} \label{sec:intro}

Supernovae and supernova remnants (SNRs) are the most significant sources of chemical enrichment in the interstellar medium (ISM) of our Galaxy. More than half of the material in the Milky Way has been processed by supernovae and their remnants \citep{Padmanabhan2001}. Thus, our knowledge of the Galaxy and its evolution is necessarily informed by our understanding of the Galactic SNR population. 

In this paper we seek to investigate the so-called ``missing supernova remnant problem,'' which refers to the discrepancy between the number of SNRs that are believed to exist in our Galaxy and the number that have been discovered \citep{Brogan2006,Helfand2006,Green2014,Green2015}. The exact size of the discrepancy is unknown, as accurately quantifying this problem is challenging due to variations in SNR density and radio visibility across the Galactic plane. Based on observations of extra-galactic supernovae, we know that in galaxies like ours a supernova should occur every 30 to 50 years \citep{Tammann1994}. We can combine this rate with the expected SNR radio lifetime to obtain an estimate of the number of SNRs that should be detectable at radio wavelengths.

\begin{figure*}
    \centering
    \subfigure{\includegraphics[width=0.8\textwidth]{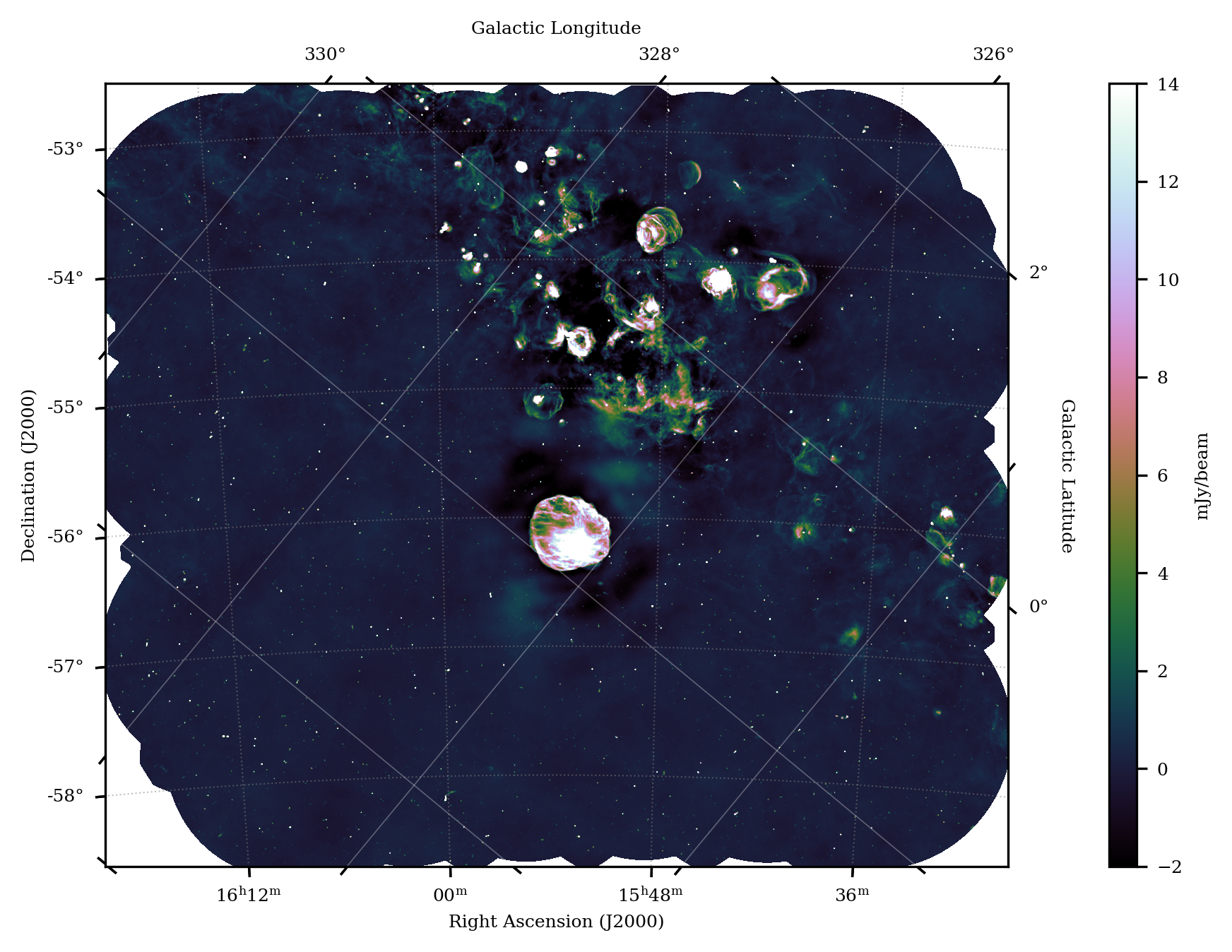}\label{fig:EMUfield}} 
    \subfigure{\includegraphics[width=0.8\textwidth]{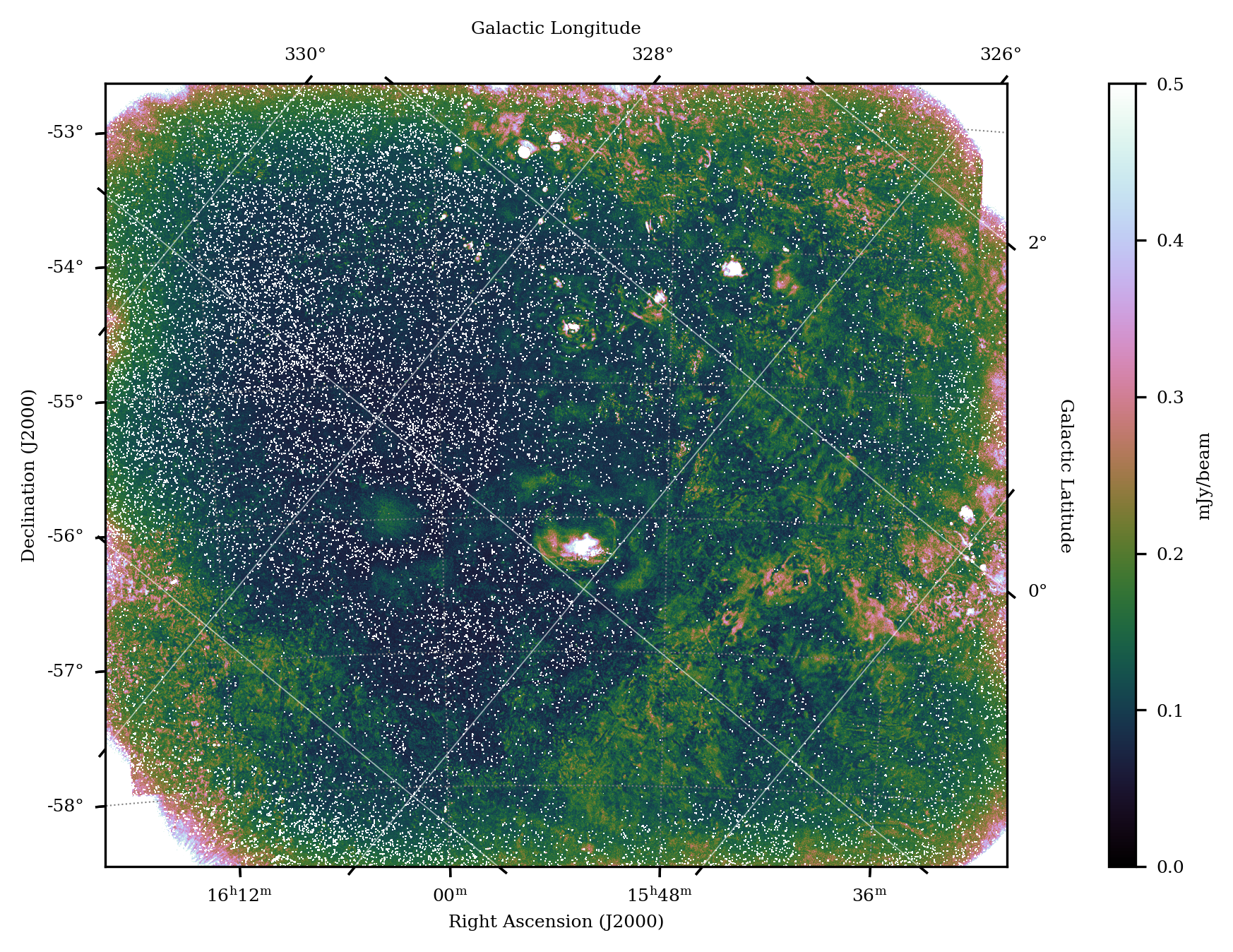}\label{fig:PIfield}} 
    \caption{The top image shows the EMU/POSSUM Galactic pilot II field as observed by ASKAP at 933 MHz in total intensity with an angular resolution of 18". The bottom image shows the same field in polarized intensity.}
    \label{fig:field}
\end{figure*}

The radio-visible lifetime of a supernova remnant is difficult to estimate, however, as it likely varies based on local conditions and depends on the frequency of the observations. By studying associations between supernova remnants and pulsars, \cite{Frail1994} estimated the mean radio SNR lifetime to be about 60,000 years. More recent work has shown that the majority of observed remnants in the Milky Way and Local Group galaxies are believed to be in the Sedov-Taylor (S-T) phase of evolution \citep{Albert2022}. Thus, in many cases the S-T lifetime may serve as a useful proxy for the radio visible lifetime with a characteristic timescale of 20 - 80 kyrs dependent on the local ISM density \citep{Sarbadhicary2017}. However, adopting these age limits may result in estimates which are too conservative. Around 25\% of known Galactic SNRs with age estimates are believed to be older than 20 kyrs \citep{Ferrand2012}. Additionally, there have been discoveries of Galactic SNRs, observable at radio wavelengths, that are well beyond the S-T phase \citep{Kothes2017b}. In comparing the \cite{Sarbadhicary2017} model predictions to observations, \cite{Leahy2022} found the model to be insufficient in reproducing observed radio emission. 

According to \cite{Ranasinghe2022}, predictions for the total number of SNRs in the Galaxy should generally be >1000, so we adopt this as a lower limit. We form a conservative upper limit based on the supernova rate and S-T lifetime and estimate that at any given time, 1000 to 2700 radio supernova remnants should be detectable in our Galaxy. So far we have only discovered somewhere in the range of 300 to 400 \citep{Green2022, Ferrand2012}. We aim to detect some of these missing SNRs and, by studying their properties, gain further insight into the nature of this discrepancy. 

The majority of supernova remnants (approximately 95\%) discovered in our Galaxy have been detected in the radio \citep{Dubner2017}. Thus, radio observations play an important role in the search for Galactic SNRs. Because of the limitations in working with radio data due to the relatively poor angular resolution and sensitivity when compared to observations at shorter wavelengths, sources with a small angular size and/or low surface brightness are more likely to be missed. It is therefore reasonable to expect that these types of sources may comprise a significant portion of the missing SNR population. This is especially true in regions of the Galaxy where radio emission is dominated by thermal sources, such as HII regions, and distinguishing SNRs becomes more difficult. X-ray observations of young Galactic SNRs are believed to be fairly complete with an implied Galactic SNR birth rate of $\sim$1/35 years, consistent with the supernova explosion rate \citep{Leahy2020}. Thus, we mostly expect to find old, faint SNRs of varying angular sizes, dependent on the distance to the source and the local environment.

\rev{Confusion with other extended radio sources, particularly HII regions, presents a significant challenge to confidently identifying SNR candidates. To address this, we adapt a commonly used methodology involving the comparison of radio and mid-infrared (MIR) fluxes. This technique has been used by many other Galactic SNR surveys, as well as in follow up studies of SNR candidates \citep{Whiteoak1996,Helfand2006,Brogan2006,Green2014,Anderson2017,Hurley-Walker2019,Dokara2021}. While SNRs and HII regions can have similar radio morphologies, HII regions produce strong MIR emission from warm dust and polycyclic aromatic hydrocarbons (PAHs). Conversely, SNRs have been found to produce little to no MIR emission \citep{Fuerst1987,Whiteoak1996,Pinheiro2011}. The absence of an MIR counterpart can therefore be used as evidence that a potential SNR candidates is not an HII region.}

The Evolutionary Map of the Universe (EMU) and the Polarization Sky Survey of the Universe's Magnetism (POSSUM) are radio surveys that will be observed together with the Australian Square Kilometre Array Pathfinder (ASKAP). Because of the improved resolution and sensitivity when compared to previous southern sky radio surveys, such as the MGPS-2 \citep{Green2014}, the EMU/POSSUM surveys should be expected to uncover some of these small and faint sources. Additionally, ASKAP's large field of view and good uv-coverage should allow for the detection of old SNRs that are large with low surface brightness. Here we utilize data from the pilot observations of these surveys to search for supernova remnants within a small field of the Galactic plane. 

In this paper, we aim to (1) validate the quality of the EMU/POSSUM Galactic pilot field data by studying the properties of known supernova remnants in the field, (2) identify new supernova remnant candidates and uncover some of the missing SNR population, and (3) develop analysis techniques that can be used to study supernova remnants and search for new candidates with the full EMU/POSSUM sky survey data as they become available. Descriptions of the data used in this paper can be found in Section~\ref{sec:methods}. In Section~\ref{sec:results} we present the data and describe how SNR candidates were identified. We also discuss the known SNRs in the field and their properties. In Section~\ref{sec:candidates} we present our SNR candidates and in Section~\ref{sec:discussion} we provide some analysis and comparison to the known Galactic SNR population. The conclusions are summarized in Section~\ref{sec:conclusions}.

\begin{figure}
    \centering
    \includegraphics[width=0.44\textwidth]{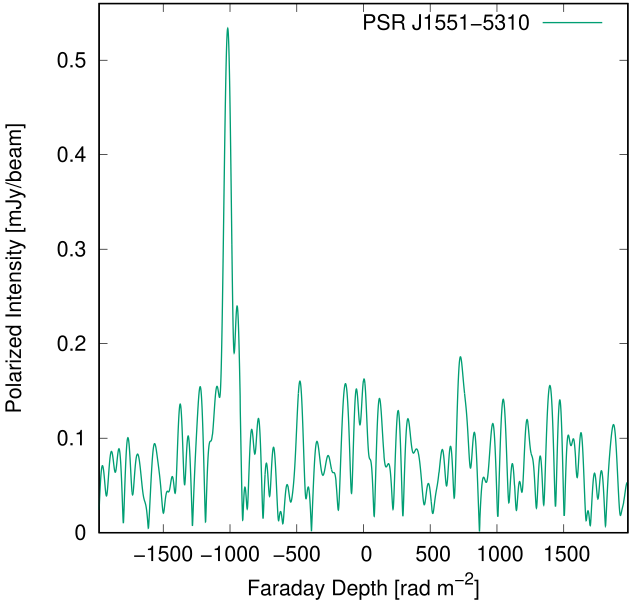}    
    \caption{Faraday depth spectrum of the pulsar J1551-5310, which is located inside SNR candidate G328.0$+$0.7.}
    \label{fig:PSRFD}
\end{figure}

\section{Methods and Observations} \label{sec:methods}

\subsection{The EMU and POSSUM Observations with ASKAP} \label{sec:observations}

We use data from the ASKAP telescope \citep{Hotan2021}, an interferometer with 36 12-meter dishes equipped with phased-array feeds (PAFs). These data were obtained during the commissioning phase of the telescope, specifically from the second pilot observations of the commensal EMU \citep{Norris2011,Norris2021} and POSSUM surveys \citep[][Gaensler in prep.]{Gaensler2010}. These observations use a full 10-hour track having full Stokes with 288 MHz bandwidth, centred at 933 MHz. Imaging is performed using ASKAPsoft and the standard commensal EMU/POSSUM imaging parameter set \citep{Norris2021}, which produces both a multi-frequency synthesis (MFS) band-averaged Stokes I image for the EMU survey, and full Stokes I, Q, U, and V frequency cubes with 1 MHz channels for POSSUM. The cubes have been convolved to a common resolution of 18” across all PAF beams and all frequency channels. The observations for this particular pilot II survey field were observed on 06 November 2021 (Scheduling block 33284). Primary beam correction in all Stokes parameters are performed using beam models derived from standard observatory holography observations from 20 June 2021 (Scheduling block 28162). The long period of time between the holography and the observations resulted in a poor leakage correction, and therefore this field was re-observed on 07 September 2022 (SB 43773). This field was corrected using holography observations from 28 July 2022 (SB 43057). This improved correction mitigates leakage from Stokes I into Stokes Q and U at around the 1\% level or less over most of the field. The top image in Figure \ref{fig:field} shows the Stokes I image from the original pilot II observation and the bottom image shows the same field in polarized intensity (PI). The polarized intensity was taken from the peak of the Faraday depth (FD) function for each pixel.

For the known supernova remnants and our candidates, we did not use the FD cube from the POSSUM pipeline, but calculated them ourselves. We used the Q and U data cubes from scheduling block 43773. Instead of a Fourier transform, we de-rotated the Q and U data in each frequency channel for each rotation measure (RM). We probed an RM range from $-2000$ to $+2000$~rad\,m$^{-2}$ with a step size of 1~rad\,m$^{-2}$. A sample FD function is shown in Figure~\ref{fig:PSRFD}. This was taken towards the pulsar PSR~J1551$-$5310 inside our SNR candidate G328.0$+$0.7. The RM for this pulsar is catalogued by \citet{psrrm2018} to be $-1023.3 \pm 6.3$~rad\,m$^{-2}$. In our data we find $RM = -1017 \pm 5$~rad\,m$^{-2}$.

\subsection{Ancillary Data} \label{sec:ancillarydata}

In addition to the radio data from ASKAP, we utilize images of the same field of the Galactic plane from two other sky surveys. Comparing radio and MIR fluxes is a commonly used technique for identifying supernova remnants. Thus, we make use of 12 $\mu$m infrared data from WISE (Widefield Infrared Survey Explorer) \citep{Wright2010} with an angular resolution of 6.5" as part of the candidate identification process, further outlined in Section~\ref{sec:radioMIR}. \rev{We use 12 $\mu$m data because it traces emission from hot dust and PAHs, both of which are expected to be abundant in HII regions \citep{Anderson2014} and largely destroyed in SNRs \citep{Slavin2015}.} Pixel values in WISE data are measured in digital number (DN) units, which are designed for relative photometric measurements \citep{Cutri2013} and are sufficient for our purposes.

We use low frequency (198 MHz) data from GLEAM (The GaLactic and Extragalactic All-sky MWA survey) \citep{Wayth2015} to calculate spectral indices for some remnants, the details of which are provided in Section~\ref{sec:indices}. The angular resolution of the GLEAM data ($\sim$169"$\times$149") is relatively poor compared to the resolution of ASKAP so while we are able to calculate spectral indices for six of the known remnants, we can obtain indices with these data for only two of our candidates. 

\subsection{Flux Integration} \label{sec:fluxint}

To calculate total intensity flux densities, we use a combined map that was created by averaging data from the two ASKAP second pilot observations. This was done to minimize the effect of background fluctuations. 

Because of the complexity of the background, different methods for calculating flux densities and performing background subtraction were explored. Integrating over radial profiles using Karma software \citep{Gooch1995} did not allow us to properly account for variations in the background or surrounding bright sources. Attempting to use a circular aperture to define the source with multiple circular apertures defining the background, as done by \cite{Anderson2017}, presented similar challenges and lacked consistency. Ultimately, flux integration was performed using the \textsc{Polygon\_Flux} software, which was developed by \cite{Hurley-Walker2019} for the GLEAM survey to deal with extended sources that have complicated backgrounds. The software calculates the flux density within a chosen region, subtracts user-selected point sources, and performs background subtraction, allowing the user to select surrounding regions that should not be included as part of the background.

For each source, the calculations were performed multiple times in order to obtain a more accurate flux value and an error estimate. Three different definitions of the background were used to estimate the systematic uncertainty due to flux aperture definition. The calculations were run using backgrounds defined as 4 to 10, 10 to 16, and 16 to 22 pixels from the source (with a pixel size of 2"). Additionally, the source and background selection process was performed at least twice for each source to account for uncertainties resulting from the definition of the source perimeter. Thus, the flux density calculations were run at least six times per source. The flux densities in Table~\ref{tab:2} were taken as the median values of these calculations and the errors were determined by the range between the extrema. Instrumental uncertainties in the fluxes were found to be relatively insignificant compared to the systematic estimates and were not included.

\section{Results} \label{sec:results}

\subsection{The EMU/POSSUM Galactic Pilot Field} \label{sec:pilotfield}

\begin{figure*}
    \centering
    \subfigure{\includegraphics[width=0.82\textwidth]{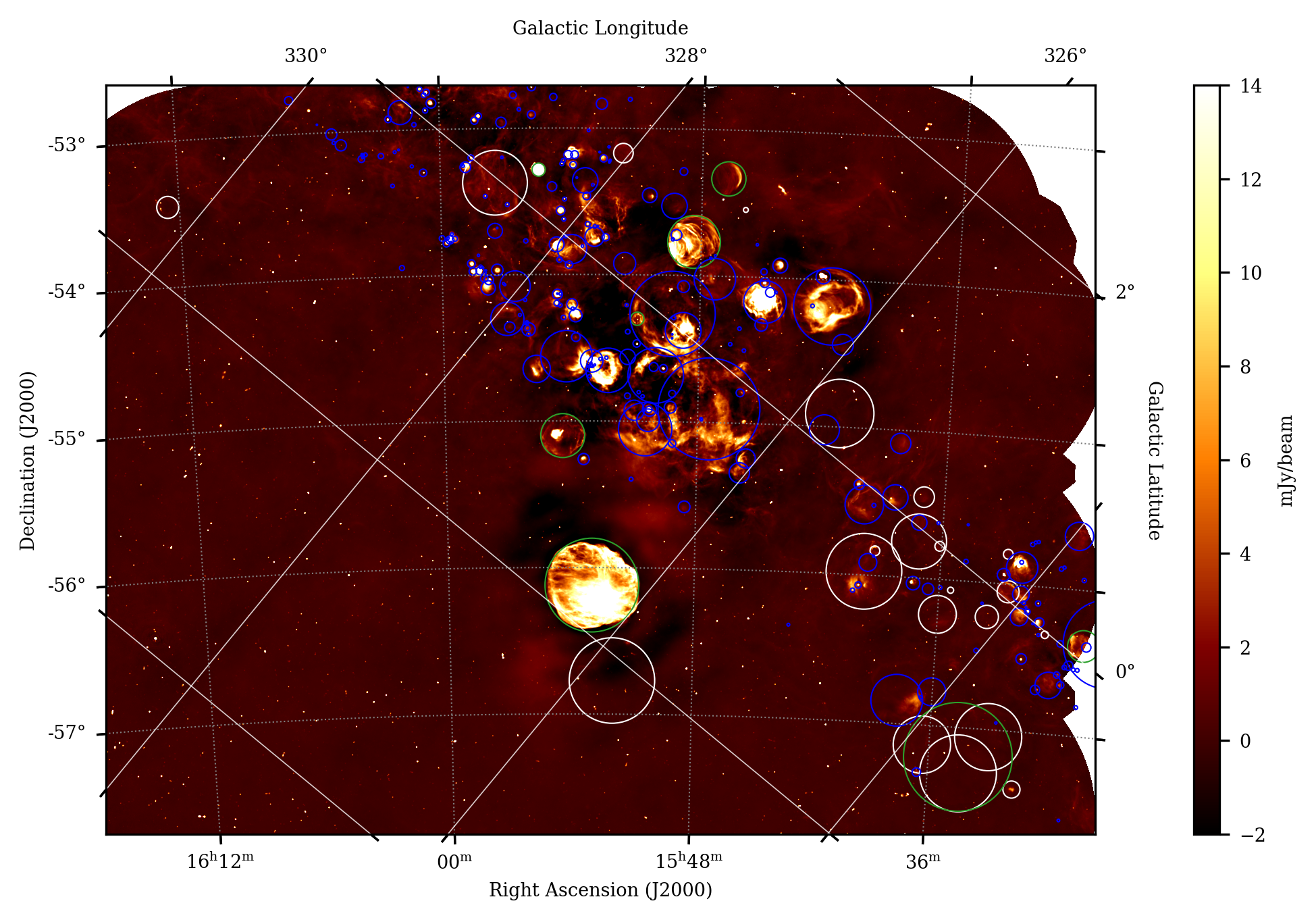}\label{fig:askapfield}} 
    \subfigure{\includegraphics[width=0.82\textwidth]{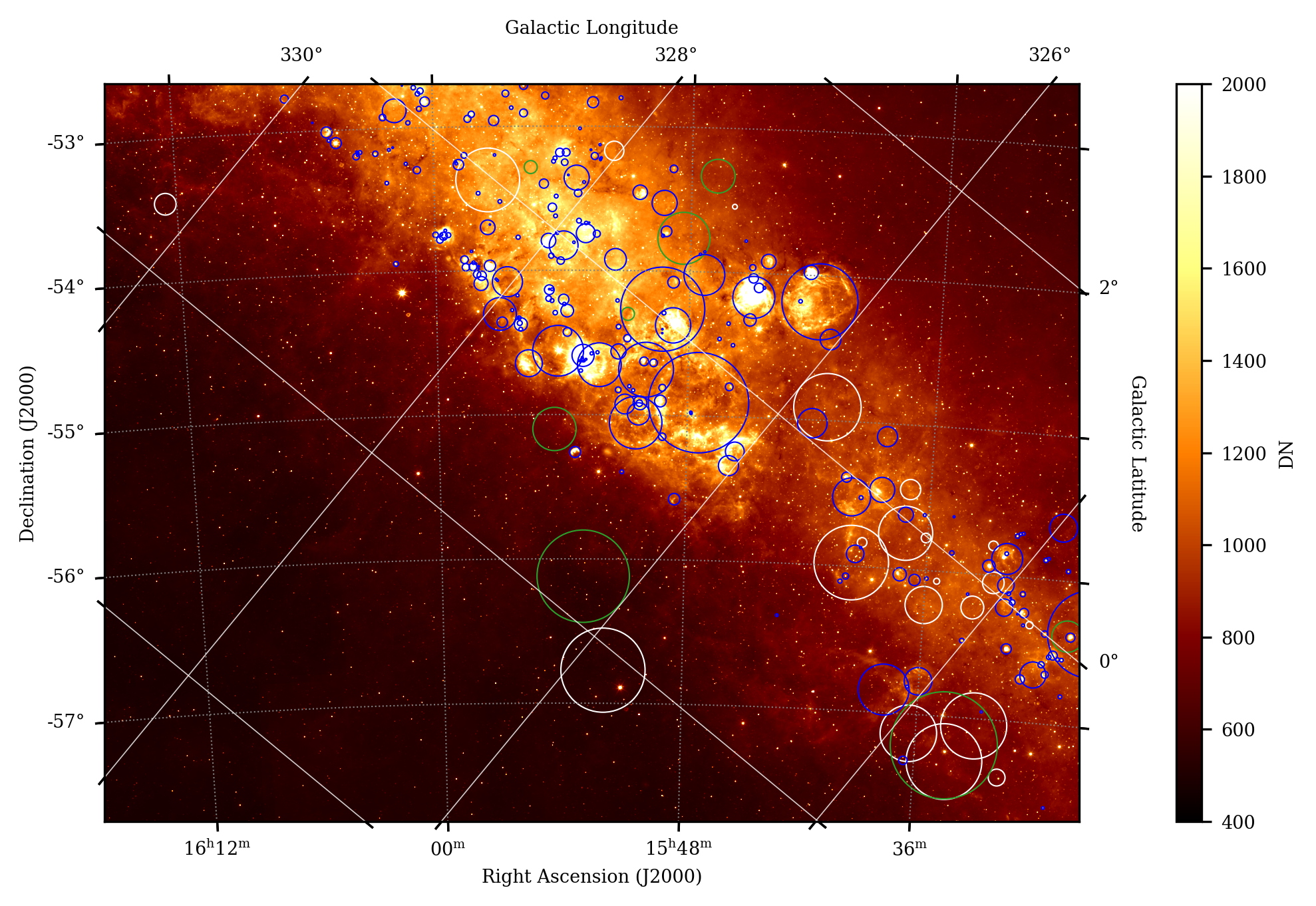}\label{fig:wisefield}} 
    \caption{The top image shows the EMU/POSSUM Galactic pilot II field from ASKAP at 933 MHz in total intensity. The bottom image shows the same field in the mid-infrared from WISE (12 $\mu$m). The green circles show the locations of known SNRs, taken from \protect\cite{Green2022}. The blue circles indicate the locations of known HII regions, taken from \protect\cite{Anderson2014}. The white circles indicate the locations of our SNR candidates.}
    \label{fig:field2}
\end{figure*}

Figure \ref{fig:field2} shows the EMU/POSSUM Galactic pilot II data as observed by ASKAP and the same field in the mid-infrared as observed by WISE \citep{Wright2010}. The field looks across the Galactic plane, with an approximate longitude and latitude of \(323^\circ\leq l \leq 330^\circ\) and \(-4^\circ\leq b \leq 2^\circ\) respectively, along a tangent to the Norma arm and across several other spiral arms. This gives us a long line of sight through the inner Galaxy, up to distances of about 18 kpc. 

The annotations shown in Figure~\ref{fig:field2} indicate the locations of the known SNRs (green) and HII regions (blue) within this field as well as the locations of our 21 SNR candidates (white). The known SNRs are taken from the \cite{Green2022} radio SNR catalogue and the HII regions come from the WISE Catalogue of Galactic HII regions \citep{Anderson2014}. There are 8 known SNRs in this field, including one that we believe should be reclassified as multiple sources (discussed further in Section~\ref{sec:candidates}). One of the known SNRs lies at the edge of the field and is only partially imaged in the combined map as shown.

This field was selected in part because it can be broken into two regions that are visually and meaningfully distinct in both the radio and MIR. The upper left part of the field looks along a tangent to the Norma arm and thus we see a high density of HII regions and thermal emission. The lower right part of the field is noticeably fainter with less background emission and a lower density of HII regions. This allows us to test our ability to detect SNR candidates in each of these regions. Since we are primarily expecting to find low surface brightness sources, it is probable that there are candidates in the upper left region of the Galactic plane that we are unable to detect due to  the high concentration of thermal emission. The locations of the candidates in Figure~\ref{fig:field2} support this as they are clearly concentrated in areas with fewer HII regions and less background emission. 

\subsection{SNR Identification and Verification} \label{sec:snridentification}

A supernova remnant is formed in the aftermath of a stellar explosion as the ejected material expands into the ISM bounded by a supersonic shock wave that sweeps up interstellar material and magnetic fields as it travels. At the shock front, electrons are accelerated to relativistic speeds and interact with the magnetic field to produce highly linearly polarized synchrotron emission that is best observed in the radio \citep{vanderLaan1962}. A supernova remnant can typically be identified by the distinctive shell-like structure that is formed through this interaction between the supernova shock and the ISM. The structural evolution of the remnant will depend on factors like the characteristics of the explosion, the density of the surrounding ISM, and the ambient magnetic field \citep{Whiteoak1968,Truelove1999,Kothes2009}. While these factors may result in  asymmetries, we generally expect to see well-defined rounded edges, produced by the shock, with fainter emission coming from the remnant's centre.

Here we identify supernova remnant candidates primarily by looking for radio-emitting shell-like structures that lack clear mid-infrared counterparts. After identifying candidates, we attempt to find further evidence that they are supernova remnants. First, the presence of a young pulsar indicates that a supernova explosion has recently occurred so spatial coincidence of a candidate with this type of star can significantly increase our confidence in its classification. Second, radio emission from SNRs is primarily non-thermal synchrotron emission that can be differentiated from thermal emission using polarization and spectral indices. Non-thermal synchrotron emission is associated with a steep negative spectral index and linear polarization while thermal optically thin free-free emission is associated with a flat, unpolarized spectrum. As we are observing at relatively low radio frequencies, we can also expect to find compact HII regions with optically thick free-free emission and \rev{steep positive spectral indices}.

\begin{figure*}
    \centering
    \subfigure{\includegraphics[width=0.75\textwidth]{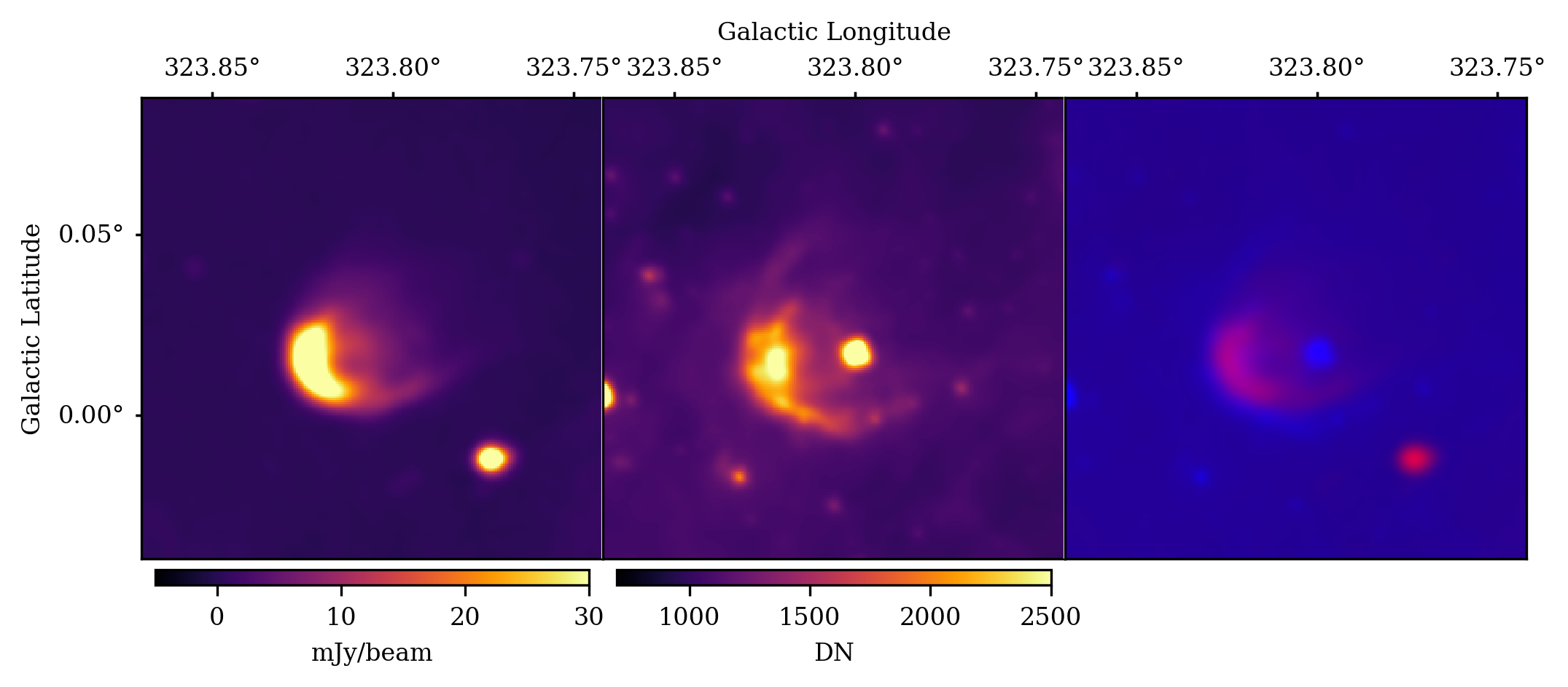}} 
    \subfigure{\includegraphics[width=0.75\textwidth]{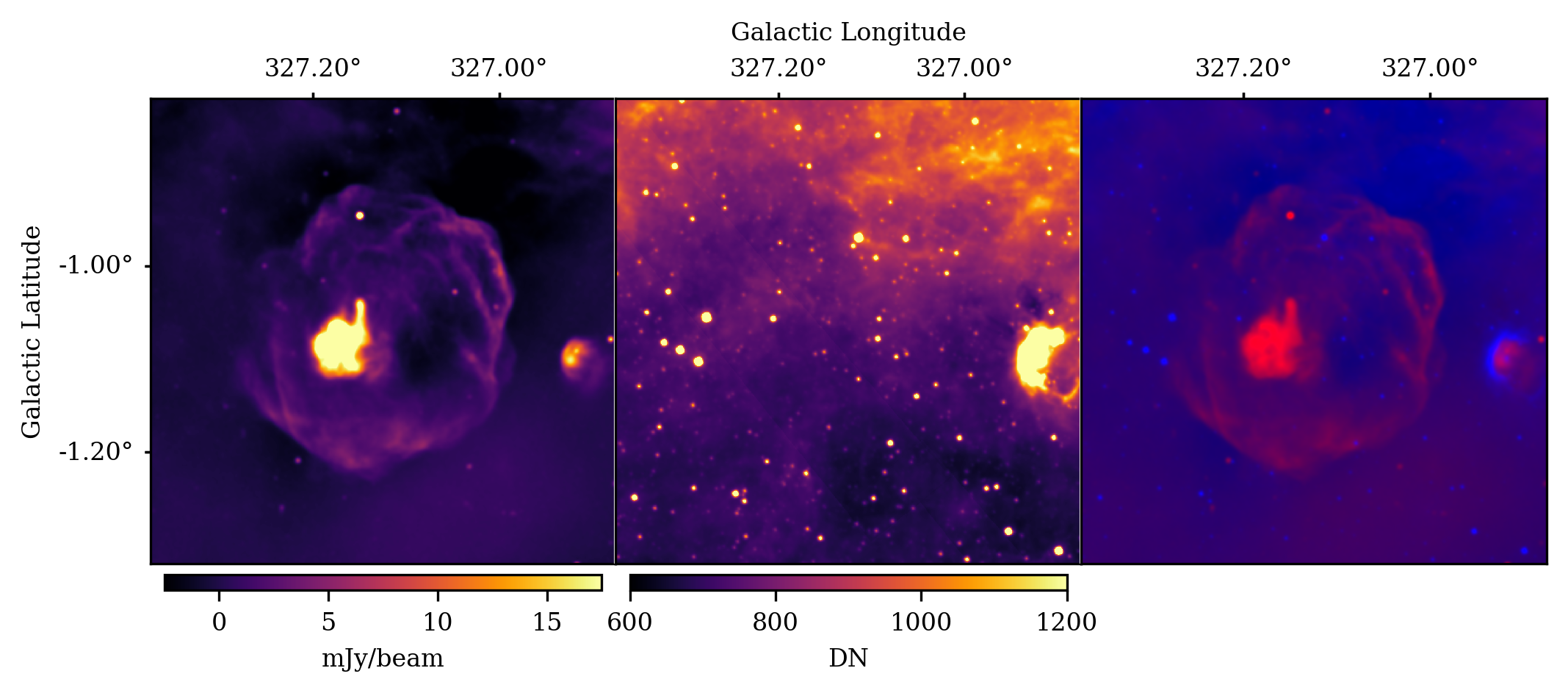}\label{fig:2}} 
    \caption{Comparing radio and MIR emission from an HII region and an SNR. From left to right, the top images show an HII region in radio (from ASKAP), MIR (from WISE), and a combined image with radio in red and MIR in blue. The bottom images show the same data for a known SNR, G327.1$-$1.1, and demonstrate the absence of an MIR counterpart.}
    \label{fig:hii}
\end{figure*}

\subsubsection{Radio and MIR Emission} \label{sec:radioMIR}

Comparing radio and MIR fluxes is a commonly used technique for distinguishing non-thermal SNR emission from thermally emitting sources like HII regions \citep{Whiteoak1996, Helfand2006, Green2014, Anderson2017}. HII regions produce MIR emission primarily through stochastic heating of small dust grains and the vibrational and bending modes of polycyclic aromatic hydrocarbons (PAHs) \citep{Draine2011}. Supernovae are known to produce significant amounts of dust and SNRs can also produce MIR emission through these thermal processes. However, this emission is relatively weak. This is in part because most dust (>90\%) in SNRs is \rev{found in the dense, cool gas phase, rather than the X-ray emitting plasmas, resulting in a spectral energy distribution that peaks at longer infrared wavelengths than studied here} \citep{Priestley2022}. Additionally, while the supernova shock can heat dust grains, it can also result in the destruction of a significant amount of dust and large molecules like PAHs \citep{Slavin2015}. According to \cite{Bianchi2007}, only 2 - 20\% of the initial dust mass survives the passage of the reverse shock, depending on the density of the surrounding ISM. Therefore, dust emission from SNRs is typically expected to be weak, if it is detectable at all, and importantly it has been shown that SNRs have significantly lower MIR to radio flux ratios when compared to HII regions \citep{Whiteoak1996, Pinheiro2011}. Thus, while some SNRs do emit in the MIR, sources that lack MIR emission are unlikely to be HII regions.

We began our search for SNR candidates by comparing the radio data from ASKAP to MIR data from WISE. The images used can be found in Figure~\ref{fig:field2}. Because of the small size of the field, we were able to perform a detailed search by eye, specifically looking for shell-like structures in the radio that do not have MIR counterparts. This was done separately by two of the authors before comparing results. To simplify the search, we eliminated sources that had already been classified as HII regions. The WISE Catalogue of Galactic HII Regions \citep{Anderson2014} was used to identify all known HII regions in the field. The catalogue was made using 12 $\mu$m and 22 $\mu$m data from WISE and includes over 8000 Galactic HII regions and HII region candidates. \rev{We use Version 2.2 of the catalogue, downloaded from \url{http://astro.phys.wvu.edu/wise/}, and include all listed entries.} We did not find any sources in our field that we believe to be HII regions that were not already part of the WISE HII region catalogue. Specifically, we did not find any new extended sources that were clearly visible in both radio and MIR. Thus, the radio to MIR comparison was not used to rule out any potential candidates but instead served primarily as evidence that our candidates are not HII regions.

\subsubsection{Spectral Indices} \label{sec:indices}

Spectral indices can help to differentiate thermal and non-thermal emission. We assume the relation \(S(\nu) \propto \nu^{\alpha}\) where \(S\) represents the flux density, \(\nu\) represents the frequency, and \(\alpha\) represents the spectral index. The synchrotron spectral index is determined by the power-law energy distribution of relativistic particles which are accelerated through multiple crossings at the shock front in a process known as diffusive shock acceleration (DSA) \citep{Bell1978, Blandford1978}. For a shell-type remnant, linear DSA predicts a spectral index of \(-0.5\) and observations show that most catalogued Galactic SNRs have a spectral index within the range \(\alpha = -0.5 \pm 0.2\) \citep{Reynolds2012,Dubner2015}. However, there is significant uncertainty in many of the measured values and outliers do exist. 

Thermally-emitting HII regions are expected to have flatter spectral indices for optically thin free-free emission, generally around \(-0.1\), or steep inverse spectra, around \(+2\), for optically thick. Pulsar wind nebulae (PWNe), or centre-filled supernova remnants, tend to have flatter spectra as well, usually within the range \(-0.3\leq\alpha\leq0.0\), though in rare cases they can be as steep as \(-0.7\) \citep{Kothes2017}. PWNe accelerate relativistic particles through a different mechanism, the interaction of the pulsar wind with the supernova ejecta, which typically results in a flatter particle energy distribution. Thus, PWNe can be more difficult to distinguish from thermal sources when they do not have a visible shell component. This is not the only inherent bias against detecting these types of sources as they are also generally smaller and have a less visually distinct morphology. In fact, only 3\% of catalogued Galactic SNRs are shell-less PWNe \citep{Dubner2017}.

We calculate spectral indices using data from GLEAM \citep{Wayth2015} for sources that are large enough and bright enough to be detected in the highest frequency band of the GLEAM survey. These appear to be sources that are at least 5' in diameter with a flux density of around 1~Jy in the GLEAM band. This includes all of the known supernova remnants in the field but only two of our SNR candidates. The GLEAM flux densities were calculated using the method outlined in Section~\ref{sec:fluxint}. These spectral indices can be found in Table~\ref{tab:2} with errors based on the uncertainties of the flux densities. Attempts were made to calculate in-band spectral indices with the ASKAP pilot II data but the uncertainties were too significant to produce meaningful results. Follow up observations at a second frequency would be valuable in allowing for the calculation of spectral indices for sources that are not visible in the GLEAM survey.

\subsubsection{Polarization} \label{sec:polarization}

Detection of linearly polarized radio emission is strong evidence that an extended Galactic radio nebula is a supernova remnant. SNRs emit highly linearly polarized synchrotron emission. The degree of polarization can be intrinsically more than 70\%. For the SNR G181.1$+$9.5, in both the Effelsberg 5~GHz observations and 1.4~GHz observations with the synthesis telescope at the Dominion Radio Astrophysical Observatory (DRAO ST), \cite{Kothes2017b} find polarization of about 70\%. G181.1$+$9.5 is a highly evolved SNR with a highly compressed and very regular magnetic field in its shell. Conversely, young SNRs typically have much lower intrinsic degrees of polarization as they display significant turbulence in their expanding shells. The lowest polarization observed at a high radio frequency in an SNR is possibly the 2\% observed in SNR G11.2$-$1.1 at a frequency of 32~GHz \citep{Kothes2001}.

EMU and POSSUM observe at radio frequencies between 800 and 1087~MHz. At these low frequencies, Faraday rotation strongly affects polarized signals. There is foreground Faraday rotation in the magneto-ionic medium between us and the nebula and there may be internal effects inside the SNR's shell. In the SNR's shell we find a mix of synchrotron emitting and Faraday rotating plasmas, which means that internally the synchrotron emission may be affected by different amounts of Faraday rotation depending on where the emission comes from within the shell. Integrating this emission along the line of sight through the emission region may lead to significant depolarization. \rev{These effects become significantly worse at low frequencies as the amount of depolarization is inversely proportional to the frequency squared.} Faraday rotation in the foreground ISM may also lead to depolarization, especially if the foreground path traverses turbulent ionized areas such as HII regions, or even spiral arms.

Because we are observing at a relatively low frequency, failure to detect polarization should not be considered evidence that a candidate is not an SNR, especially considering that many of our candidates are small or faint sources that may be located at far distances across the Galactic plane. The probability of detecting polarization from our SNR candidates is higher for high latitude sources, as their foreground likely does not contain any turbulent ionized regions. SNRs located close to the plane of the Galaxy may suffer from foreground depolarization caused by overlapping HII regions in the foreground or parts of a spiral arm between the SNR and us as spiral arms contain enhanced electron densities and magnetic fields.

Distinguishing between real and instrumental polarization is an additional challenge. As shown by the polarized intensity data in Figure~\ref{fig:PIfield}, we see evidence of instrumental effects at the positions of known HII regions. Particularly for bright sources, we believe that polarization with a smooth structure that closely resembles what is seen in total power is potentially the result of Stokes I leakage. Polarization that has a speckled appearance is more likely to be real as this indicates a changing rotation measure, or intrinsic polarization angle, on small scales. For further evidence of real polarization, we look for structures that are contained entirely within the SNR and that appear similar in the polarized intensity and rotation measure maps. Real polarization from the SNR should be distinct from what is seen in the surrounding background in intensity and rotation measure. 

We detect polarized signal significantly above the noise from all known SNRs in our field, but we are not convinced that all of it is real. Further details for each source are provided in Section~\ref{sec:knownSNRs}. \rev{Similarly to \cite{Dokara2018}, we are mostly unsuccessful in detecting polarization from Galactic SNR candidates.} We are only able to detect what we believe to be real polarization from one of our candidates, G328.0$+$0.7. The details of this can be found in the candidate description in Section~\ref{sec:snrCandidates}.

\begin{figure*}
    \centering
    \subfigure{\includegraphics[width=0.22\textwidth]{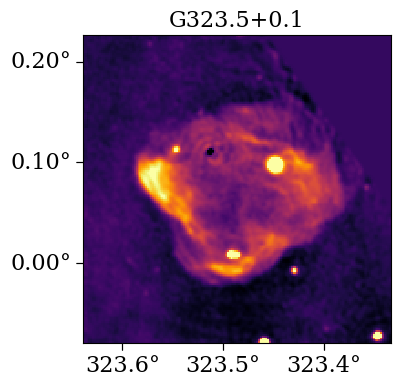}}
    \subfigure{\includegraphics[width=0.22\textwidth]{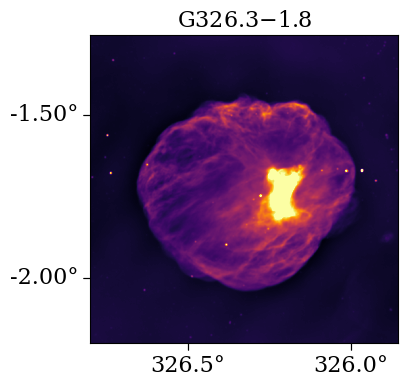}}
    \subfigure{\includegraphics[width=0.22\textwidth]{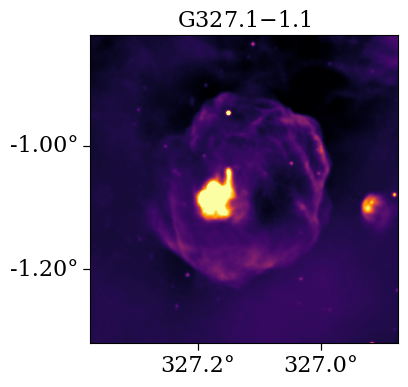}}
    \subfigure{\includegraphics[width=0.22\textwidth]{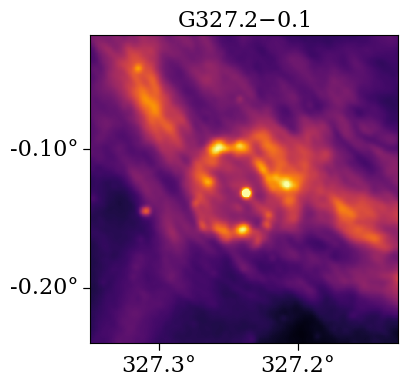}}
    \subfigure{\includegraphics[width=0.22\textwidth]{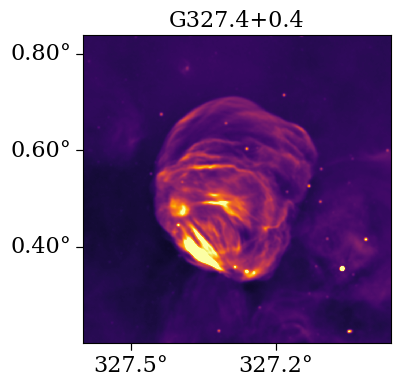}}
    \subfigure{\includegraphics[width=0.22\textwidth]{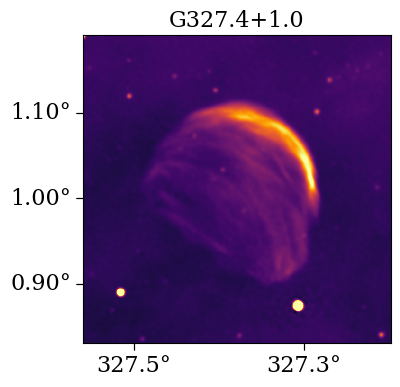}}
    \subfigure{\includegraphics[width=0.22\textwidth]{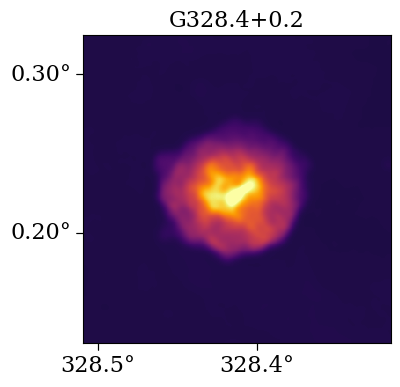}}
    \caption{933 MHz images of the 7 known SNRs with Galactic longitude on the x-axis and Galactic latitude on the y-axis. For more detailed images with colour bars, see Appendix~\ref{sec:appendixA}.}
    \label{fig:snrs}
\end{figure*}

\subsection{Characteristics of Known Supernova Remnants} \label{sec:knownSNRs}

 As discussed in Section~\ref{sec:radioMIR}, supernova remnants can be identified by looking for extended radio sources that lack a mid-infrared counterpart. The top images in Figure~\ref{fig:hii} show a known HII region that could potentially be misidentified as an SNR based solely on its radio morphology. The presence of a clear counterpart in the MIR helps to correctly identify this source as an HII region. The bottom images in Figure~\ref{fig:hii} show a known SNR, G327.1$-$1.1, and its clear lack of MIR emission. A small HII region can be seen along the right edge of the SNR images in both the radio and the MIR, further illustrating this distinction. 

There are eight known supernova remnants in the EMU/POSSUM Galactic pilot II field that appear in the \cite{Green2022} catalogue. We believe G323.7$-$1.0 should be reclassified as three separate sources so it is discussed under the candidates section. Our observations of the other seven known SNRs are discussed here. The 933 MHz ASKAP images of the SNRs can be found in Figure~\ref{fig:snrs}. More detailed images of the known SNRs in both radio and MIR, which include coordinate axes and colour bars, can be found in Appendix~\ref{sec:appendixA}.

\begin{figure*}
\centering
\subfigure{
    \includegraphics[width=0.49\textwidth]{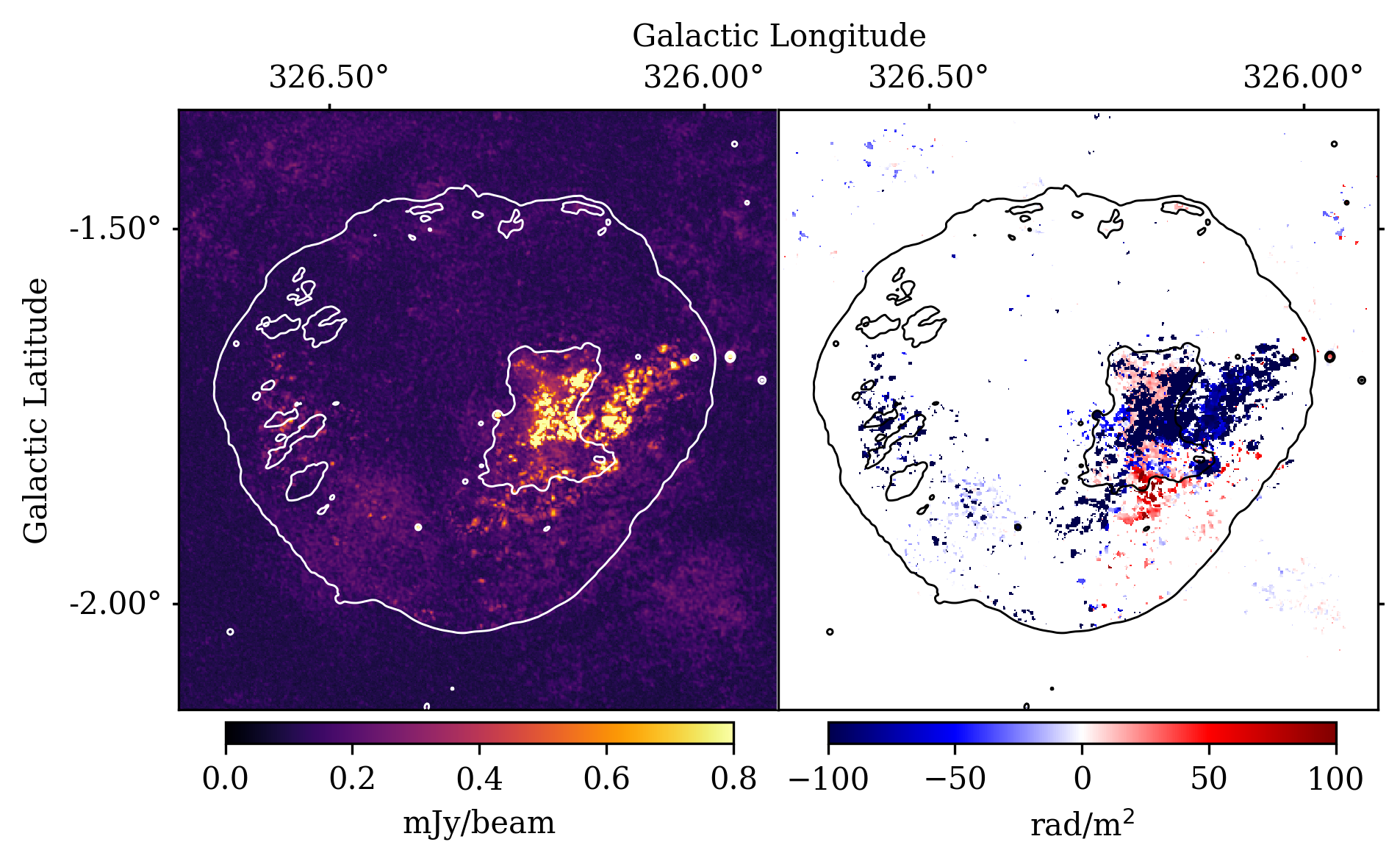}\label{fig:pol1}}
\subfigure{
    \includegraphics[width=0.49\textwidth]{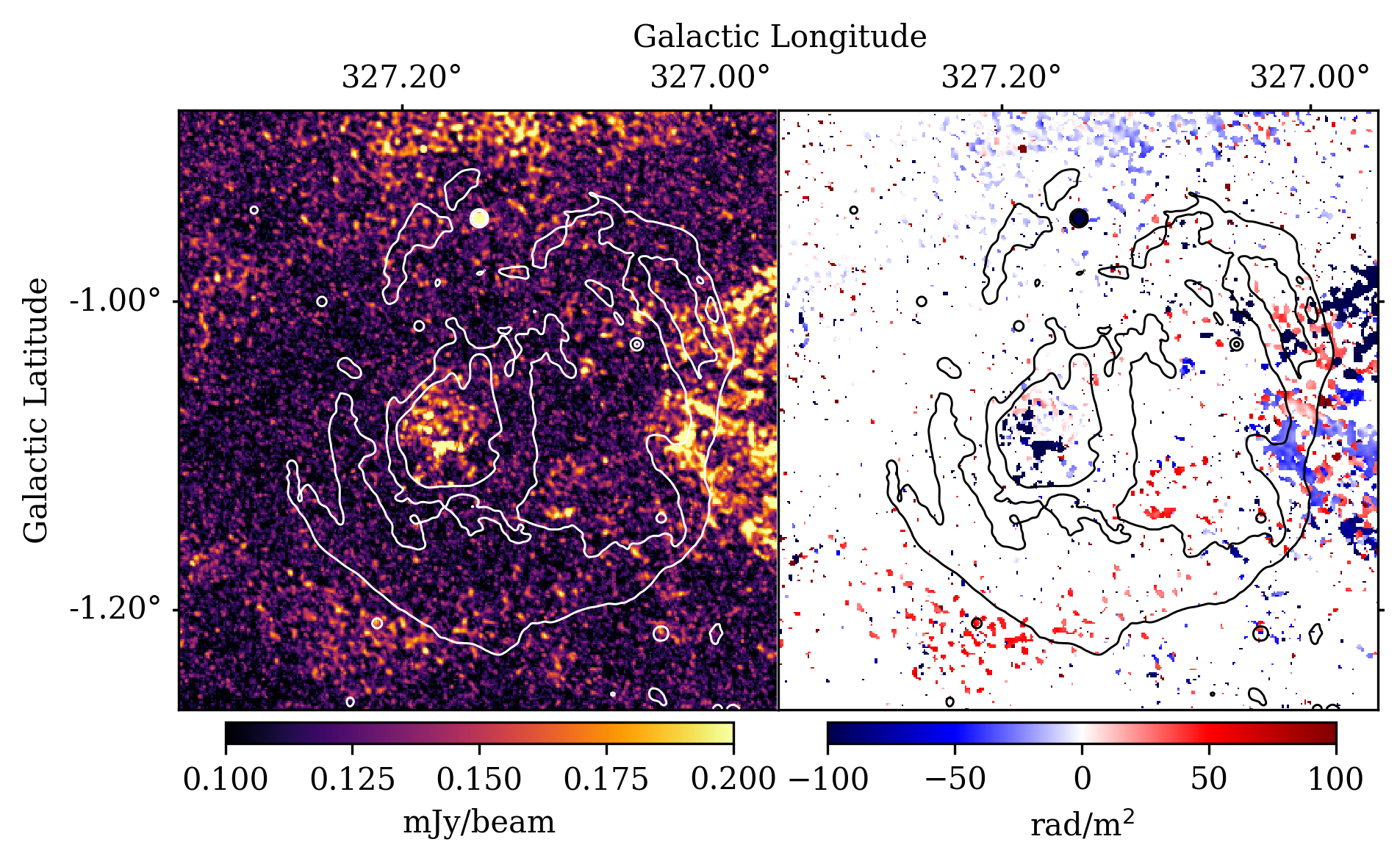}\label{fig:pol2}}
\subfigure{
    \includegraphics[width=0.49\textwidth]{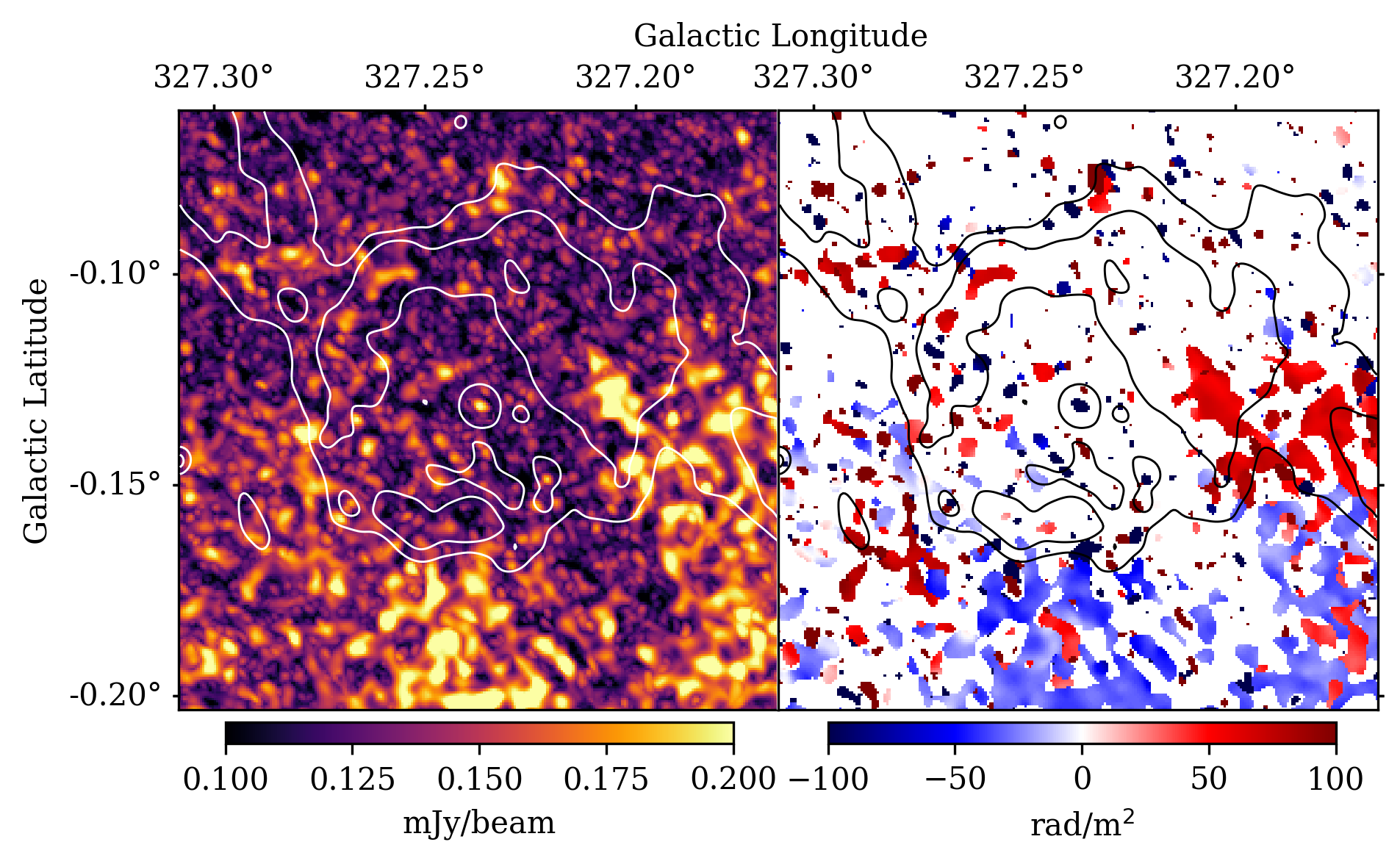}\label{fig:pol3}}
\subfigure{
    \includegraphics[width=0.49\textwidth]{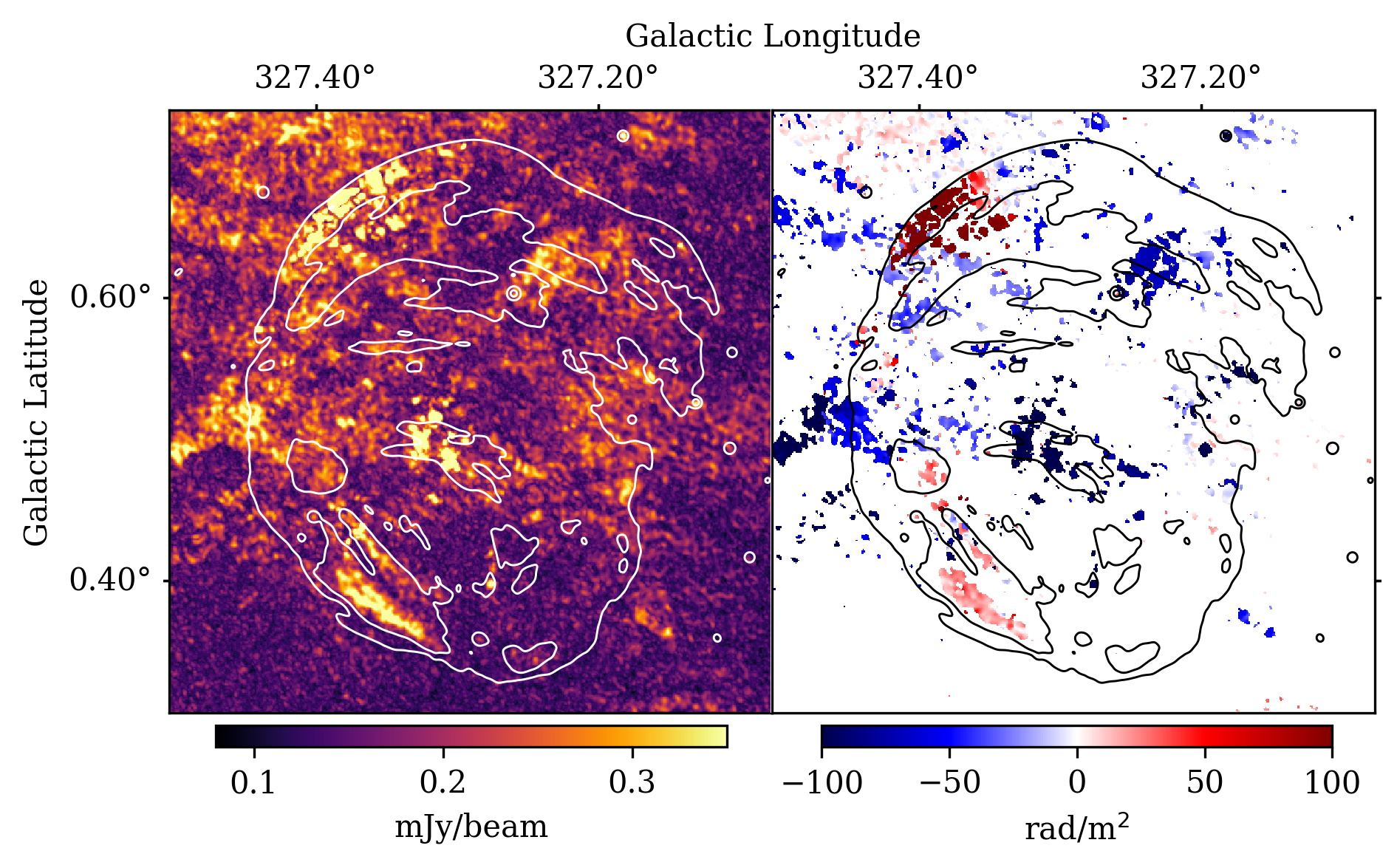}\label{fig:pol4}}
\subfigure{
    \includegraphics[width=0.49\textwidth]{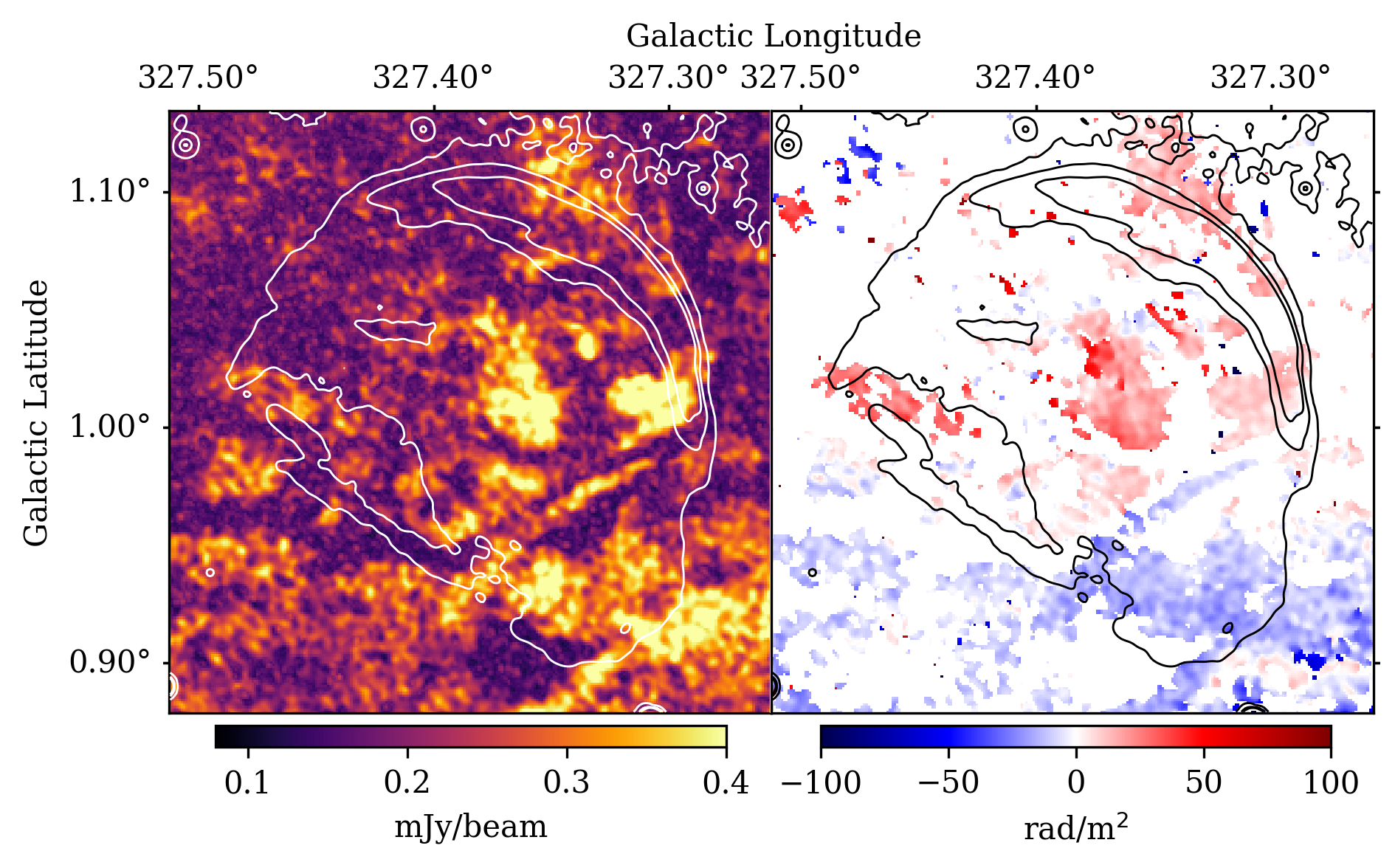}\label{fig:pol5}}
\subfigure{
    \includegraphics[width=0.49\textwidth]{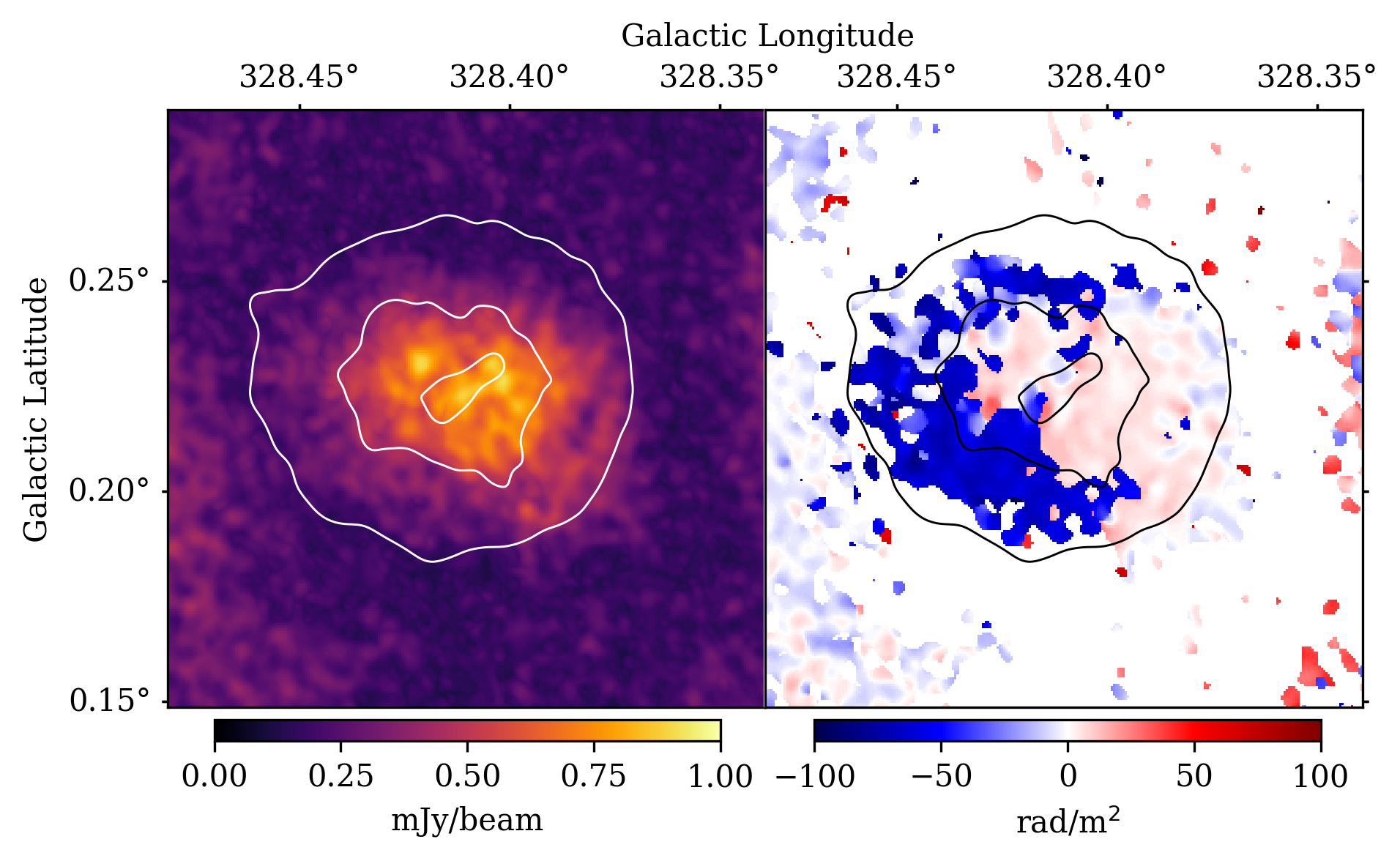}\label{fig:pol6}}
\caption{For each source, the left image shows polarized intensity and the right image shows Faraday rotation measure. The contours are used to indicate the total power structures. From left to right, Top row: G326.3$-$1.8, G327.1$-$1.1. Middle row: G327.2$-$0.1, G327.4$+$0.4. Bottom row: G327.4$+$1.0, G328.4$+$0.2.}
\label{fig:pol}
\end{figure*}

\subsubsection{G323.5$+$0.1}
This source is a shell-type supernova remnant that lies at the edge of our field. It is only fully imaged in the ASKAP data at low frequencies so it has not been studied here in-depth and thus does not appear in Table~\ref{tab:2}. Figure~\ref{fig:7} was made using a 48 MHz wide channel centred at 823.5 MHz since the source is not fully visible in the higher frequency channels. The source overlaps with a bright HII region that can be seen in MIR.

\subsubsection{G326.3$-$1.8, MSH\,15$-$56}
 This SNR is a composite source, a bright pulsar wind nebula with a well-defined radio shell. It is the largest supernova remnant in the field with a size of \(38'\). Bright filamentary emission can be seen coming from the shell. The PWN is offset to the west of the remnant's centre and is elongated in the north-south direction (Figure~\ref{fig:1}). The source is estimated to be at a distance of \(3.5-5.8\) kpc \citep{Ranasinghe2022}. There is clear polarization, mostly concentrated around and extending from the PWN (Figure~\ref{fig:pol}). This can be seen in both the polarized intensity and rotation measure maps. There is also some polarization coming from the east side of the shell. We believe all of this polarization is real since it is not smooth and aligns with what is expected from total power without mirroring it exactly, which could indicate leakage. It also has a high negative RM and there is almost no polarized emission coming from the background in this part of the field it could be confused with. 

\subsubsection{G327.1$-$1.1}
This SNR is another composite source with a bright pulsar wind nebula and a relatively faint shell that is likely missing some short spacings in the ASKAP data. This is evidenced by the negative bowls of emission, seen in Figure~\ref{fig:2}. There is a bright HII region located to the west of the SNR. The remnant is believed to be located at a distance of \(4.5-9\) kpc \citep{Wang2020,Sun1999}, indicating it is likely located within or near the Norma arm. There is obvious polarization coming from the PWN but no clear polarization coming from the shell (Figure~\ref{fig:pol}). We believe this polarization to be real because it is contained entirely within the PWN and is not smooth.

\subsubsection{G327.2$-$0.1}
This source is believed to be a shell-type remnant associated with a young magnetar, J1550-5418, located near its centre \citep{Gelfand2007}. The magnetar has a characteristic age of 1.4 kyrs and a rotation measure of \(-1860\pm 20\) rad/m$^{2}$ \citep{Camilo2007,Camilo2008}. Distance estimates for the shell are between \(4-5\) kpc \citep{Tiengo2010} while the magnetar has been estimated to be at a distance of 9 kpc \citep{Camilo2007}. The source lies on a larger filament of emission that is likely unrelated to the SNR (Figure~\ref{fig:3}). While we were not able to detect any clear real polarization coming from the shell due to confusion with the background, there does seem to be real polarization coming from the centre (Figure~\ref{fig:pol}). Specifically, there are two peaks in the rotation measure for the central point source. We believe the first peak, around \(-1820\) rad/m$^{2}$, comes from the pulsar and we speculate that the second peak, around \(15-30\) rad/m$^{2}$, may come from a previously undetected pulsar wind nebula.

\begin{figure*}
    \subfigure{\includegraphics[width=0.33\textwidth]{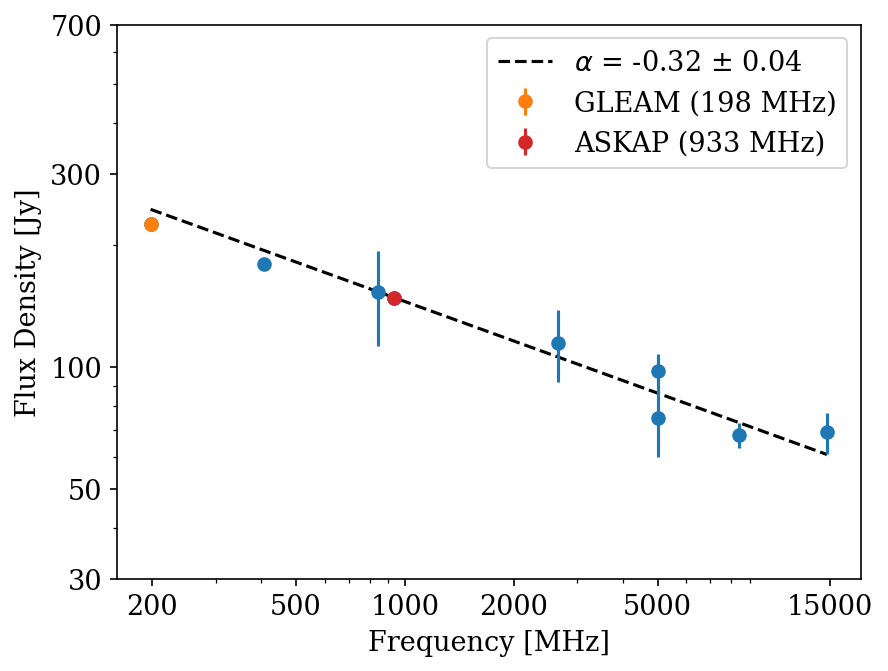}\label{fig:index1}} 
    \subfigure{\includegraphics[width=0.32\textwidth]{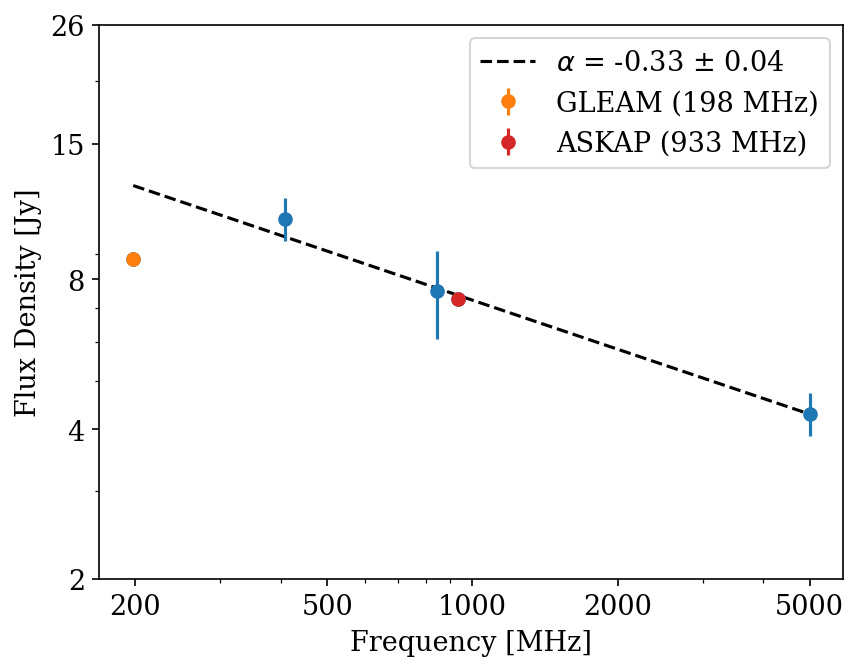}\label{fig:index2}} 
    \subfigure{\includegraphics[width=0.33\textwidth]{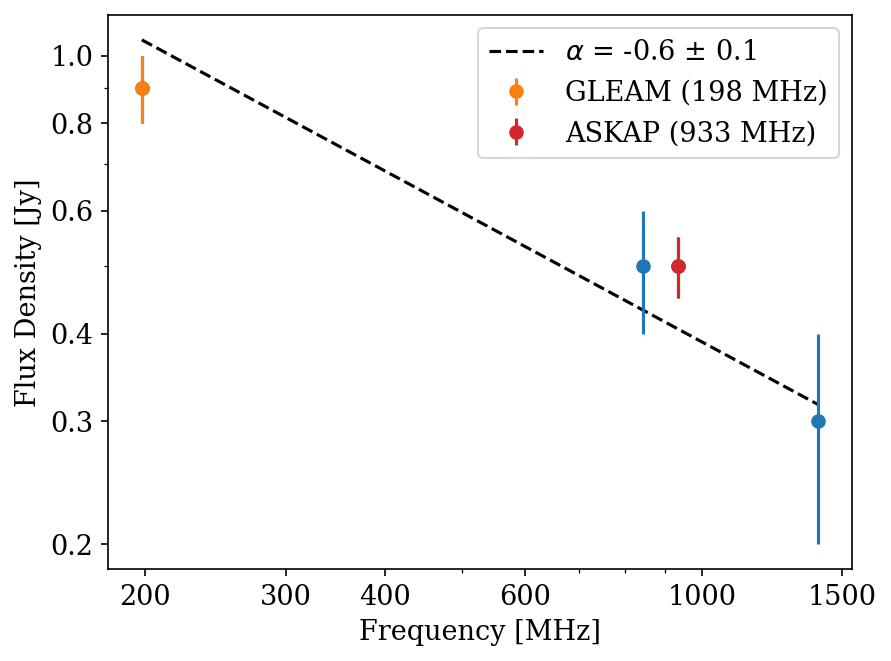}\label{fig:index3}} 
    \subfigure{\includegraphics[width=0.33\textwidth]{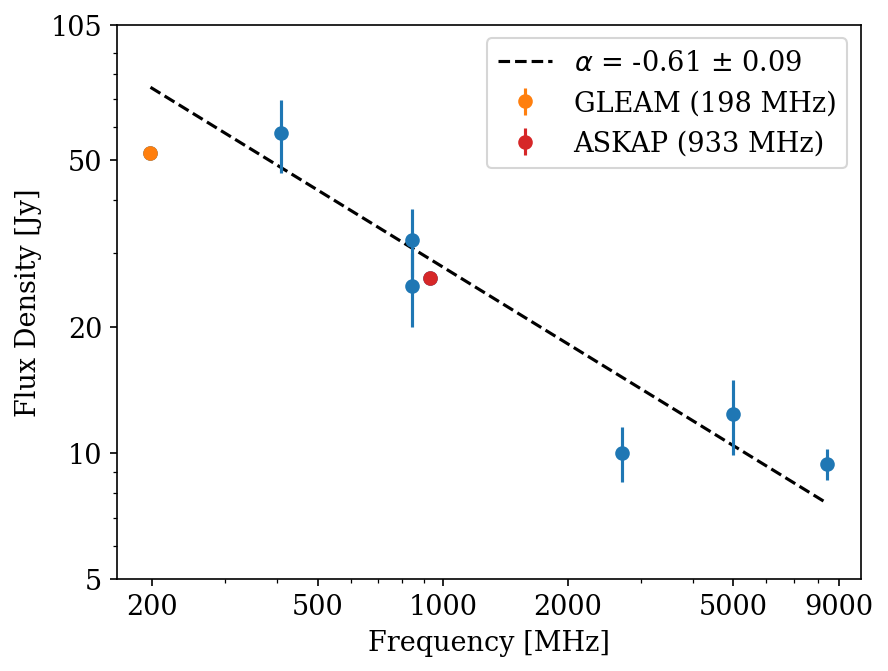}\label{fig:index4}} 
    \subfigure{\includegraphics[width=0.315\textwidth]{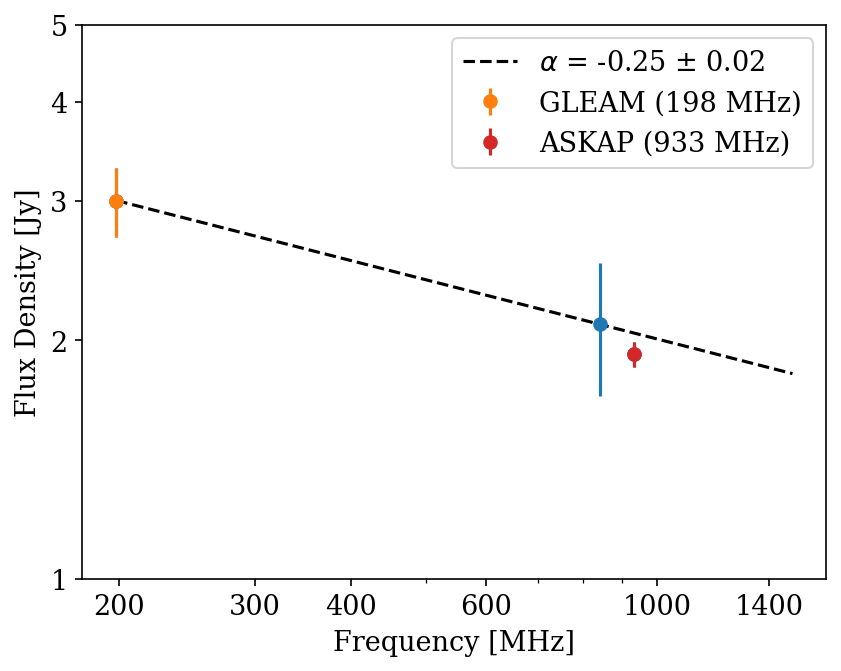}\label{fig:index5}} 
    \subfigure{\includegraphics[width=0.32\textwidth]{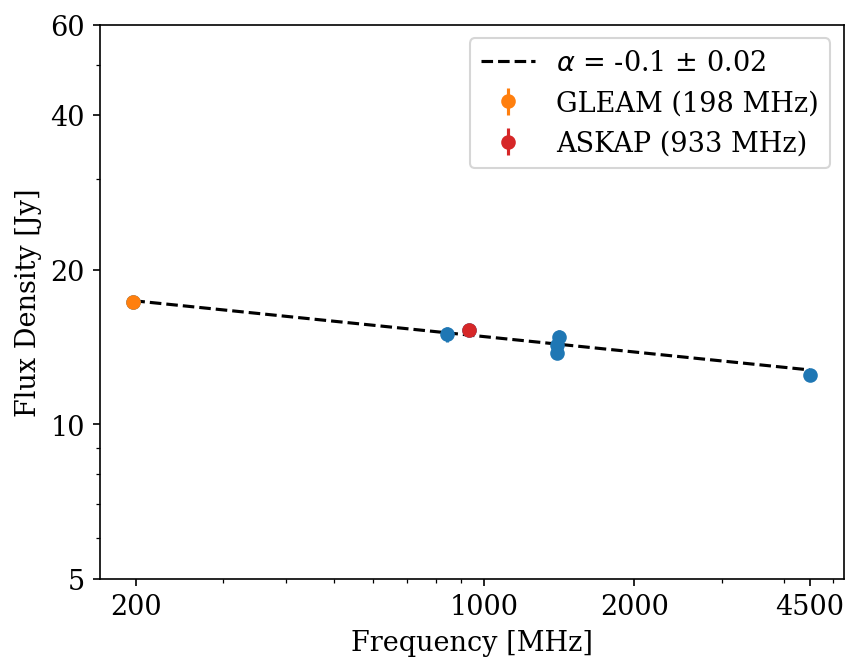}\label{fig:index6}}
    \caption{Flux densities of known SNRs. From left to right, Top row: G326.3$-$1.8, G327.1$-$1.1, G327.2$-$0.1. Bottom row: G327.4$+$0.4, G327.4$+$1.0, G328.4$+$0.2. The red and orange data points are the values calculated in this paper using ASKAP and GLEAM data respectively. The blue data points represent the flux densities of known SNRs taken from \protect\cite{Green2019} and references therein. 20$\%$ errors are assumed if none were given. $\alpha$ represents the spectral index as determined by the slope of the linear fit.}
    \label{fig:indices}
\end{figure*}

\subsubsection{G327.4$+$0.4, Kes\,27}
This shell-type SNR exhibits multiple shell structures and many internal filaments. There is a small overlapping HII region that can be seen in both radio and MIR (Figure~\ref{fig:4}). HI absorption suggests the SNR is located at a distance of \(4.3-5.4\) kpc \citep{McClure2001} while optical extinction suggests a distance of 2.8 kpc \citep{Wang2020}. We detect polarization coming from several parts of the remnant (Figure~\ref{fig:pol}). We believe the polarization seen along the northeast edge of the remnant to be real as it is distinct from the total power structure, and thus cannot be leakage, and has a high positive rotation measure that is distinct from the RM found in the rest of the image. The polarization in the southeast may be real as well but this is not certain as the smoother appearance and flatter RM may indicate that it is instrumental. There is also some polarization near the centre of the remnant that has a high negative RM and may be real but this is also unclear. 

\subsubsection{G327.4$+$1.0}
This source is an asymmetrical shell that is brightest along the northwestern edge. There is also some faint central emission with filaments that curve in the same direction as the shell (Figure~\ref{fig:5}). There is potential polarization coming from the centre of the remnant but it is not strong enough to be definitively distinct from the background features seen in the south (Figure~\ref{fig:pol5}). However, the positive RM seems to be mostly confined to the source with some extending to the north of the remnant. This extension is somewhat mirrored in total power, possibly indicating that some of this polarization is real but this is not conclusive.

\subsubsection{G328.4$+$0.2, MSH\,15$-$57}
 This SNR is believed to be the largest and most radio luminous pulsar wind nebula in our galaxy \citep{Gaensler2000}. It has no visible shell but there is a central bar structure that runs in the southeast to northwest direction (Figure~\ref{fig:6}). It is believed to be located at a distance of over 16.7 kpc \citep{Ranasinghe2022} placing it along the outer edge of the Galaxy. The source appears to be polarized but because it is so bright we believe this is likely the result of leakage. In RM we see two components, a high negative component and a low positive component. The high negative component extends over the remnant and has a similar structure to what is seen in PI. The low positive component seems to be concentrated to the west side of the PWN (Figure~\ref{fig:pol}).

\subsection{Spectral Indices of Known Remnants}  \label{sec:knownIndices}

For each of the known SNRs, we calculate a spectral index using the 933 MHz flux density from ASKAP and the 198 MHz flux density value from GLEAM. These indices can be found in Table~\ref{tab:2}. Figure~\ref{fig:indices} shows the ASKAP and GLEAM flux densities calculated in this paper plotted with flux densities taken from \cite{Green2019} and references therein. The spectral indices seen on the plots are calculated using the slopes of the fit lines, which are made using the literature values as well as the ASKAP and GLEAM values. We use this second method of calculating spectral indices to evaluate whether or not the indices found using only the ASKAP and GLEAM data are reliable. 

G326.3$-$1.8 is a composite source that has been shown to have an intermediate index ($\sim$$-$0.3) with a flatter component coming from the PWN and a steeper component coming from the shell \citep{Dickel2000}. The values we obtain are in moderate agreement with each other and are consistent with the expected value for this source. 

G327.1$-$1.1 is also a composite source that is believed to have an intermediate index \citep[$-$0.36,][]{Clark1975}. The flux value we obtain from the GLEAM data appears to be too low, as demonstrated by the plot, resulting in a spectral index that is likely too flat. The value obtained from the fit is consistent with what is expected and is likely more reliable. 

G327.2$-$0.1 is believed to be a shell-type remnant though it is associated with a known magnetar and could be a composite source. The values we obtain are in moderate agreement though they have relatively large errors. Both are consistent with what is expected for non-thermal emission from a shell-type remnant. 

G327.4$+$0.4 is also a shell-type remnant. The values we obtain are not in agreement though both are consistent with the expected index of a shell-type remnant. The plot appears to indicate that the GLEAM flux value may be too low. 

G327.4$+$1.0 is a shell-type remnant for which there was previously only one flux measurement \citep{Whiteoak1996} so the spectral index derived from the linear fit may be less reliable. The values we obtain are consistent with each other and although they are flat for a shell-type remnant, they are not unreasonable for this type of source. Given that many of the GLEAM fluxes seem to be lower than expected, and because our fit is based on only three data points, the spectral index of this source may be steeper than the values we have calculated here. 

G328.4$+$0.2 is a shell-less PWN that has been shown to have a flat spectral index \citep{Gaensler2000,Gelfand2007b}. The values we obtain are consistent with each other and with what is expected for a typical pulsar wind nebula. 

While the spectral indices we calculate using these two methods are generally in agreement with each other and with the expected values for the corresponding SNR types, many of these values have relatively large errors. The values obtained from the linear fit should be considered to be more reliable than the values provided in Table~\ref{tab:2} as the deviations seem to mostly be the result of lower than expected values for the GLEAM fluxes, particularly for the fainter remnants. Thus, the spectral indices we calculate for our candidates using this data should be viewed with a level of skepticism. High resolution data at a second frequency will likely be required to produce reliable spectral indices for the SNR candidates.

\section{SNR Candidates} \label{sec:candidates}

The locations of our SNR candidates are indicated in Figure~\ref{fig:field2}. Notably, the candidates are highly concentrated in the lower right corner of the image. We believe we were able to detect more candidates in this part of the field because of the relatively low density of HII regions compared to the upper left half of the image. This may seem to contradict what is expected as the true SNR density is likely to be higher within spiral arms, not away from them. Thus, we believe it is highly probable that there are faint sources within the Norma arm region of the field that we were unable to detect due to the high concentration of thermal emission and HII regions. 

Data collected for known and candidate supernova remnants is shown in Table~\ref{tab:2}. Right ascension and declination are determined by fitting an ellipse to the source using CARTA software \citep{Comrie2021} and taking the central coordinates of the ellipse. The sizes of the sources are determined by taking the major and minor axes of these ellipses.

\begin{center}
\begin{table*}
  \begin{tabular}{ |cccccccc| }
    \hline
    Name & RA & Dec & Size & Flux Density  & Surface Brightness & Polarization & Spectral Index\\
     & (J2000) & (J2000) & ['] & [Jy at 933 MHz] & \([10^{-21} \rm W m^{-2} Hz^{-1} sr^{-1}]\) & Detected & (with GLEAM)\\
    \hline
    \multicolumn{8}{c}{\rev{Known SNRs}}\\
    \hline
    G326.3$-$1.8 & 15:52:59 & $-$56:07:27 & 38 & \(148 \pm 3\) & \(15.4\pm0.3\)& Y & \(-0.27 \pm 0.02\)\\
    G327.1$-$1.1 & 15:54:26 & $-$55:05:59 & 18 & \(7.3 \pm 0.2\) & \(3.4\pm0.1\) & Y & \(-0.1 \pm 0.1\)\\
    G327.2$-$0.1 & 15:50:57 & $-$54:18:00 & 5 & \(0.50 \pm 0.05\) & \(3.0\pm0.3\) & Y? & \(-0.4 \pm 0.1\)\\
    G327.4$+$0.4 & 15:48:23 & $-$53:46:13 & 21 & \(26.1 \pm 0.4\)  & \(8.9\pm0.1\) & Y & \(-0.44 \pm 0.01\)\\
    G327.4$+$1.0 & 15:46:53 & $-$53:19:51 & 14 & \(1.92 \pm 0.07\)& \(1.47\pm0.05\)& Y? & \(-0.3 \pm 0.1\)\\
    G328.4$+$0.2 & 15:55:32 & $-$53:17:02 & 5 & \(15.3 \pm 0.3\) & \(92\pm2\)& N/A** & \(-0.1 \pm 0.1\)\\
    \hline
    \multicolumn{8}{c}{\rev{SNR Candidates}}\\
    \hline
    G323.2$-$1.0 & 15:31:40 & $-$57:23:42 & 7 & \(0.32 \pm 0.01\) & \(0.97\pm0.04\) & N & \\
    G323.6$-$1.1* & 15:34:03 & $-$57:21:42 & \(34 \times 20\) & N/A & N/A & N & \\ 
    G323.6$-$0.8* & 15:32:56 & $-$57:02:53 & \(28 \times 23\) & \(< 1\) & \(< 0.3\) & N & \\
    G323.7$+$0.0 & 15:30:39 & $-$56:19:26 & 3 & \(0.07 \pm 0.02\) & \(1.1\pm0.3\)& N & \\
    G323.9$-$1.1* & 15:36:24 & $-$57:08:45 & \(29 \times 18\) & \(< 1\) & \(< 0.3\) & N & \\
    G324.1$-$0.2 & 15:33:34 & $-$56:13:59 & \(10 \times 9\) & \(0.26 \pm 0.03\) & \(0.43\pm0.06\)& N & \\
    G324.1$+$0.0 & 15:32:37 & $-$56:03:05 & \(11 \times 7\) & \(1.0 \pm 0.1\) & \(1.9\pm0.2\) & N & \(-0.3 \pm 0.2\)\\
    G324.3$+$0.2 & 15:32:45 & $-$55:47:45 & 4 & \(0.120 \pm 0.008\) & \(1.13\pm0.08\)& N & \\
    G324.4$-$0.4 & 15:36:00 & $-$56:14:08 & \(18 \times 13\) & \(0.44 \pm 0.09\) & \(0.28\pm0.06\)& N & \\
    G324.4$-$0.2 & 15:35:26 & $-$56:04:08 & \(3 \times 2\) & \(0.006 \pm 0.001\) & \(0.14\pm0.02\)& N & \\
    G324.7$+$0.0 & 15:36:06 & $-$55:46:29 & 4 & \(0.055 \pm 0.005\) & \(0.52\pm0.04\)& N & \\
    G324.8$-$0.1 & 15:37:07 & $-$55:45:03 & \( 25 \times 20\) & \(< 1\) & \(< 0.3\) & N & \\
    G325.0$-$0.5 & 15:39:43 & $-$55:58:29 & \(34 \times 28\) & \(< 6\) & \(< 0.9\) & N & \\
    G325.0$-$0.3 & 15:39:14 & $-$55:49:52 & 4 & \(0.19 \pm 0.01\) & \(1.7\pm0.1\)& N & \\
    G325.0$+$0.2 & 15:37:02 & $-$55:26:56 & 5 & \(0.11 \pm 0.01\) & \(0.66\pm0.08\) & N & \\
    G325.8$-$2.1 & 15:52:03 & $-$56:46:04 & \(36 \times 30\) & N/A & N/A & N & \\
    G325.8$+$0.3 & 15:41:15 & $-$54:54:26 & 28 & N/A & N/A & N & \\
    G327.1$+$0.9 & 15:45:59 & $-$53:32:33 & 2 & \(0.019 \pm 0.004\) & \(0.7\pm0.1\)& N & \\
    G328.0$+$0.7 & 15:51:41 & $-$53:10:11 & 8 & \(0.50 \pm 0.06\) & \(1.2\pm0.1\) & Y? & \\
    G328.6$+$0.0 & 15:57:31 & $-$53:22:05& \(34 \times 19\) & \(3.5 \pm 0.9\) & \( 0.8\pm0.2\) & N & \(-0.75 \pm 0.06\)\\
    G330.2$-$1.6 & 16:12:32 & $-$53:27:01 & 9 & \(0.08 \pm 0.02\) & \(0.14\pm0.04\)& N & \\
    \hline
  \end{tabular}
  \caption{Known SNRs and SNR candidates in the EMU/POSSUM Galactic pilot II field. Surface brightnesses are given at 933 MHz. Spectral indices are calculated using the 933 MHz flux value from ASKAP and the 198 MHz flux value from GLEAM. *Currently classified as a single source (G323.7$-$1.0). **Leakage.}
  \label{tab:2}
\end{table*}
\end{center}

\subsection{Characteristics of SNR Candidates} \label{sec:snrCandidates}

\begin{figure*}
    \centering
    \subfigure{\includegraphics[width=0.195\textwidth]{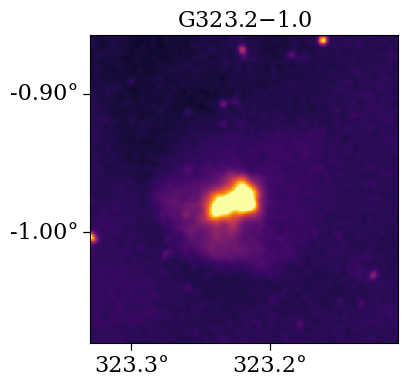}}
    \subfigure{\includegraphics[width=0.195\textwidth]{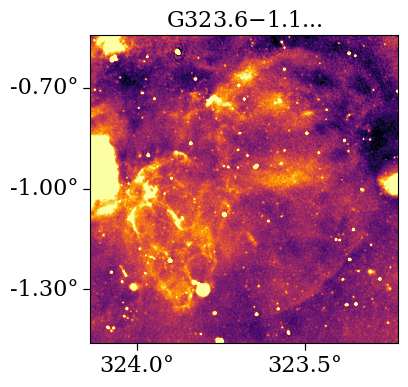}}
    \subfigure{\includegraphics[width=0.195\textwidth]{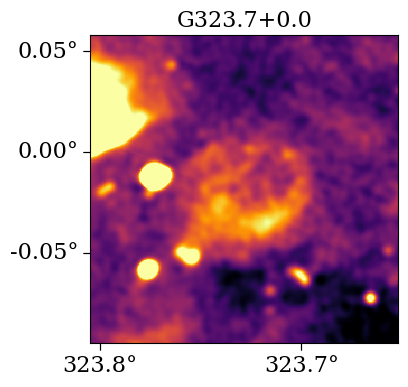}}
    \subfigure{\includegraphics[width=0.195\textwidth]{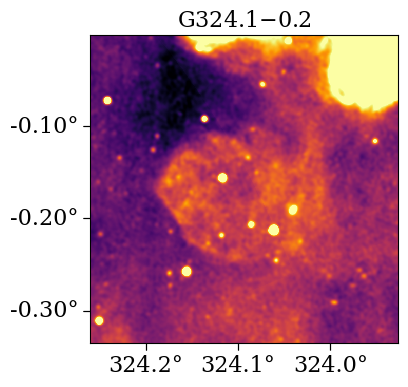}}
    \subfigure{\includegraphics[width=0.195\textwidth]{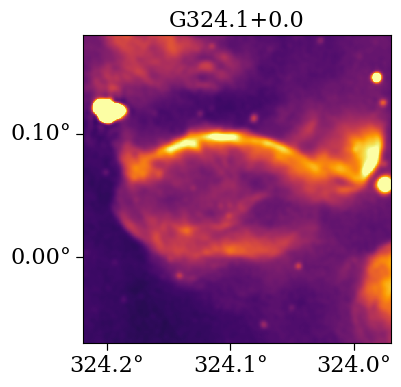}}
    \subfigure{\includegraphics[width=0.195\textwidth]{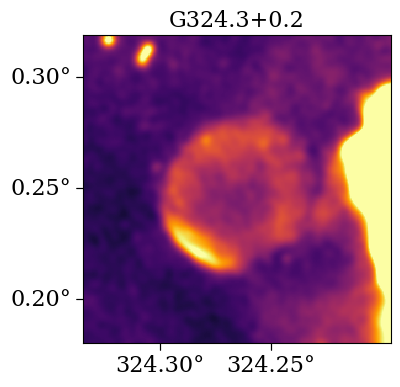}}
    \subfigure{\includegraphics[width=0.195\textwidth]{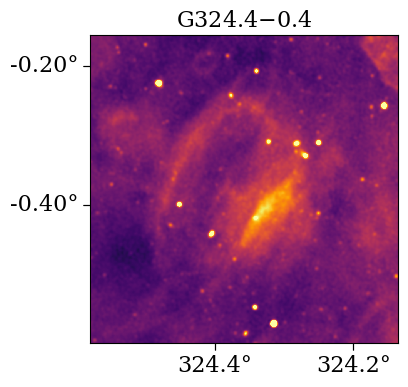}}
    \subfigure{\includegraphics[width=0.195\textwidth]{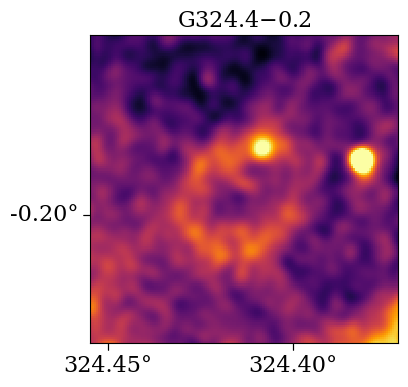}}
    \subfigure{\includegraphics[width=0.195\textwidth]{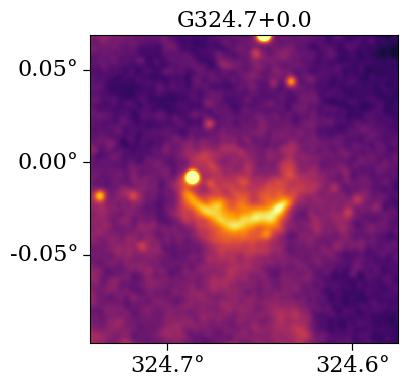}}
    \subfigure{\includegraphics[width=0.195\textwidth]{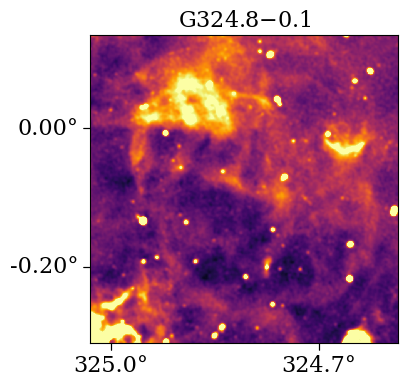}}
    \subfigure{\includegraphics[width=0.195\textwidth]{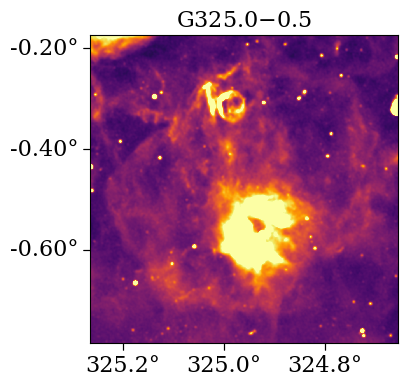}}
    \subfigure{\includegraphics[width=0.195\textwidth]{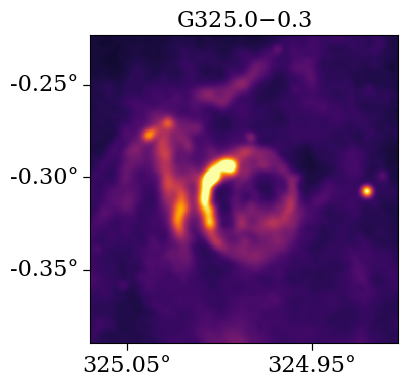}}
    \subfigure{\includegraphics[width=0.195\textwidth]{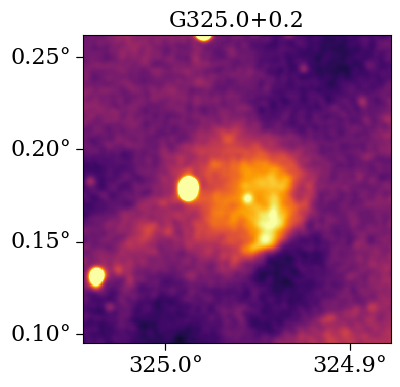}}
    \subfigure{\includegraphics[width=0.195\textwidth]{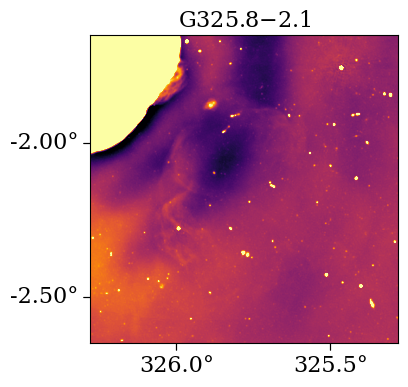}}
    \subfigure{\includegraphics[width=0.195\textwidth]{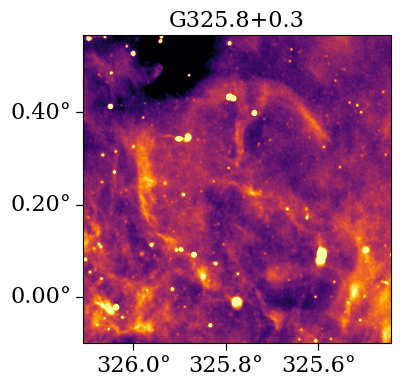}}
    \subfigure{\includegraphics[width=0.195\textwidth]{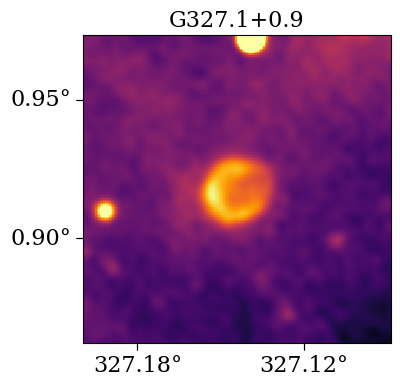}}
    \subfigure{\includegraphics[width=0.195\textwidth]{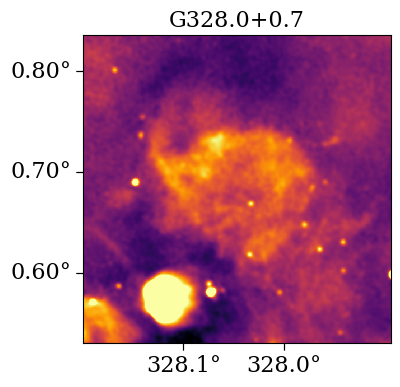}}
    \subfigure{\includegraphics[width=0.195\textwidth]{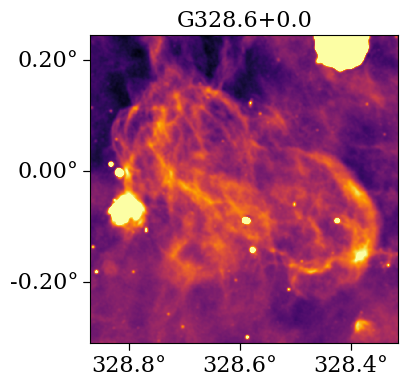}}
    \subfigure{\includegraphics[width=0.195\textwidth]{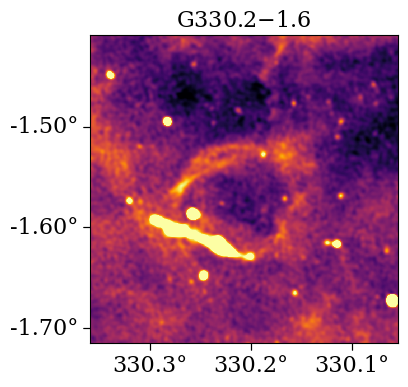}}
    \caption{933 MHz images of the SNR candidates with Galactic longitude on the x-axis and Galactic latitude on the y-axis. For more detailed images with colour bars, see Appendix~\ref{sec:appendixB}.}
    \label{fig:candidates}
\end{figure*}

Here we list 21 supernova remnant candidates, three times the number of known SNRs in this field. We believe some of these sources to be strong SNR candidates while others are weaker and will require further observations to determine if they are indeed SNRs. We define the strength of our candidates based on whether or not we are able to find evidence of other expected SNR properties, such as polarization or a steep negative spectral index. Only two of the candidates were visible in the GLEAM data, allowing us to calculate spectral indices, and only one shows clear evidence of real polarization. The 933 MHz images of the candidates can be found in Figure~\ref{fig:candidates}. More detailed images of the SNR candidates in both radio and MIR, which include coordinate axes and colour bars, can be found in Appendix~\ref{sec:appendixB}.

\paragraph*{G323.2$-$1.0}
 This source has the morphology of a pulsar wind nebula with a very faint shell. As shown in Figure~\ref{fig:13}, there is an infrared source that overlaps with the PWN but the IR source has a different morphology and has been identified as a star by \cite{Cutri2003}. The shell is roughly circular and the PWN has a similar shape to the PWN of G326.3$-$1.8, but elongated along the east-west axis. This source was listed as an SNR candidate by \cite{Whiteoak1996} but was not confirmed and no image was provided. 
 
\paragraph*{G323.6$-$1.1, G323.6$-$0.8, G323.9$-$1.1}
 This object is currently classified as a single source in Green's catalogue, G323.7$-$1.0 \citep{Green2014}. The ASKAP observations have revealed faint filamentary structures that were previously not visible, leading us to believe that it is actually three separate overlapping sources. As shown in Figure~\ref{fig:8}, there are two brighter, elliptically shaped sources, one located to the northwest (G323.6$-$0.8) and the other to the southeast (G323.9$-$1.1). The third source (G323.6$-$1.1) appears to lie behind the other two sources and only a faint southwestern edge can be seen. Because this source is very faint and overlaps with the other sources, it was not possible to obtain a flux estimate. For the two brighter sources, only rough upper limits could be obtained for the flux densities. It is unclear if all three sources are supernova remnants but none of them has an obvious MIR counterpart.
 
 \paragraph*{G323.7$+$0.0}
 The source shown in Figure~\ref{fig:21} is a small shell-like structure that is roughly circular and brightest along the southern edge. There is a small amount of overlap with a known HII region in the southeast and a larger HII region can be seen to the northeast. There is some faint MIR emission to the east that could be related to the radio emission. Since the emission is faint and the relation is not entirely clear, we include the source in our list of candidates.
 
 \paragraph*{G324.1$-$0.2}
 This candidate does not have a clear shell-structure but it has a roughly circular shape with well-defined rounded edges, characteristic of a shock front (Figure~\ref{fig:15}). There are several overlapping point sources and a bright HII region can be seen in the northwest. There is no clear MIR counterpart.
 
 \paragraph*{G324.1$+$0.0}
 This source was originally identified as an SNR candidate by \cite{Whiteoak1996} but an image was not included. \cite{Green2014} have also listed it as a candidate and provided an image but there was insufficient evidence for it to be included in the \cite{Green2022} catalogue. The source, seen in Figure~\ref{fig:10}, is an elliptical shell, elongated in the east-west direction, with the brightest emission coming from the north and a fainter shell visible in the south. Multiple HII regions can be seen in the image and there is some overlap between the SNR candidate and HII regions in the northwest and northeast. This candidate is visible in the GLEAM data and a spectral index of \(-0.3 \pm 0.2\) was determined. The index indicates the emission may be nonthermal but the uncertainty is too large to be conclusive. Because this source has been studied previously and has a very clear shell-like morphology, we believe it should be classified as an SNR.
 
 \paragraph*{G324.3$+$0.2}
 The source shown in Figure~\ref{fig:12} has a shell structure that is almost perfectly circular, with some brightening towards the southeast. Based on the distinct morphology and clear lack of an MIR counterpart, we believe this candidate should be classified as an SNR. A large, bright HII region can be seen to the west of the source. 
 
 \paragraph*{G324.4$-$0.4}
 In Figure~\ref{fig:14} we see an elliptical shell with brightening along the southwest edge. There are many overlapping point sources and a couple of known HII regions located to the northeast. There is overlapping emission in the MIR but nothing that clearly mirrors the shell structure seen in the radio.
 
 \paragraph*{G324.4$-$0.2}
 The source shown in Figure~\ref{fig:22} is a very small, faint shell-like structure with an overlapping point source. It is located just north of G324.4$-$0.4 and can also be seen in Figure~\ref{fig:14}. There is no obvious MIR counterpart.
 
 \paragraph*{G324.7$+$0.0}
 This candidate is a partial shell that arcs in the southern direction with no clear northern counterpart though some faint emission can be seen extending north from the southern shell (Figure~\ref{fig:18}). There is a bright overlapping point source in the east. Several bright point sources can be seen in the MIR but the shell has no counterpart. 
 
 \paragraph*{G324.8$-$0.1}
 As shown in Figure~\ref{fig:20}, this source consists of faint filaments that form a large ellipse. The source overlaps with G324.7$+$0.0 and with a large HII region in the northeast. Because of this overlap, only a rough upper limit could be obtained for the flux. The rounded filaments do not appear to have an MIR counterpart. 
 
 \paragraph*{G325.0$-$0.5}
 This candidate, shown in Figure \ref{fig:19}, is composed of filaments that form a roughly elliptical structure. Many overlapping sources can be seen, including G325.0$-$0.3 and several HII regions. A long filament, which can be seen in MIR and is thus likely thermal, runs from the northeast to the southwest. The filaments that form the edges of the ellipse are not visible in MIR. Because of the overlapping emission, only an upper limit could be obtained for the flux.

\begin{figure*}
     \centering
     \includegraphics[width=0.75\textwidth]{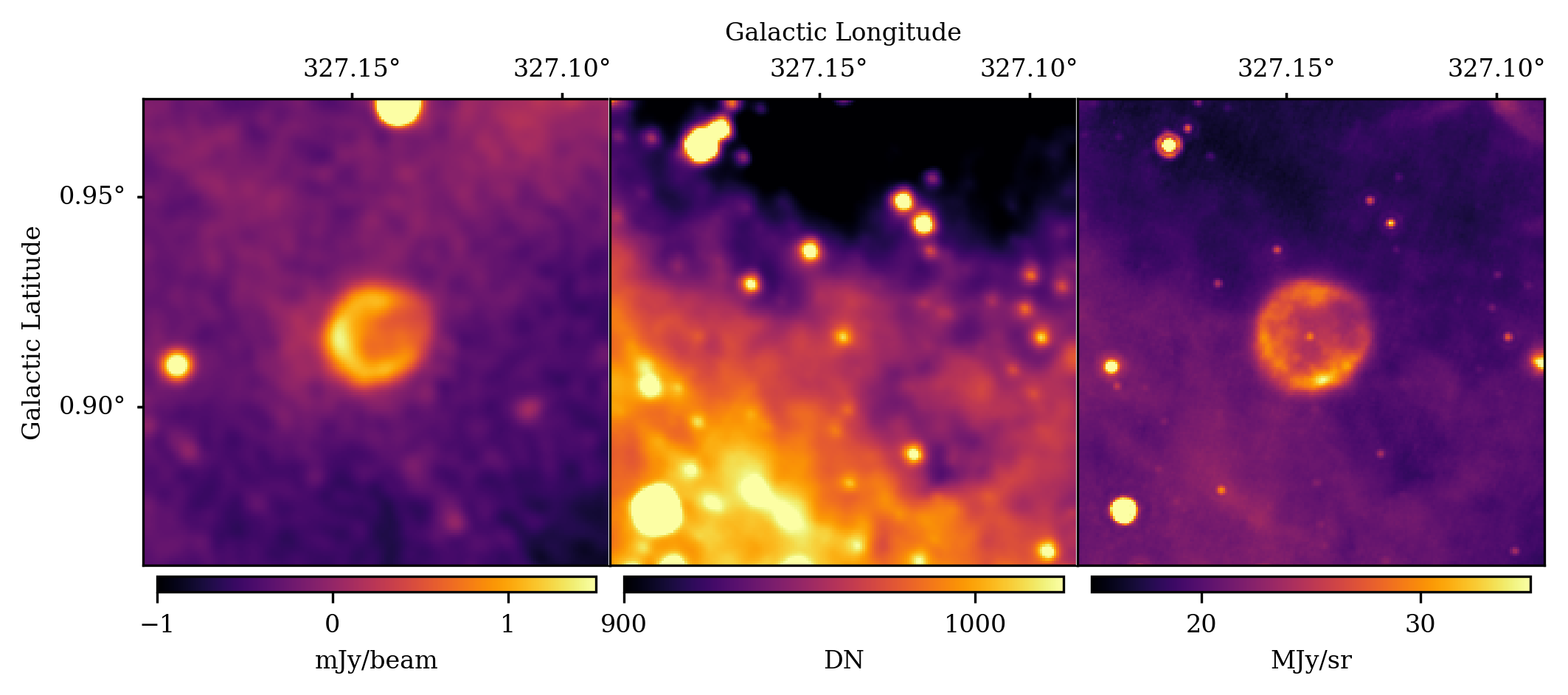}
     \caption{933 MHz ASKAP data, 12 $\mu$m WISE data, and 24 $\mu$m MIPS data for G327.1$+$0.9.}
     \label{fig:mips16}
 \end{figure*}
 
 \paragraph*{G325.0$-$0.3} This candidate was originally identified by \cite{Whiteoak1996} (with no image) and later studied by \cite{Green2014} who provided an image but the source is not included in the \cite{Green2022} catalogue. In Figure~\ref{fig:9} a clear shell structure can be seen that is brightest to the east. A filament of emission can be seen to the east of the candidate but it is likely thermal and unrelated. The candidate has no MIR counterpart. Based on our observations and the previous studies, we believe this source should be classified as an SNR.

 \paragraph*{G325.0$+$0.2}
 This source consists of roughly circular emission that is brightened to the west where the edge of the source is the most well-defined (Figure~\ref{fig:27}). A bright overlapping point source can be seen to the east. Emission can be seen in the MIR but it does not have the same rounded structure as the radio source.
 
 \paragraph*{G325.8$-$2.1}
 This object is composed of very faint filaments of emission found to the southwest of the bright known SNR G326.3$-$1.8. As shown in Figure~\ref{fig:23} the source overlaps with negative bowls of emission, likely caused by missing short spacings. The filaments can be seen most clearly in the southeast but some faint structures are also visible in the northwest. This source is only visible because it is located at a far distance from the Galactic plane and there is little thermal emission. The source was too faint to obtain an estimate of the flux density.
 
 \paragraph*{G325.8$+$0.3}
 This candidate is composed of rounded filamentary structures that can be seen to the north and southeast in Figure~\ref{fig:24}. The source overlaps with a large HII region and many point sources. Missing short interferometer spacings generate a negative bowl around these strong emitters. Because of this, and because the filaments are so faint, we were unable to obtain a flux estimate for the source.

 \begin{figure*}
     \centering
     \includegraphics[width=0.75\textwidth]{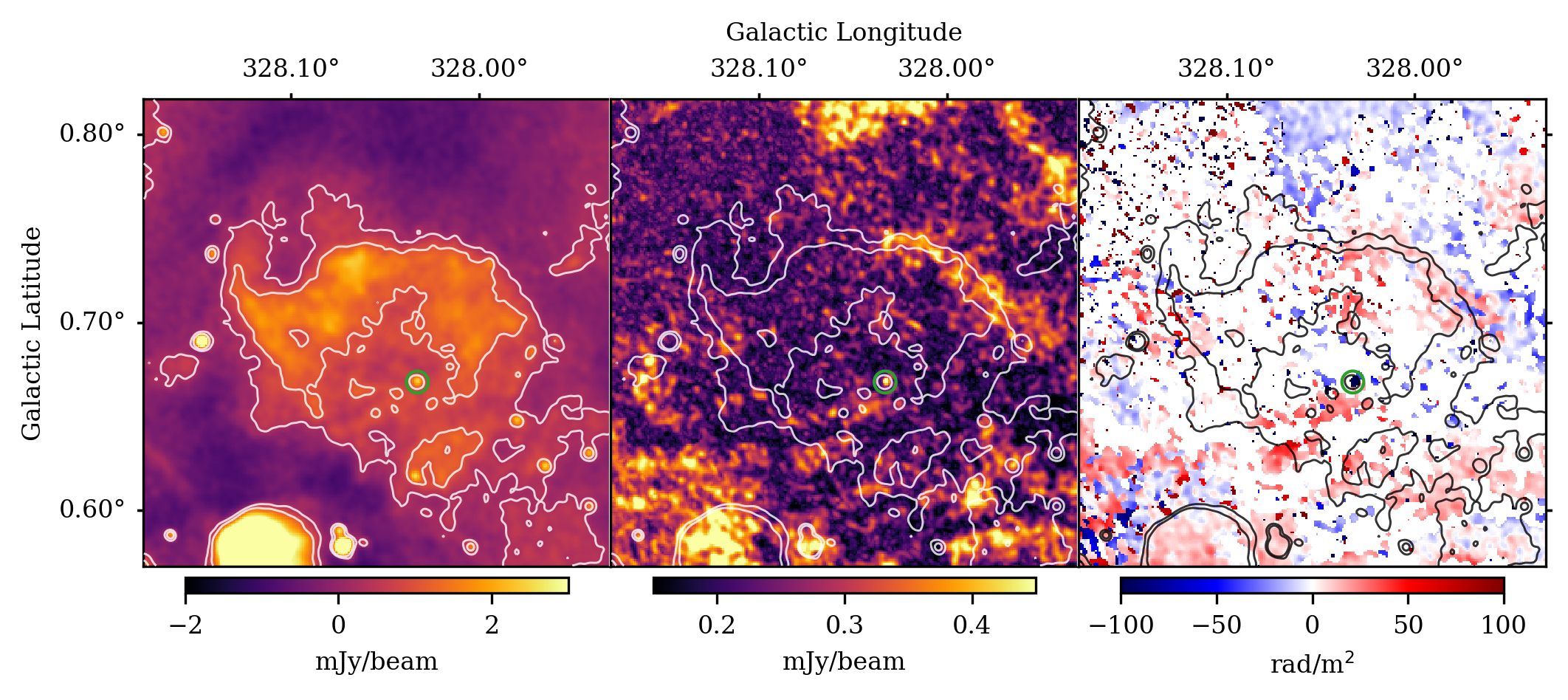}
     \caption{Total power, PI, and RM maps for G328.0$+$0.7 with total power contours to indicate where emission overlaps. The green circle indicates the location of the young pulsar, which is polarized and has a high rotation measure.}
     \label{fig:pol17}
 \end{figure*}
 
 \paragraph*{G327.1$+$0.9}
  This candidate is a small, roughly circular source that appears to have a brightened shell-like edge to the east (Figure~\ref{fig:16}). This is our smallest source with a size of 2' and we believe that this marks the approximate lower size limit of sources in which we would be able to detect a shell-like structure, given the resolution of our data. This source appears in \cite{Gvaramadze2010} in their table of 24 $\mu$m nebulae compiled using data from the Multiband Imaging Photometer for Spitzer (MIPS).  There is a circular MIR nebula at 24 $\mu$m (diameter of $1.6'$) with a total flux of 1.4 Jy \citep{mizuno10}, with a compact source in the infrared visible at the centre. Since there is no counterpart at the WISE 12 $\mu$m image (see Figure~\ref{fig:mips16}, $F_{12}<0.05~\mathrm{Jy}$), the source meets our criteria for a candidate.  However, \citet{Gvaramadze2010} identify this source as a possible wind driven bubble around a Wolf-Rayet (WR) star. The dusty nebulae around WR stars can be bright at 22 $\mu$m and relatively faint at $\sim12~\mu$m, with spectroscopy studies suggesting that most emission toward WR nebulae at $\sim12~\mu$m comes from material along the line of sight \citep[e.g.,][]{toala15}. Since the winds from Wolf-Rayet stars can be detected in the 1.4 GHz radio continuum \citep[e.g.,][]{grudel02}, this candidate could be a dusty WR nebula. However, G327.1$+$0.9 is an extended shell-like radio source and the radio emission from Wolf-Rayet winds is typically centrally peaked and does not extend much further than 1000 stellar radii \citep{grudel02}. In this case we propose that there are \rev{three} possible explanations for G327.1$+$0.9. The first would be a late stage HII region produced by the central Wolf-Rayet star. The second would be \rev{a planetary nebula \citep{Ingallinera2016}. Finally, it could be }the SNR of a supernova that exploded in a binary system. Follow-up polarimetric observations at high radio frequencies with ATCA should help to solve this mystery, as only the SNR would show linearly polarized radio emission.

 \paragraph*{G328.0$+$0.7}
 The source shown in Figure~\ref{fig:17} is composed of roughly circular emission with a well-defined edge that is brightest in the northern part of the shell. There is a point source located near the centre of the candidate that coincides with the location of the known young ($\sim$ 37 kyrs) pulsar J1551$-$5310 \citep{Hobbs2004,Manchester2005}. Several HII regions can be seen in the south. There is possibly polarization associated with the northwest edge of the shell but it cannot be definitely differentiated from the background emission, as shown in Figure~\ref{fig:pol17}. We do see clear polarization coming from the central pulsar with a rotation measure of \(-1017\pm5\) rad/m$^2$, consistent with the catalogued value of \(-1023.3 \pm 6.3\) rad/m$^2$ \citep{psrrm2018}. Because of the possible association with a young pulsar and detectable polarization, we believe this source should be classified as a supernova remnant.

 \paragraph*{G328.6$+$0.0}
 This source was observed by \cite{McClure2001} and is listed by \cite{Green2022} as an SNR candidate but it does not appear in the actual catalogue. Here we see elongated filaments that form a roughly elliptical shape, which could be two separate sources (Figure~\ref{fig:11}). There is a well-defined shell-like edge in the southwest and less defined shell-like filaments in the northeast. There are several overlapping HII regions and an unidentified bright point source located near the geometric centre of the southwest arc that may be polarized. The very bright source to the northwest of the image is G328.4$+$0.2. This source is visible in the GLEAM data and a steep negative spectral index of \(-0.75 \pm 0.06\) was determined, indicating the emission is likely nonthermal. We believe this source should also be classified as a supernova remnant. 

 \paragraph*{G330.2$-$1.6}
 This source, shown in Figure~\ref{fig:25}, is very faint but it is located at a far distance from the Galactic plane making it detectable. It has a clear circular shell-like structure with little to no emission coming from the centre and no obvious MIR counterpart. The filamentary emission seen in the southeast appears to be unrelated and may instead be a radio galaxy, with a bright core, two external bright spots, and internal lobes, indicating jet precession.

\section{Discussion} \label{sec:discussion}

\subsection{Estimating the SNR Density of the Galactic Disk} \label{sec:SNRdensity}

Based on our estimate of the size of the Galactic SNR population, we can roughly estimate the theoretical SNR surface density of the Galactic disk. Since most SNRs should be found within the star-forming disk, we assume a Galactic radius of 12.5 kpc, which encompasses 95\% of the stellar mass of the Milky Way disk \citep{binney23}. We estimate that the Milky Way should have at least \(1000-2700\) radio-bright SNRs. This corresponds to an average Galactic SNR surface density between \(2.0-5.5\) SNRs/kpc$^2$. We can compare this range of values to the density of known Galactic SNRs. The University of Manitoba's SNRcat \citep{Ferrand2012} currently lists 383 SNRs and SNR candidates. Of these, 369 have been detected in the radio. Thus, the Galaxy has an average known SNR surface density of only 0.75 radio-bright SNRs/kpc$^2$. 

To gain further insight into the missing SNR population, we can analyze how the known SNRs are distributed within the Galaxy by quadrant. Quadrants I and IV look towards the Galactic centre and encompass a larger, but denser, part of the Galactic plane. Quadrants II and III look away from the Galactic centre with shorter lines of sight and less emission. The known SNR surface density distribution can be broken down by quadrant as follows:
\begin{itemize}
    \item Quadrant I (\(0^\circ\leq l<90^\circ\)): 0.74 SNRs/kpc$^2$
    \item Quadrant II (\(90^\circ\leq l<180^\circ\)): 1.23 SNRs/kpc$^2$
    \item Quadrant III (\(180^\circ\leq l<270^\circ\)): 0.63 SNRs/kpc$^2$
    \item Quadrant IV (\(270^\circ\leq l<360^\circ\)): 0.72 SNRs/kpc$^2$
\end{itemize}
We see that Quadrant II has a noticeably higher density of known SNRs than the other quadrants. It is likely higher than Quadrants I and IV because SNRs in Quadrant II are closer (on average) and there is less thermal emission in this direction. Quadrant II is also relatively well-surveyed which may explain why it has a higher density of known SNRs than Quadrant III.

The EMU/POSSUM Galactic pilot II field (\(323^\circ\leq l<330^\circ\)) has 7 known SNRs giving it an SNR surface density of 0.34 SNRs/kpc$^2$, around a factor of 2 lower than the Quadrant IV average. To achieve the theoretical average Galactic SNR density of \(2.0-5.5\) SNRs/kpc$^2$ we would need to find \(34-106\) new SNRs in this field. If we include our 21 SNR candidates, the density of known SNRs is brought up to 1.36 SNRs/kpc$^2$, which is comparable to the SNR density we see in Quadrant II. 

\begin{figure*}
    \centering
    \includegraphics[width=0.98\textwidth]{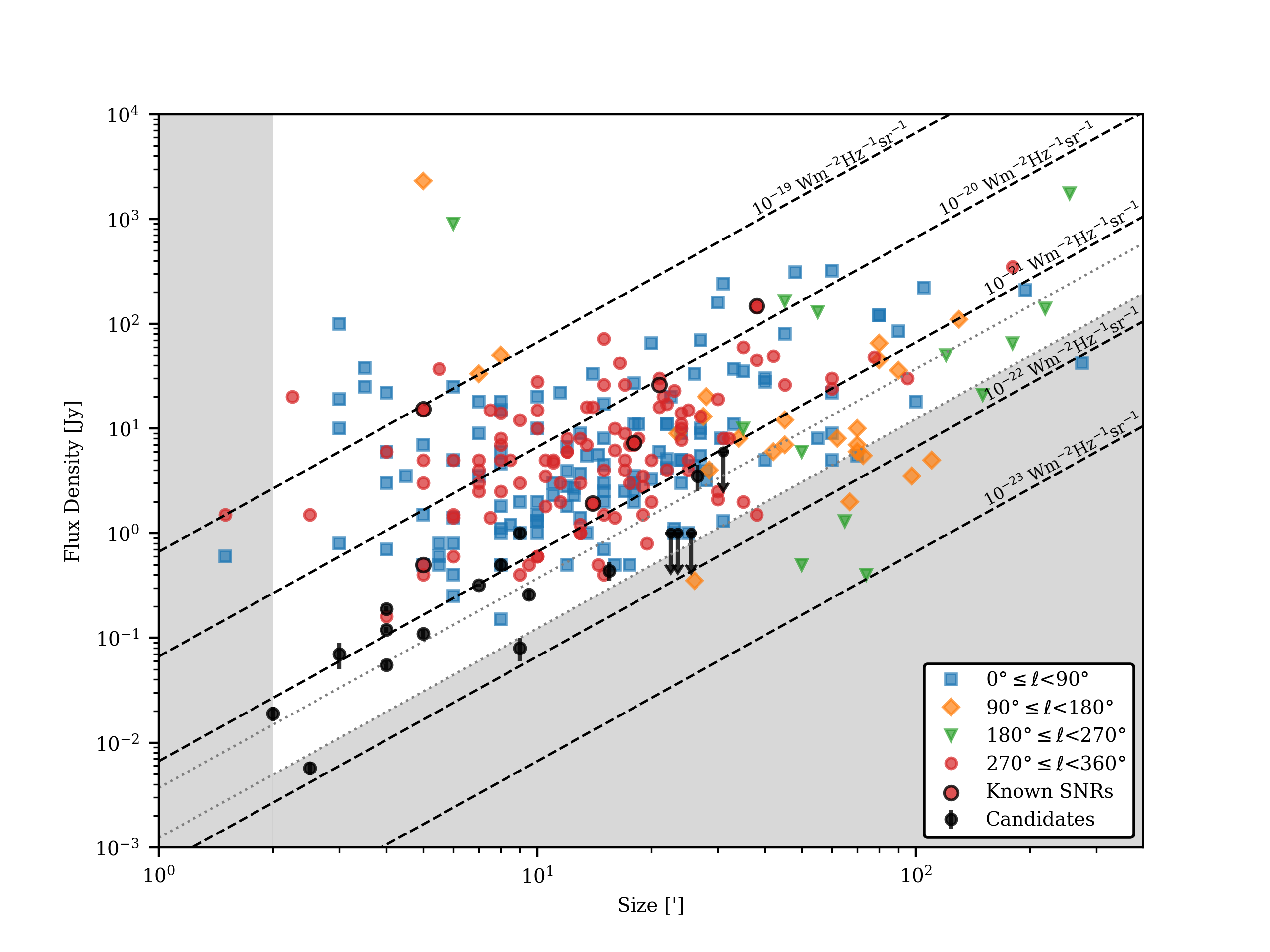}
    \caption{Sizes and flux densities of our SNR candidates compared with the known SNR population as a function of quadrant (data from \protect\cite{Green2022}, sources without flux estimates omitted). The "Known SNRs" are the six SNRs that are fully imaged in our field. The "Candidates" are the 18 SNR candidates listed in this paper for which we were able to obtain flux estimates. Note that for some candidates the error bars are too small to be visible and others have arrows to indicate that the values should be taken as upper limits. The dashed lines are lines of constant surface brightness. Shaded regions indicate the approximate limits of what we believe we should be able to detect with ASKAP at this frequency. The grey dotted lines represent the $\sigma$ and 3$\sigma$ thermal noise limits.}
    \label{fig:SB}
\end{figure*}

\subsection{The Known Galactic SNR Population} \label{sec:knownSNRpop}

Figure~\ref{fig:SB} shows how the angular sizes and flux densities of our SNR candidates compare to the known Galactic SNR population, not including the candidates for which we were unable to determine a reliable flux density. This plot was made using data from \cite{Green2022}, omitting sources where no flux estimate was given. It should be noted that 1 GHz flux densities are provided in Green's catalogue and our 933 MHz values were not adjusted to this frequency since we do not have spectral indices for most of the candidates. However, given the small frequency difference relative to the scale of the plot, the changes in the positions of the points are insignificant, even for sources with steep spectra. 

The shaded regions of the plot indicate what we estimate to be the limits of what can be detected with our data. The minimum size was chosen to be 2' as this is the size of our smallest source and we believe it would be difficult to detect shell-like structures for smaller sources given the 18" resolution. The minimum surface brightness was based on an estimate of the minimal thermal noise (taken to be \(\sigma = 40\) \(\mu\)Jy/beam), though variations across the image likely set different limits for different regions. The grey dotted lines in the plot represent the \(\sigma\) and \(3\sigma\) surface brightness limits. Because we are studying extended sources, rather than point sources, we are able to detect sources below the theoretical thermal noise limit. 

Our results support that there is likely still a large population of undetected supernova remnants within our Galaxy. Figure~\ref{fig:SB} demonstrates that many of our candidates are smaller and fainter than most known SNRs. This may be evidence that some of the missing SNR population was missing because it was previously undetectable. Improvements in the angular resolution and sensitivity of radio telescopes should then allow for the detection of some of these SNRs in future sky surveys. It may be that improvements in angular resolution are more important as the technological surface brightness lower limit is approaching the limit set by the background emission \citep{Dokara2022, Anderson2017}. However, future improvements in angular resolution may also have limited impact in Galactic SNR observations. Given that we have shown we are capable of detecting shell-like structures in sources as small as 2', even at the farthest Galactic distances within this field ($\sim$18 kpc) we should be able to detect sources with linear sizes as small as 10 pc. Based on observations of other Local Group galaxies \citep{Badenes2010,Long2010} and given that X-ray observations of young (small) Galactic SNRs are believed to be fairly complete \citep{Leahy2020}, it is unlikely that there are many undetected sources below this size limit. Thus, any remaining missing SNRs in this field are most likely obscured by superimposed radio-bright sources, particularly HII regions.

The impact of background emission on SNR detection can be further demonstrated by the fact that we were not able to detect new sources in the \rev{brighter} part of the field \rev{that we believe to be associated with the Norma arm}. In fact, almost all of our candidates are in the lower right corner of the field, as shown in Figure~\ref{fig:field2}. Theoretically, there should be a higher density of supernova remnants within and near spiral arms. If we compare the bright \rev{upper left} half of the image to the lower right half, including our candidates the former has an SNR density of $\sim$1 SNR/kpc$^2$ while the latter has a density of $\sim$2 SNRs/kpc$^2$, which is the theoretical lower limit we derived in Section~\ref{sec:SNRdensity}. While the actual SNR density should be higher in the upper left half, we find the observed SNR density to be higher in the lower right half. This indicates there is likely still a population of SNRs within the \rev{bright part} of the field that we were unable to detect. Further, this demonstrates the impact of background emission in setting detectability limits and supports that technological improvements in sensitivity may no longer be as effective for detecting new radio SNRs in these types of regions. 

\section{Conclusions} \label{sec:conclusions}

In this paper we used pilot data from ASKAP to study the known supernova remnant population in a small field of the Galactic plane. We also found 21 SNR candidates, three times the number of known SNRs in this field. \rev{Of the candidates, 13 have not been previously studied, 4 have been studied as SNR candidates, 3 classed as a single SNR, and 1 studied as an MIR nebula}. For most candidates, observations at a second, ideally higher, frequency are required to confirm the sources as SNRs as these observations would likely provide more information about polarization and spectral indices. 

The results of this paper demonstrate the potential for the full EMU/POSSUM surveys, taking place over the next few years, to fill in some of the missing Galactic supernova remnant population through the detection of small and/or faint sources. We were able to detect sources that seem to have been missed in previous surveys due to their small angular size and/or low surface brightness. Comparing the properties of our candidates to the known Galactic SNR population further supports this. Future work using ASKAP data to expand upon the size of the surveyed Galactic field will likely detect more of these types of sources, allowing for a better characterization of the Galactic SNR population. In this field we have uncovered approximately 1 new SNR/kpc$^{2}$. The full EMU/POSSUM surveys will cover roughly 60\% of the Galactic plane meaning they could hypothetically uncover around 300 new SNRs. However, it is important to note \rev{the impact of background emission and the challenges we faced in detecting low surface brightness SNRs in complex regions. Because} a significant portion of the surveyed area will be in the direction of the Galactic centre, it is unlikely that we will be as successful in these denser, brighter fields. 

The missing supernova remnant problem and the size of the Galactic SNR population remain difficult problems to exactly quantify. While technological advancements have continued to lead to new SNR detections, there is likely a population of SNRs that will never be observed due to detection limits set by the local background. Estimates of the size of the Galactic SNR discrepancy should then exclude these types of sources as they may be considered undetectable. In future work we hope to further explore this problem, utilizing the full EMU/POSSUM surveys to study variations in SNR detectability across the Galactic plane. 

\section*{Acknowledgements}

\rev{We are grateful to an anonymous referee whose comments improved the quality of the paper.} This scientific work uses data obtained from Inyarrimanha Ilgari Bundara / the Murchison Radio-astronomy Observatory. We acknowledge the Wajarri Yamaji People as the Traditional Owners and native title holders of the Observatory site. CSIRO’s ASKAP radio telescope is part of the Australia Telescope National Facility (https://ror.org/05qajvd42). Operation of ASKAP is funded by the Australian Government with support from the National Collaborative Research Infrastructure Strategy. ASKAP uses the resources of the Pawsey Supercomputing Research Centre. Establishment of ASKAP, Inyarrimanha Ilgari Bundara, the CSIRO Murchison Radio-astronomy Observatory and the Pawsey Supercomputing Research Centre are initiatives of the Australian Government, with support from the Government of Western Australia and the Science and Industry Endowment Fund. The POSSUM project (https://possum-survey.org) has been made possible through funding from the Australian Research Council, the Natural Sciences and Engineering Research Council of Canada (NSERC), the Canada Research Chairs Program, and the Canada Foundation for Innovation. This publication makes use of data products from the Wide-field Infrared Survey Explorer, which is a joint project of the University of California, Los Angeles, and the Jet Propulsion Laboratory/California Institute of Technology, funded by the National Aeronautics and Space Administration. RK acknowledges the support of NSERC, funding reference number RGPIN-2020-04853. ER acknowledges the support of NSERC, funding reference number RGPIN-2022-03499. The Dunlap Institute is funded through an endowment established by the David Dunlap family and the University of Toronto. B.M.G. acknowledges the support of NSERC through grant RGPIN-2022-03163, and of the Canada Research Chairs program. DL acknowledges the support of NSERC.

\section*{Data Availability}

ASKAP data are available on the CSIRO ASKAP Science Data Archive (CASDA). (https://research.csiro.au/casda/)
 


\bibliographystyle{mnras}
\bibliography{main} 


\appendix

\section{Known Supernova Remnants} \label{sec:appendixA}
\begin{figure*}
    \centering
    \includegraphics[width=0.85\textwidth]{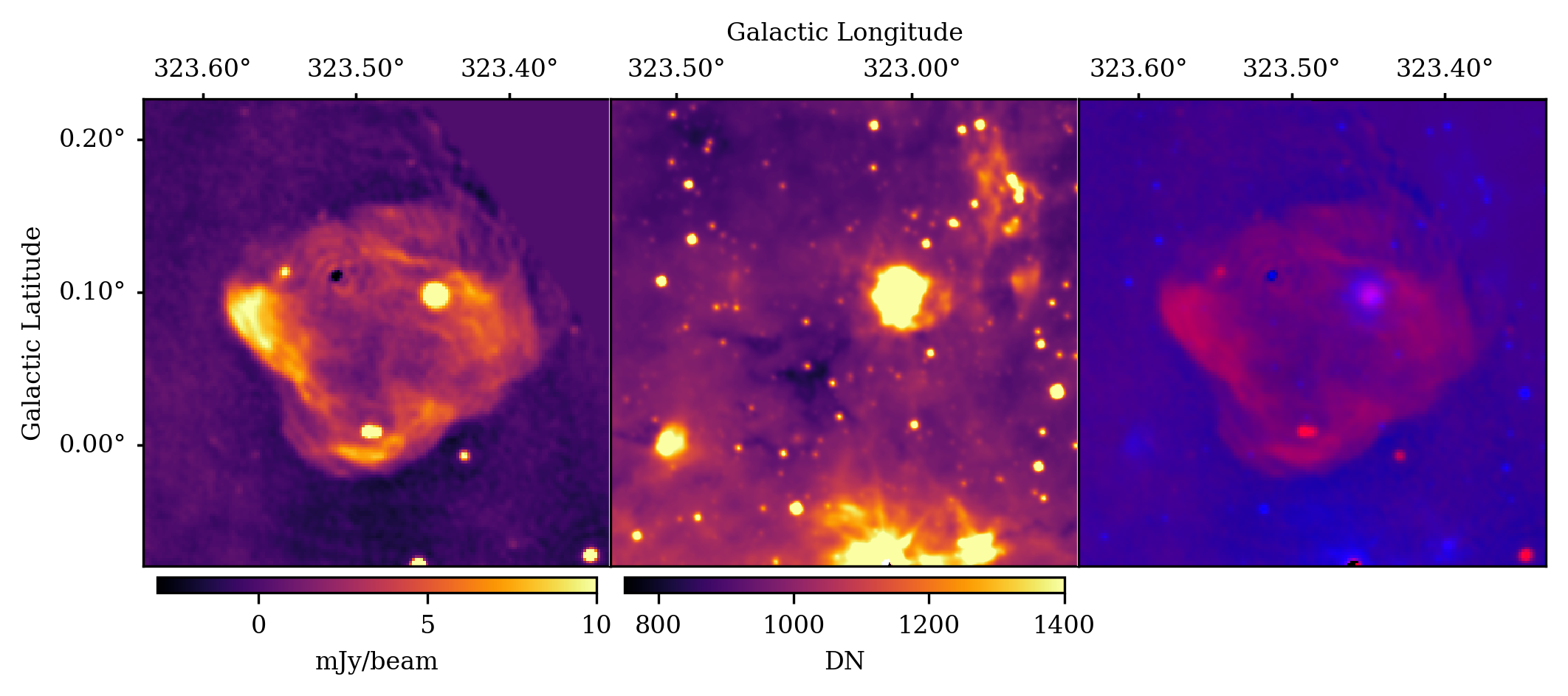}
    \caption{G323.5$+$0.1}
    \label{fig:7}
\end{figure*}

\begin{figure*}
    \centering
    \includegraphics[width=0.85\textwidth]{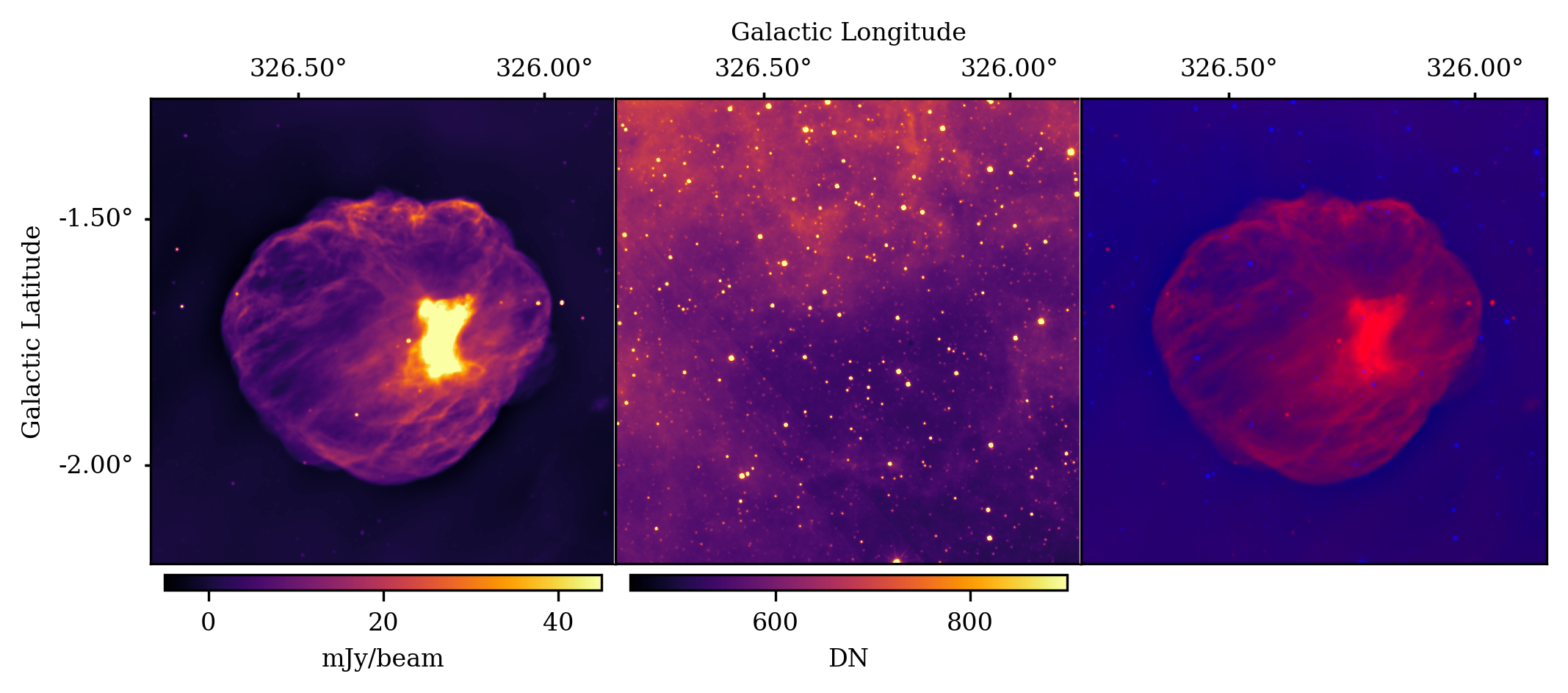}
    \caption{G326.3$-$1.8}
    \label{fig:1}
\end{figure*}

\begin{figure*}
    \centering
    \includegraphics[width=0.85\textwidth]{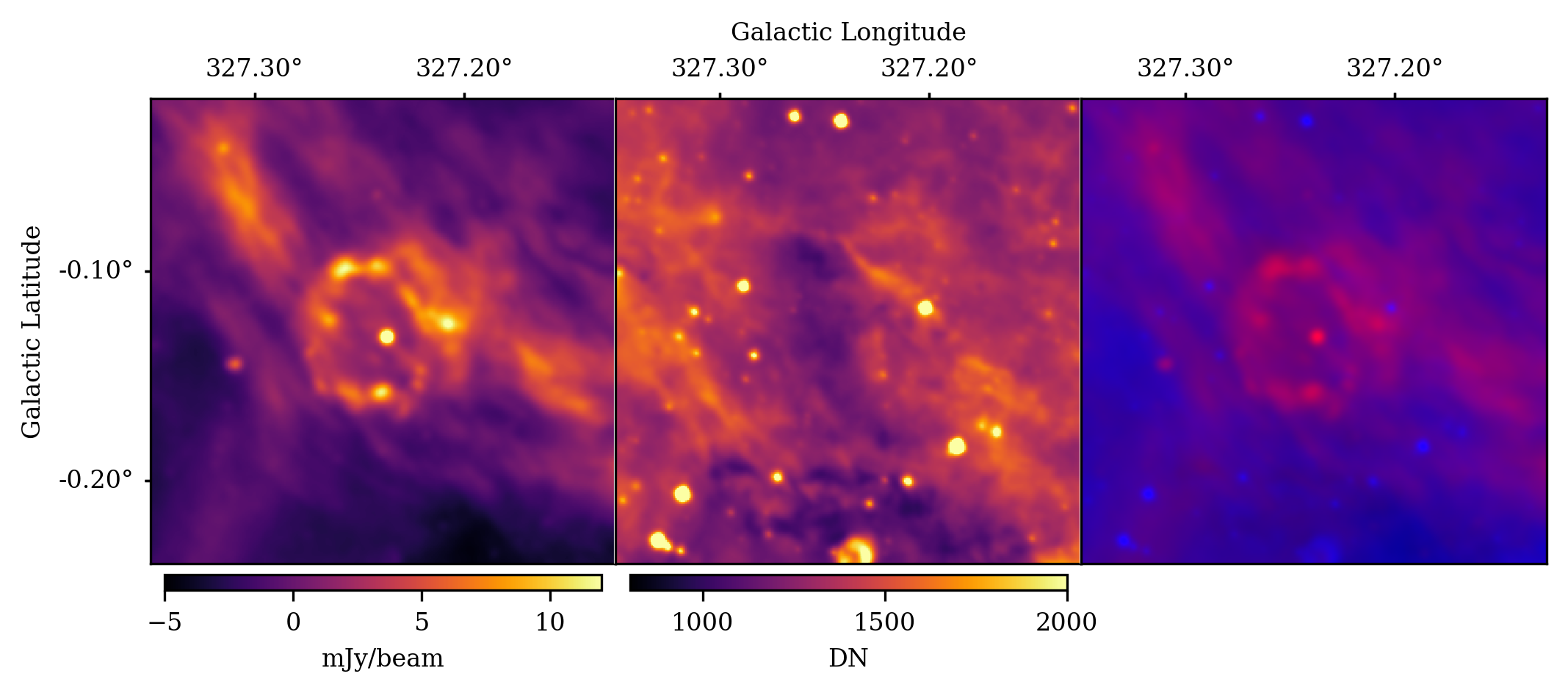}
    \caption{G327.2$-$0.1}
    \label{fig:3}
\end{figure*}

\begin{figure*}
    \centering
    \includegraphics[width=0.85\textwidth]{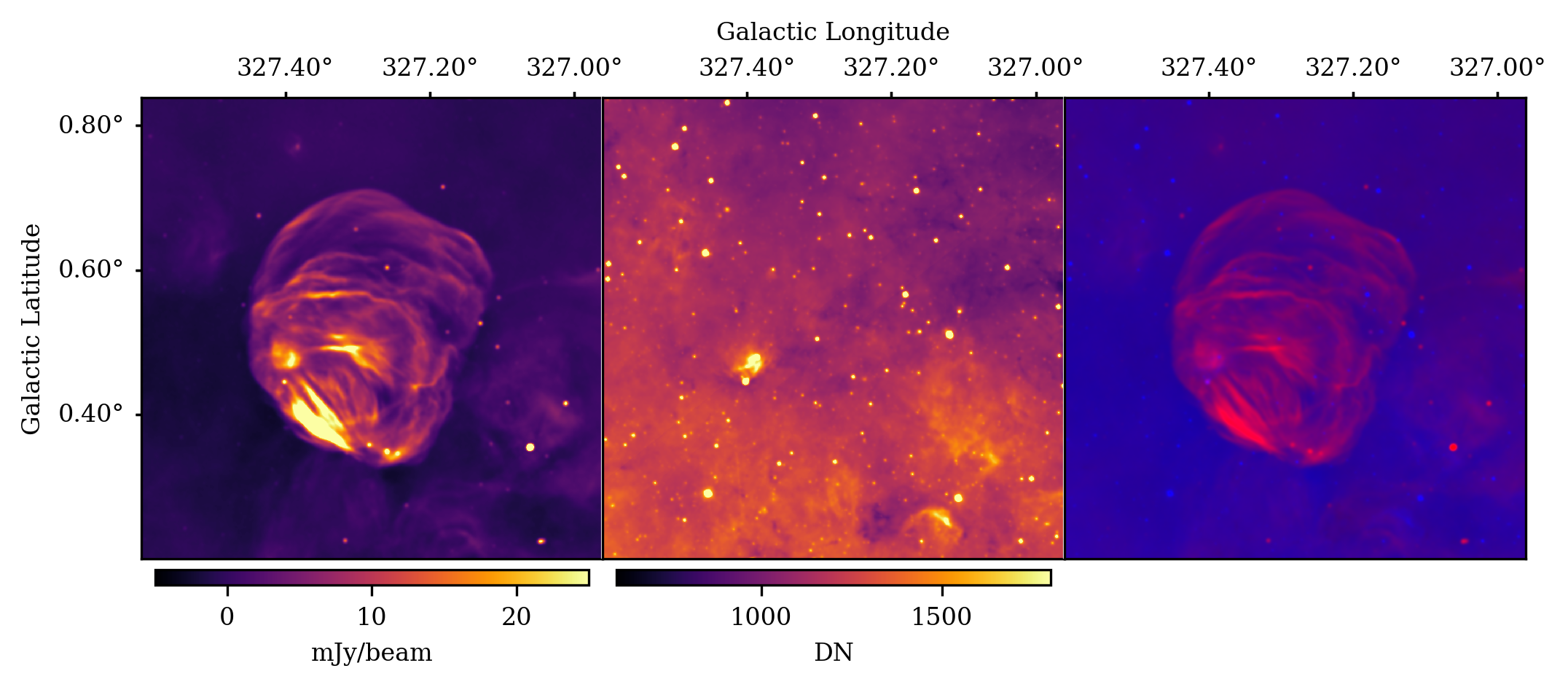}
    \caption{G327.4$+$0.4}
    \label{fig:4}
\end{figure*}

\begin{figure*}
    \centering
    \includegraphics[width=0.85\textwidth]{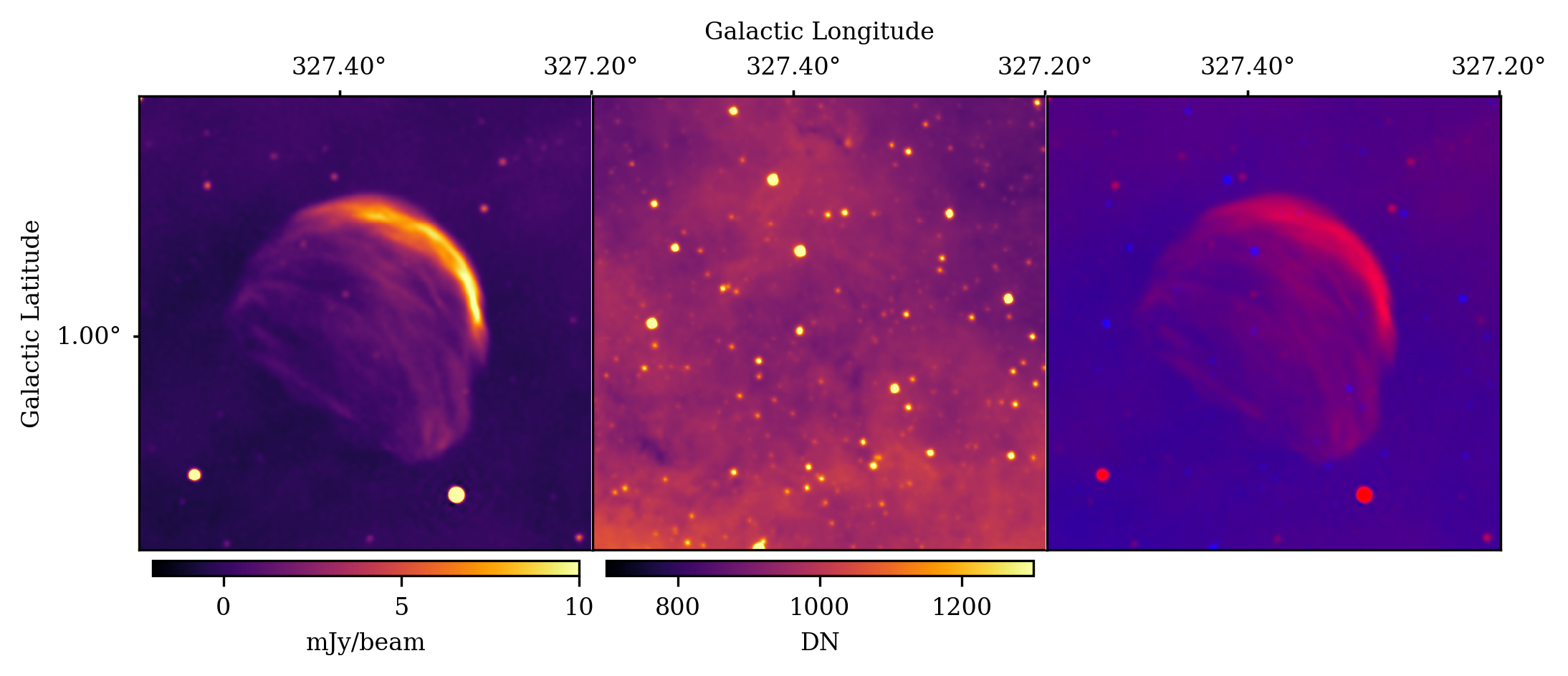}
    \caption{G327.4$+$1.0}
    \label{fig:5}
\end{figure*}

\begin{figure*}
    \centering
    \includegraphics[width=0.85\textwidth]{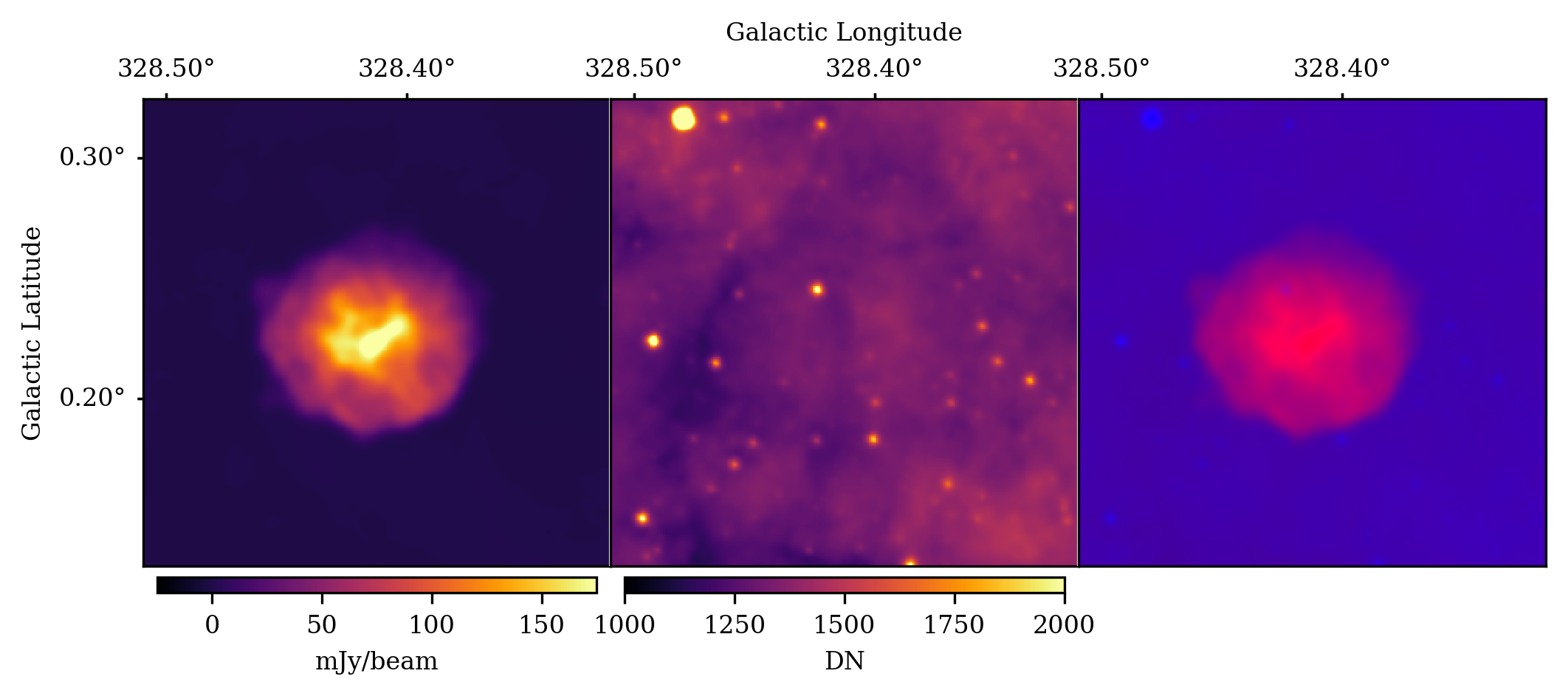}
    \caption{G328.4$+$0.2}
    \label{fig:6}
\end{figure*}

In Figures~\ref{fig:7} to \ref{fig:6} we present images of the known SNRs in the EMU/POSSUM Galactic pilot II field. Here we show the 933 MHz radio images from ASKAP, the same region in the MIR at 12 $\mu$m using data from WISE, and a composite image with radio emission shown in red and MIR in blue.


\section{Supernova Remnant Candidates}  \label{sec:appendixB}

\begin{figure*}
    \centering
    \includegraphics[width=0.85\textwidth]{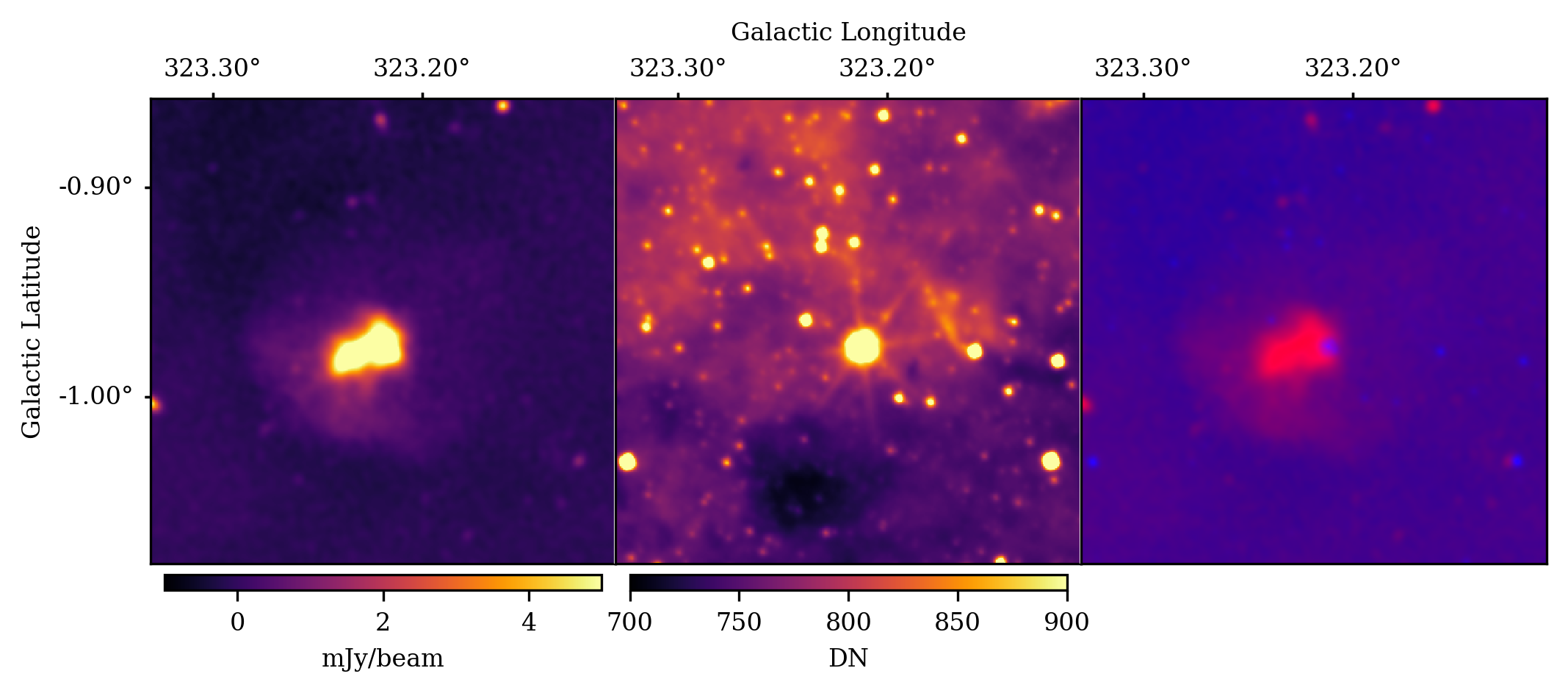}
    \caption{G323.2$-$1.0}
    \label{fig:13}
\end{figure*}

\begin{figure*}
    \centering
    \includegraphics[width=0.85\textwidth]{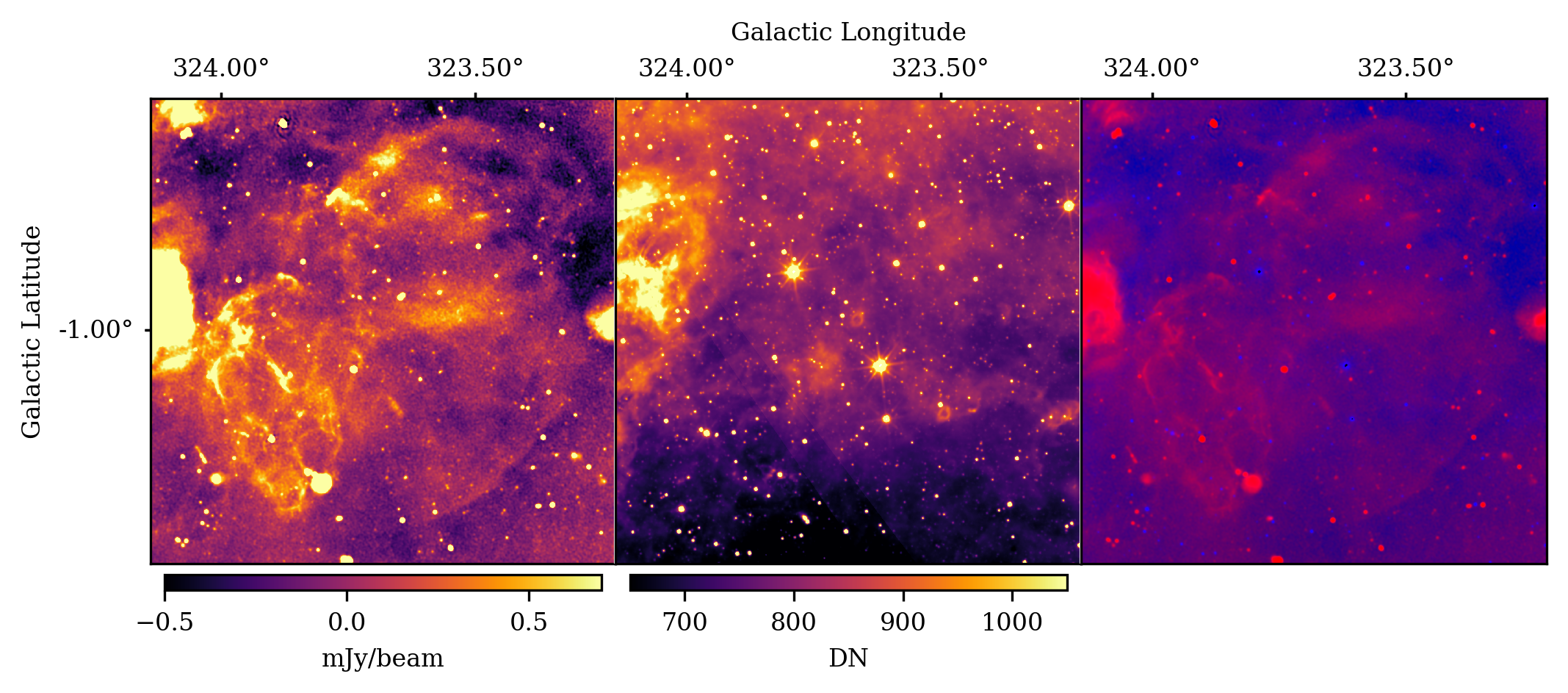}
    \caption{G323.6$-$1.1, G323.6$-$0.8, G323.9$-$1.1}
    \label{fig:8}
\end{figure*}

\begin{figure*}
    \centering
    \includegraphics[width=0.85\textwidth]{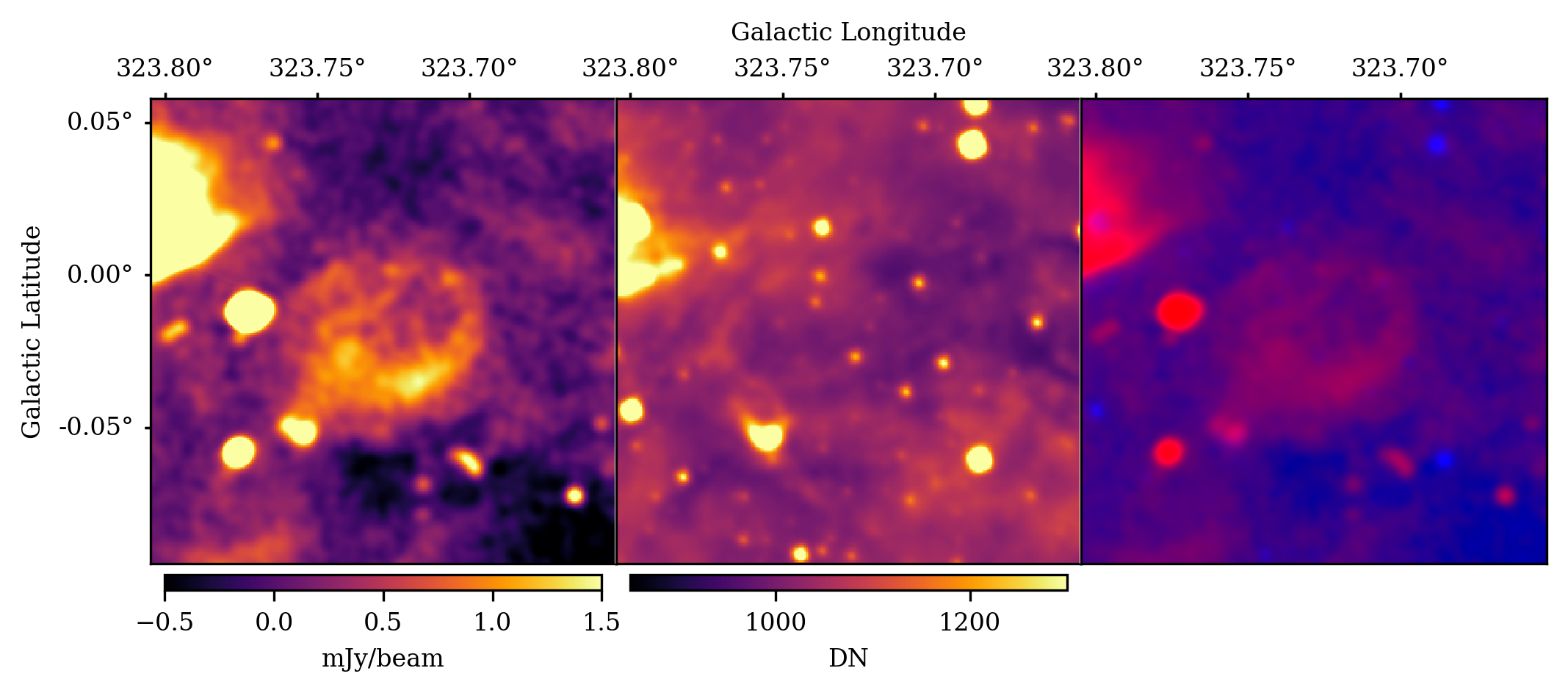}
    \caption{G323.7$+$0.0}
    \label{fig:21}
\end{figure*}

\begin{figure*}
    \centering
    \includegraphics[width=0.85\textwidth]{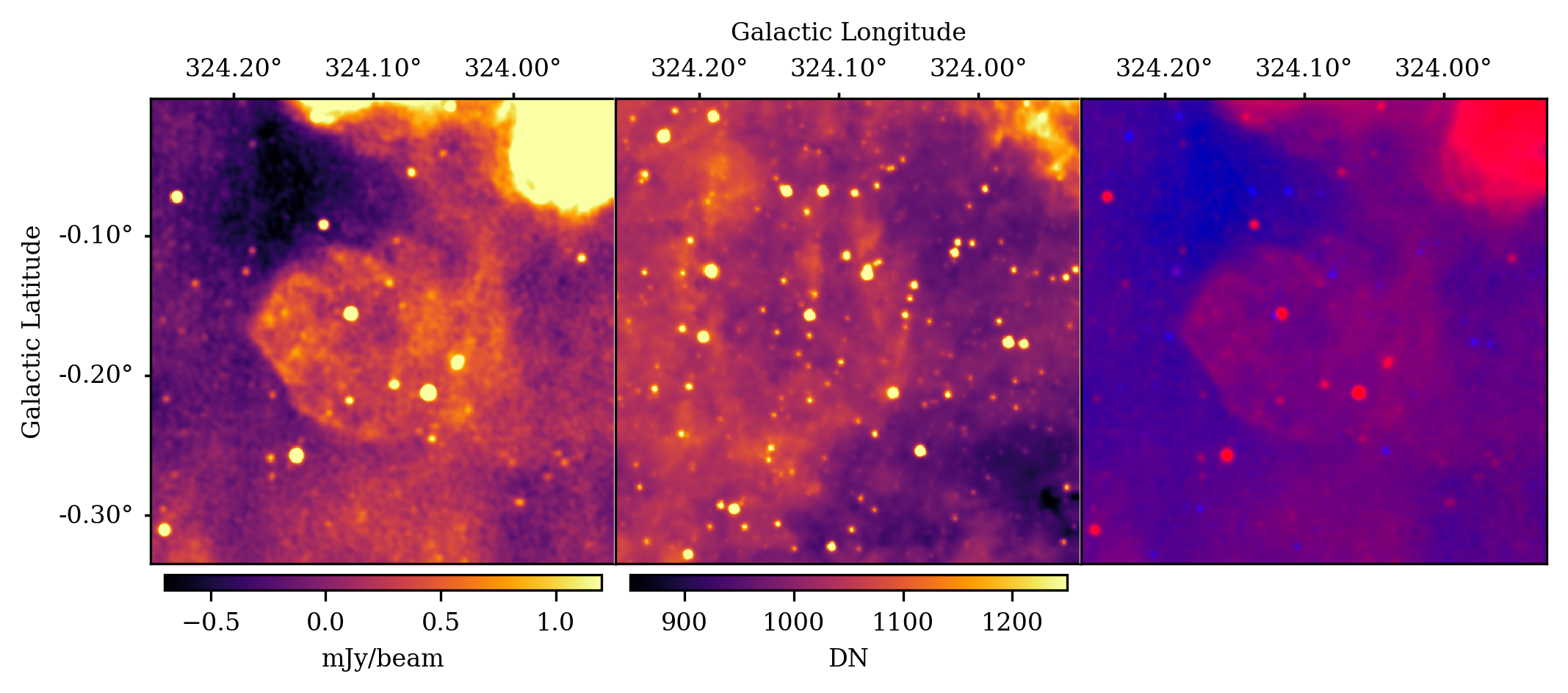}
    \caption{G324.1$-$0.2}
    \label{fig:15}
\end{figure*}

\begin{figure*}
    \centering
    \includegraphics[width=0.85\textwidth]{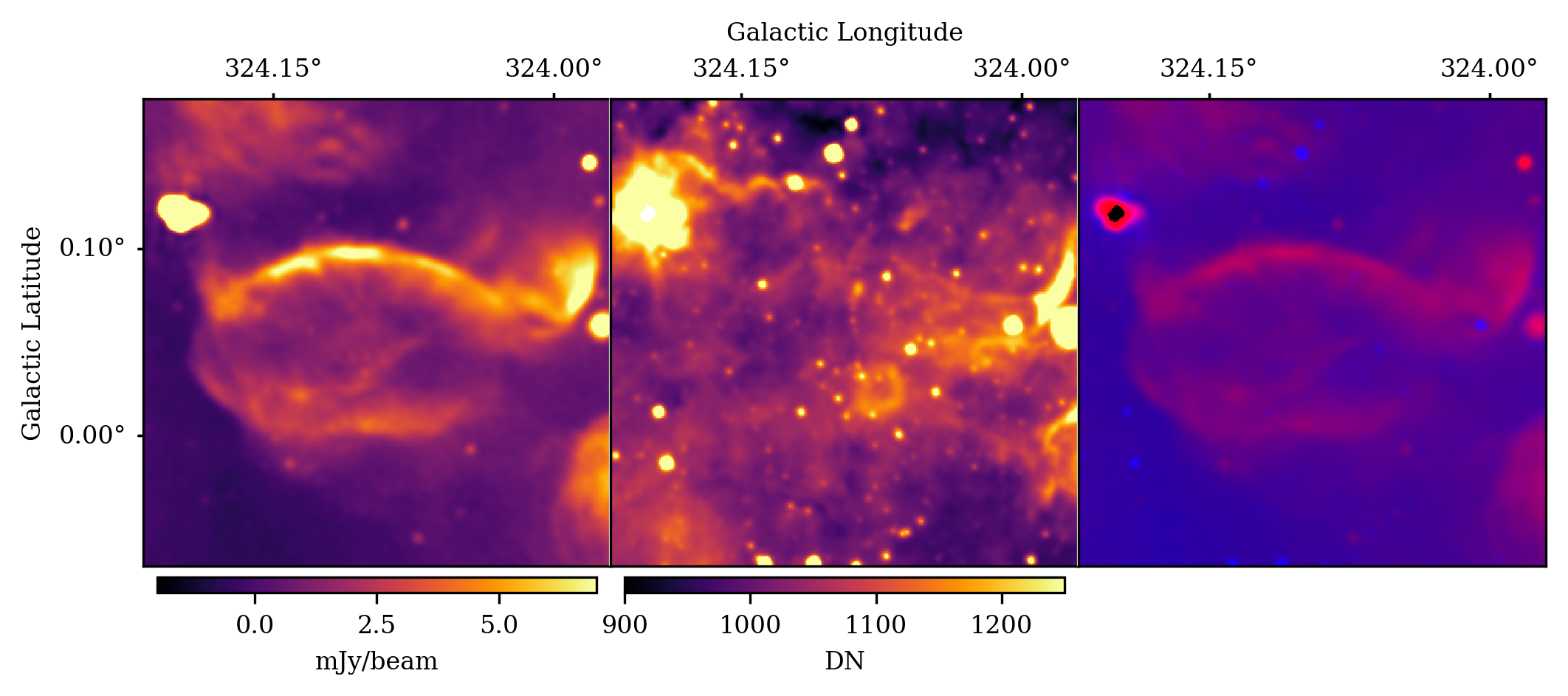}
    \caption{G324.1$+$0.0}
    \label{fig:10}
\end{figure*}

\begin{figure*}
    \centering
    \includegraphics[width=0.85\textwidth]{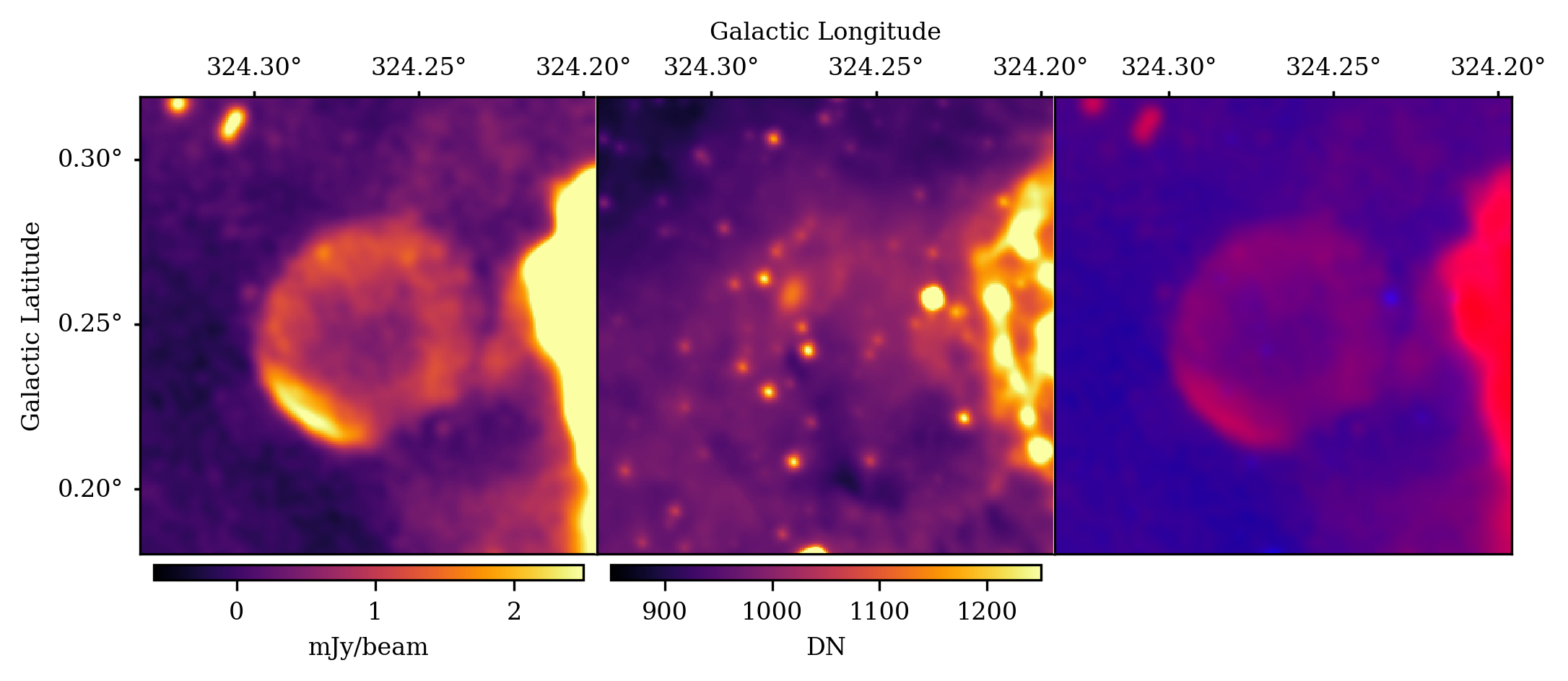} 
    \caption{G324.3$+$0.2}
    \label{fig:12}
\end{figure*}

\begin{figure*}
    \centering
    \includegraphics[width=0.85\textwidth]{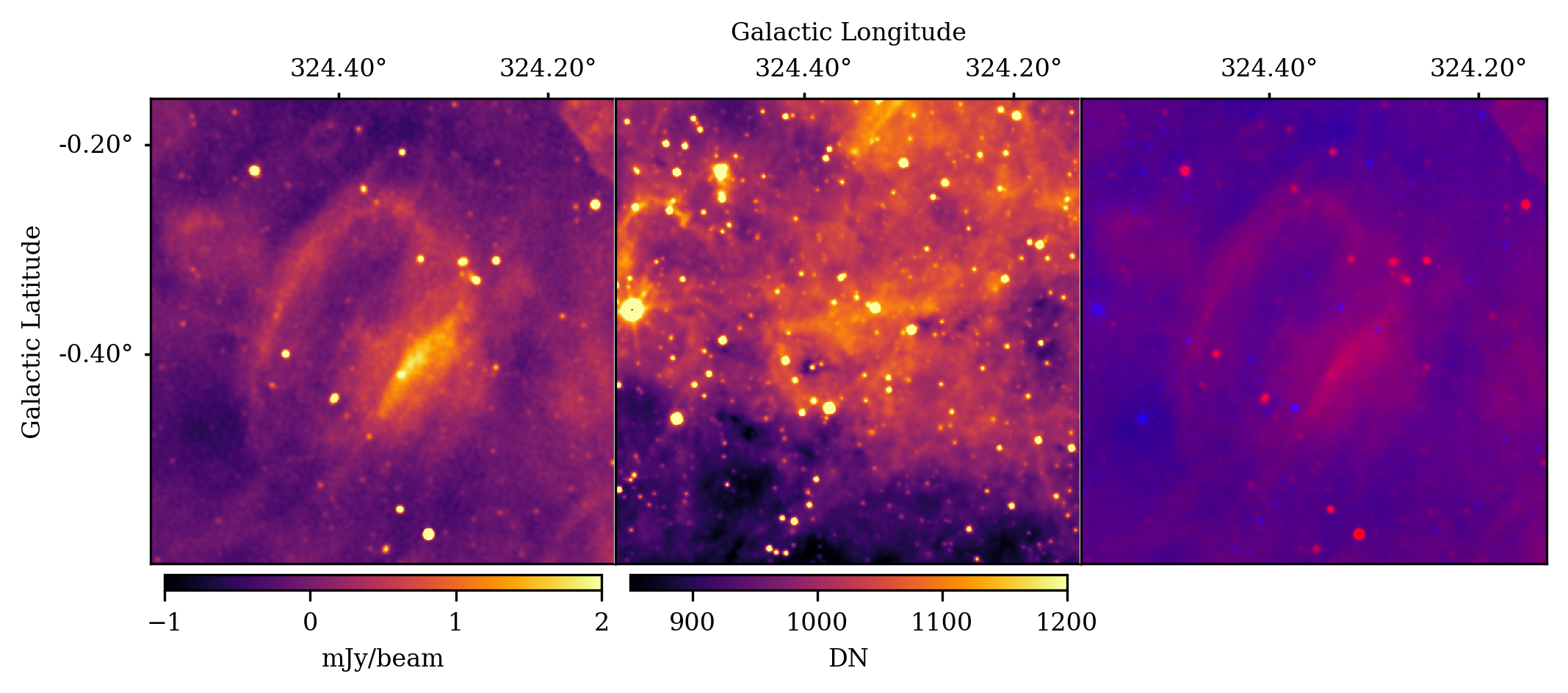}
    \caption{G324.4$-$0.4}
    \label{fig:14}
\end{figure*}

\begin{figure*}
    \centering
    \includegraphics[width=0.85\textwidth]{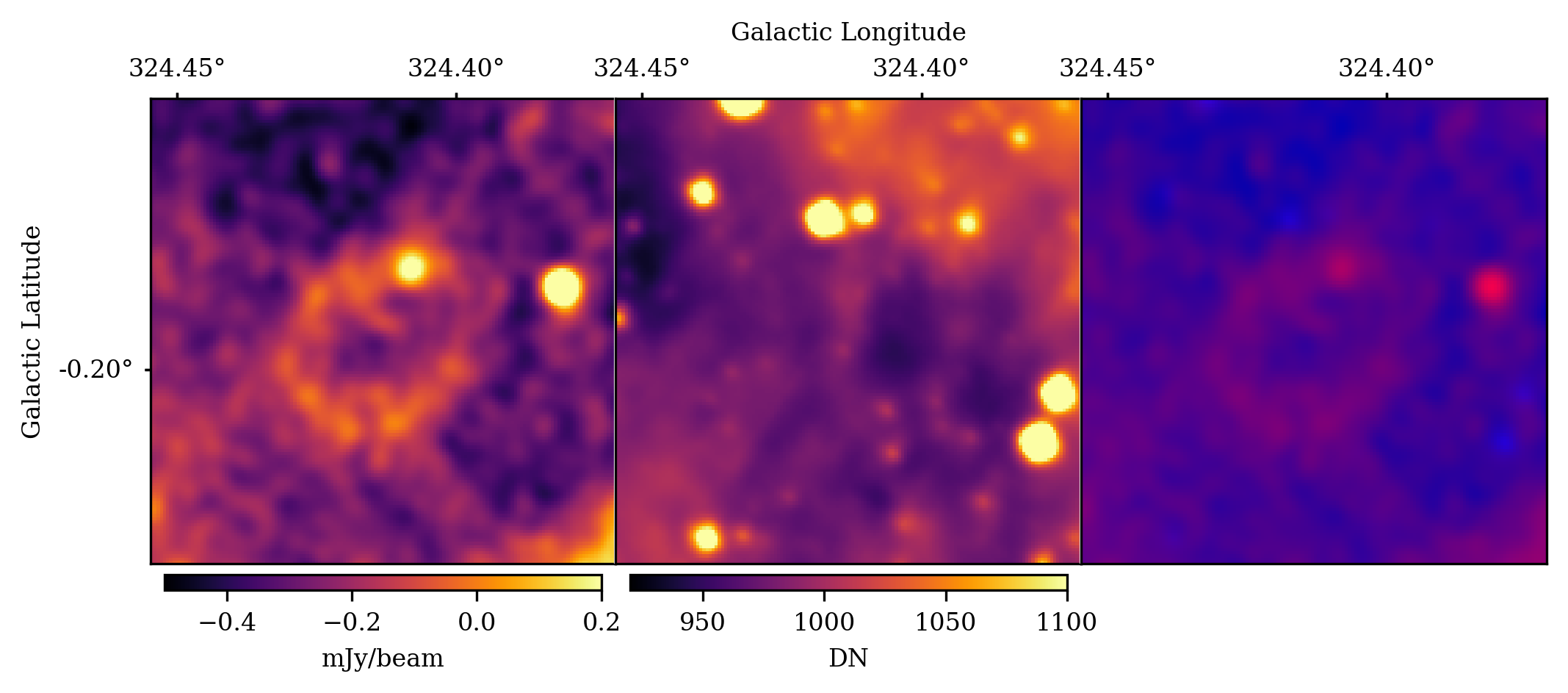}
    \caption{G324.4$-$0.2}
    \label{fig:22}
\end{figure*}

\begin{figure*}
    \centering
    \includegraphics[width=0.85\textwidth]{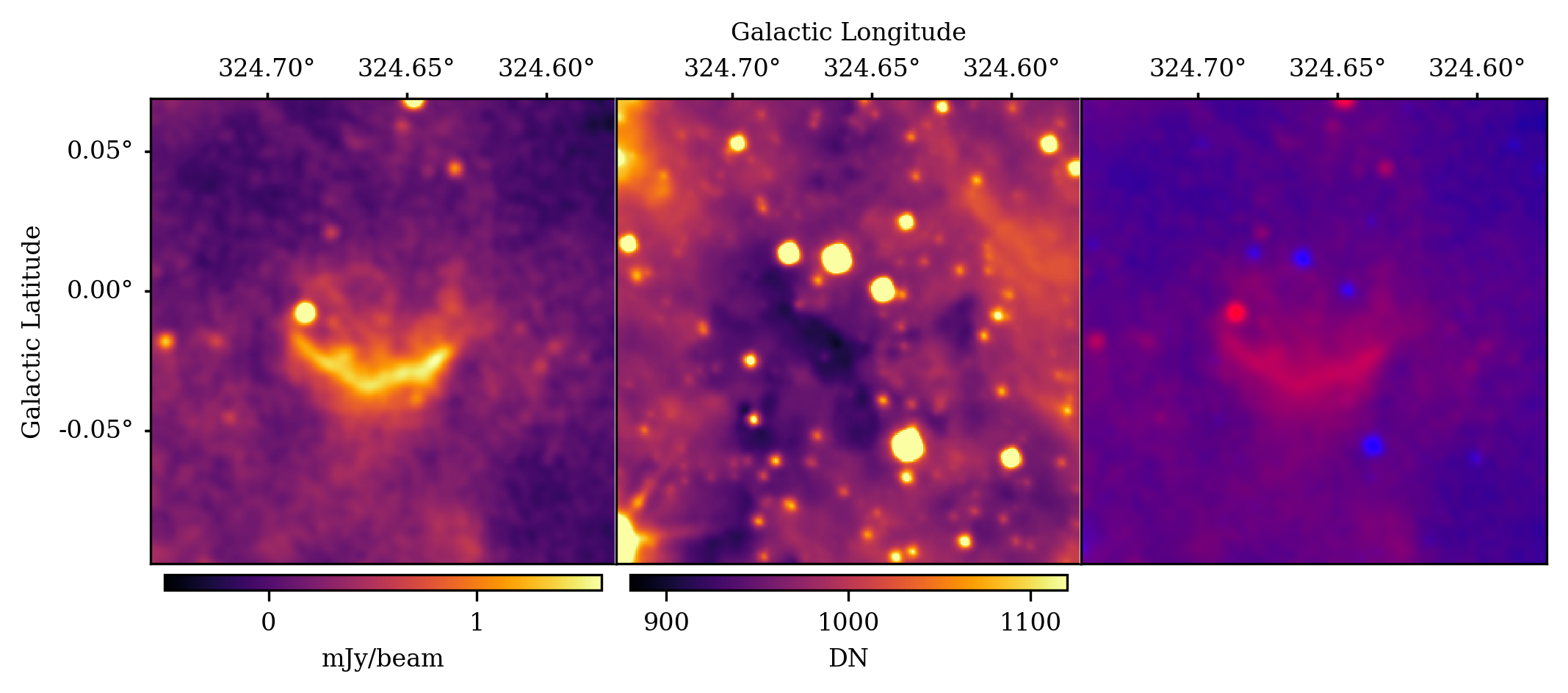}
    \caption{G324.7$+$0.0}
    \label{fig:18}
\end{figure*}

\begin{figure*}
    \centering
    \includegraphics[width=0.85\textwidth]{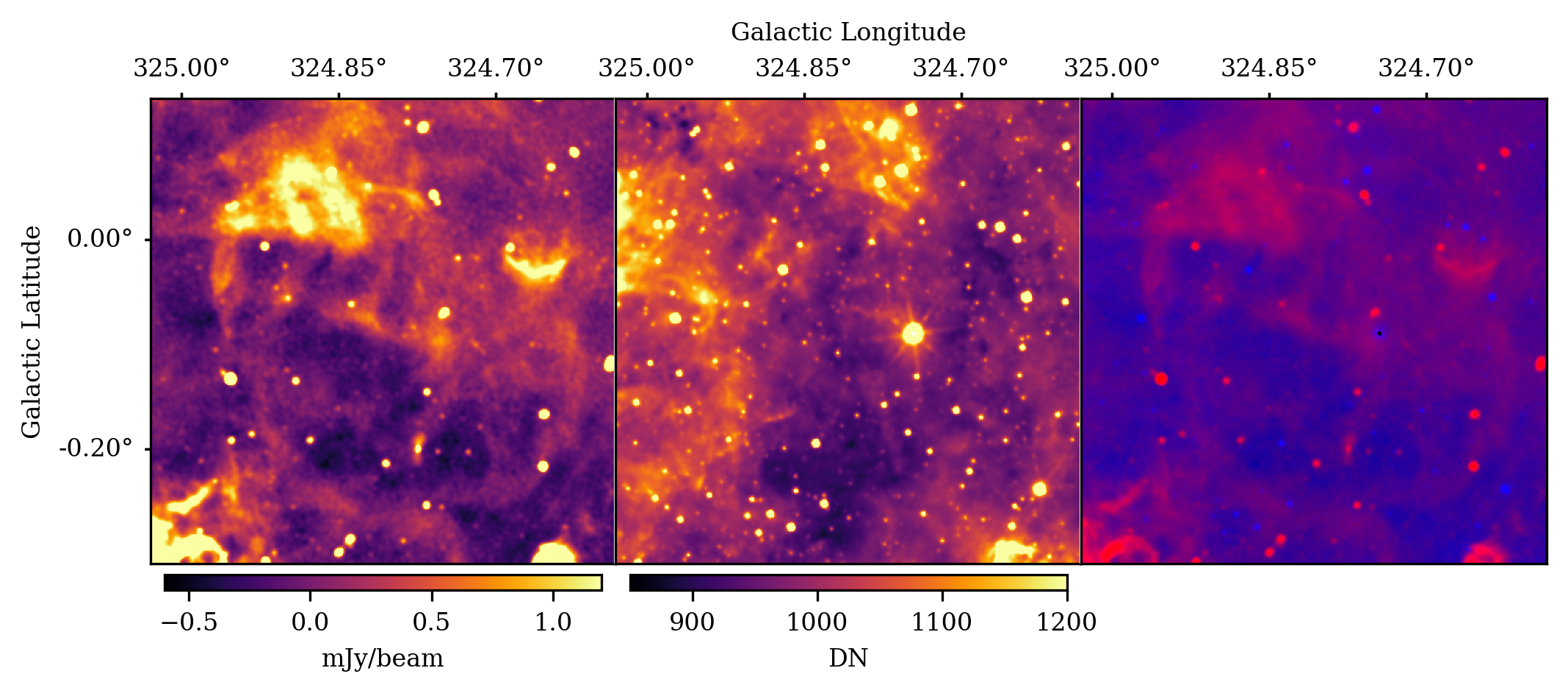}
    \caption{G324.8$-$0.1}
    \label{fig:20}
\end{figure*}

\begin{figure*}
    \centering
    \includegraphics[width=0.85\textwidth]{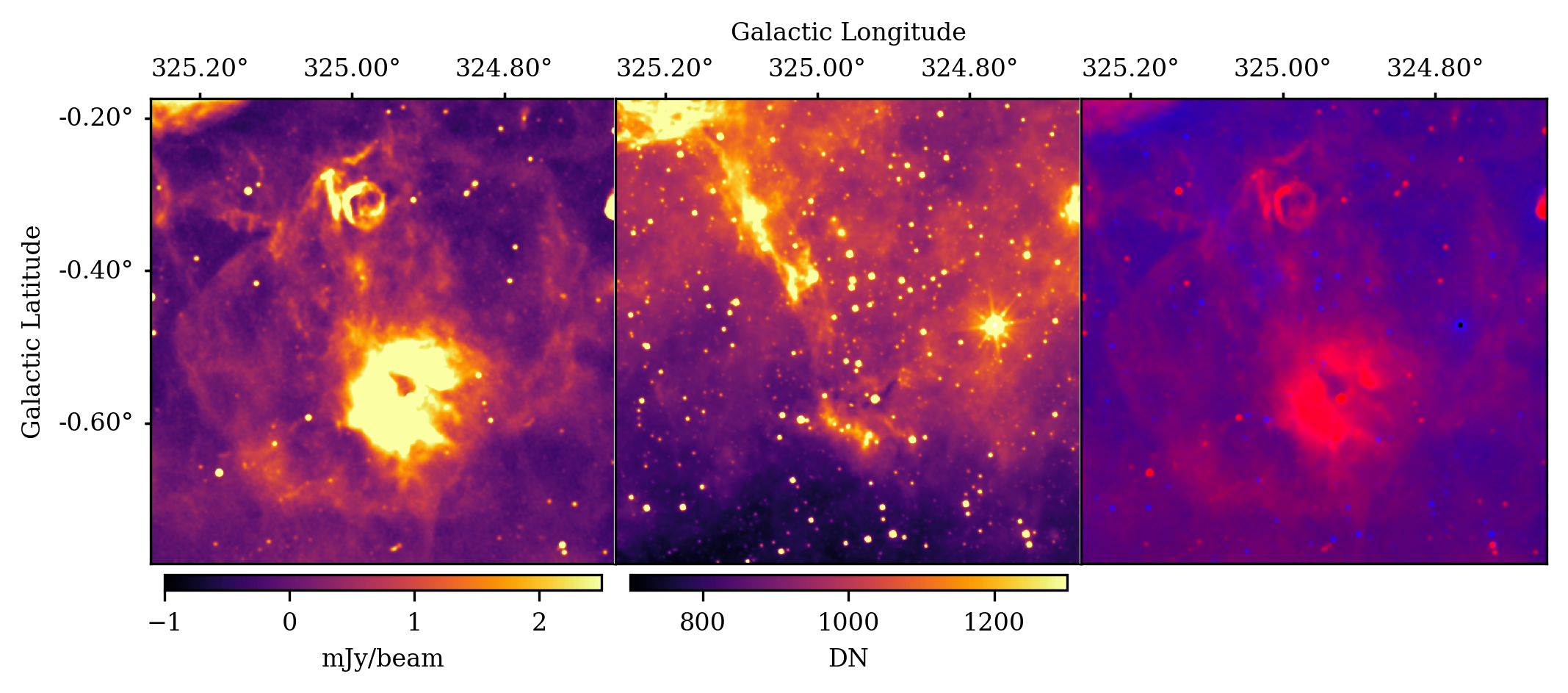}
    \caption{G325.0$-$0.5}
    \label{fig:19}
\end{figure*}

\begin{figure*}
    \centering
    \includegraphics[width=0.85\textwidth]{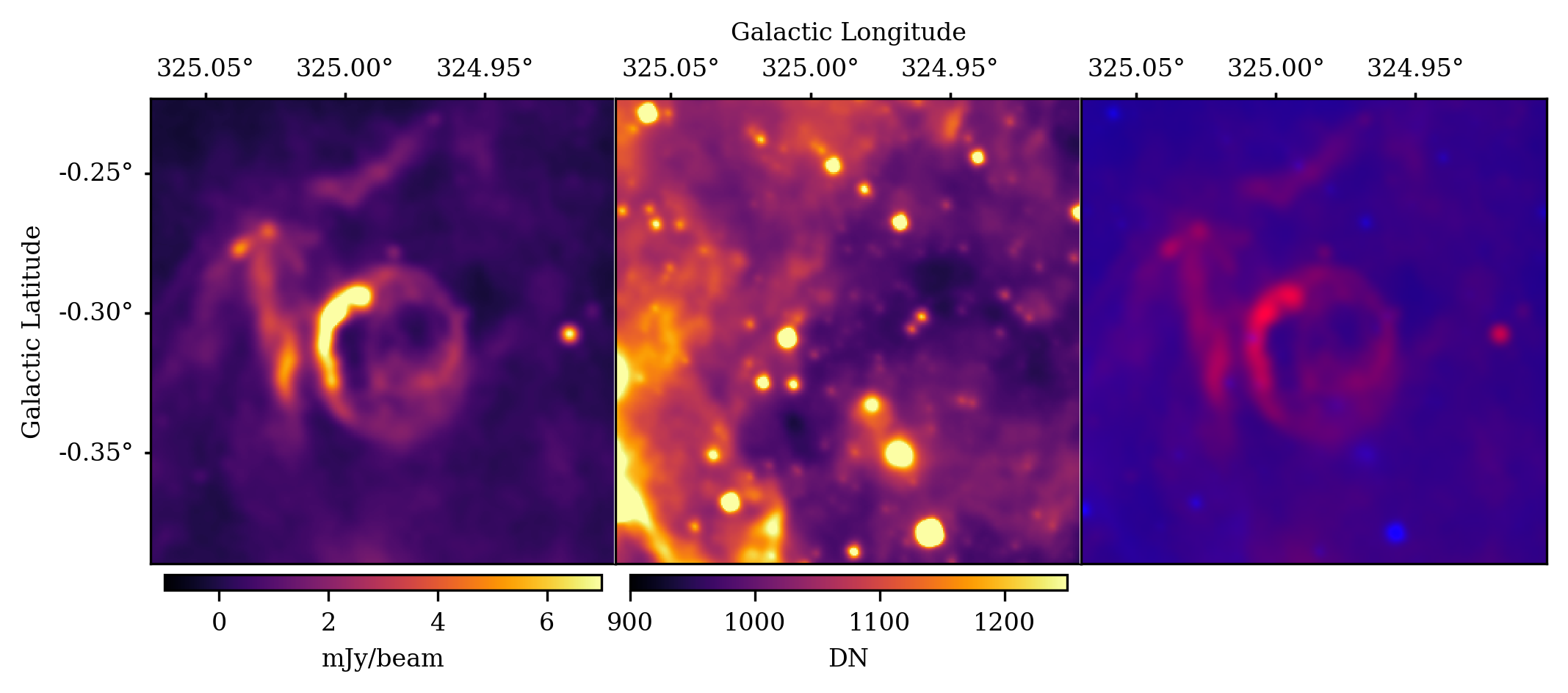}
    \caption{G325.0$-$0.3}
    \label{fig:9}
\end{figure*}

\begin{figure*}
    \centering
    \includegraphics[width=0.85\textwidth]{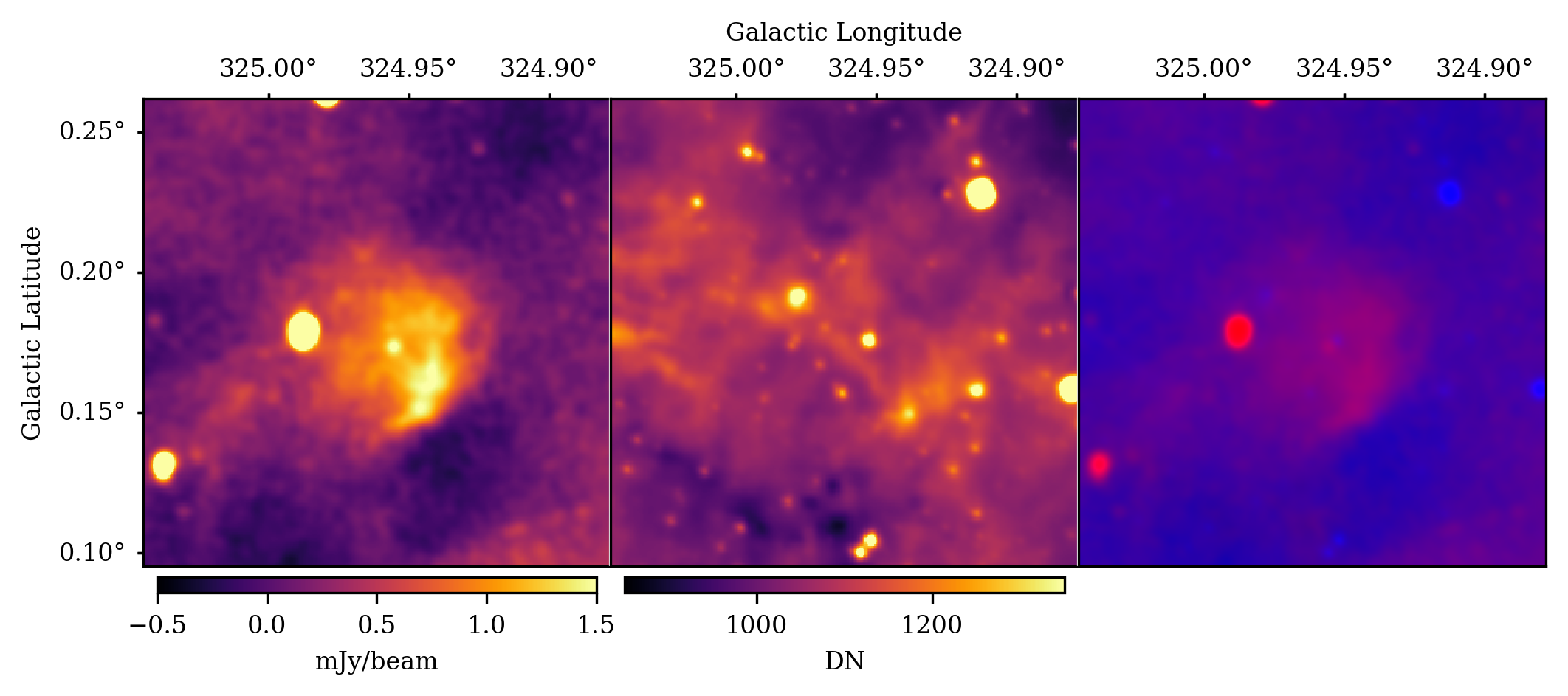}
    \caption{G325.0$+$0.2}
    \label{fig:27}
\end{figure*}

\begin{figure*}
    \centering
    \includegraphics[width=0.85\textwidth]{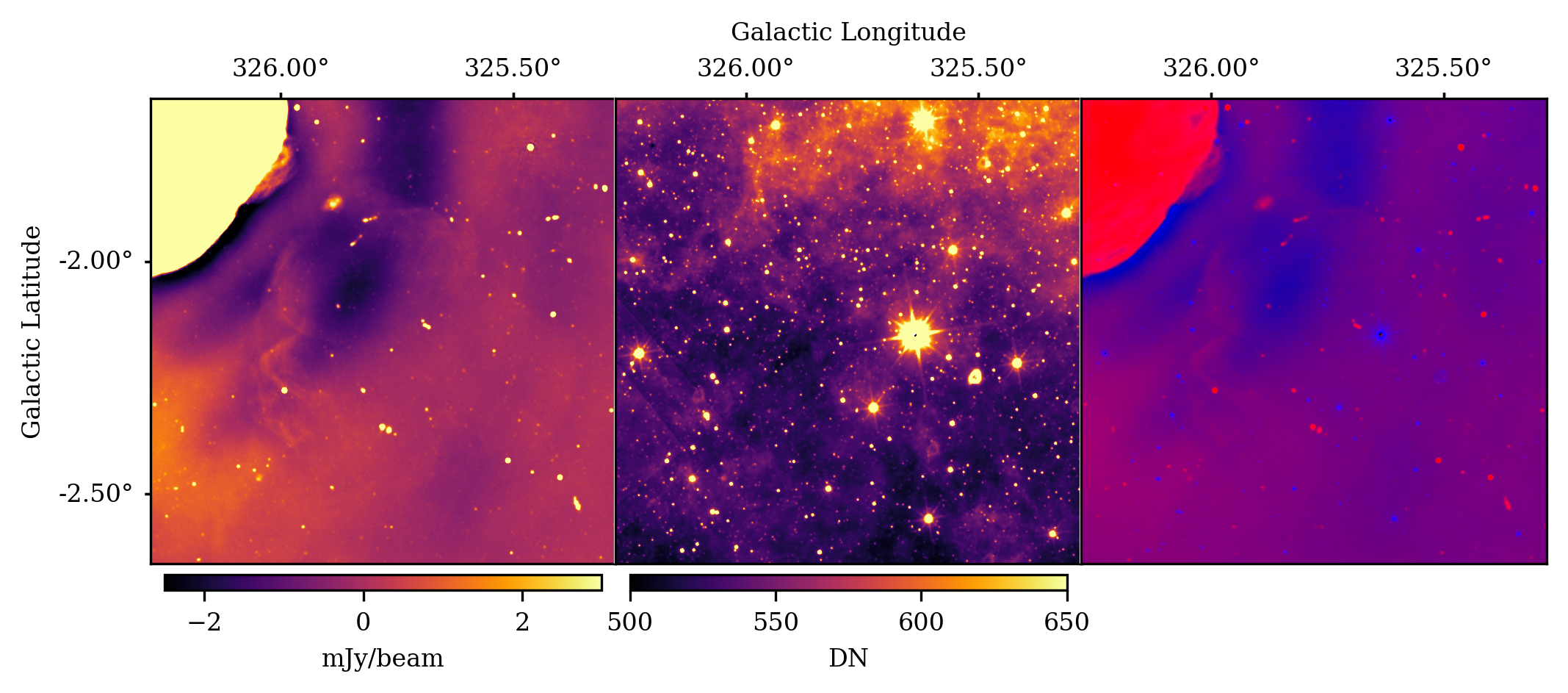}
    \caption{G325.8$-$2.1}
    \label{fig:23}
\end{figure*}

\begin{figure*}
    \centering
    \includegraphics[width=0.85\textwidth]{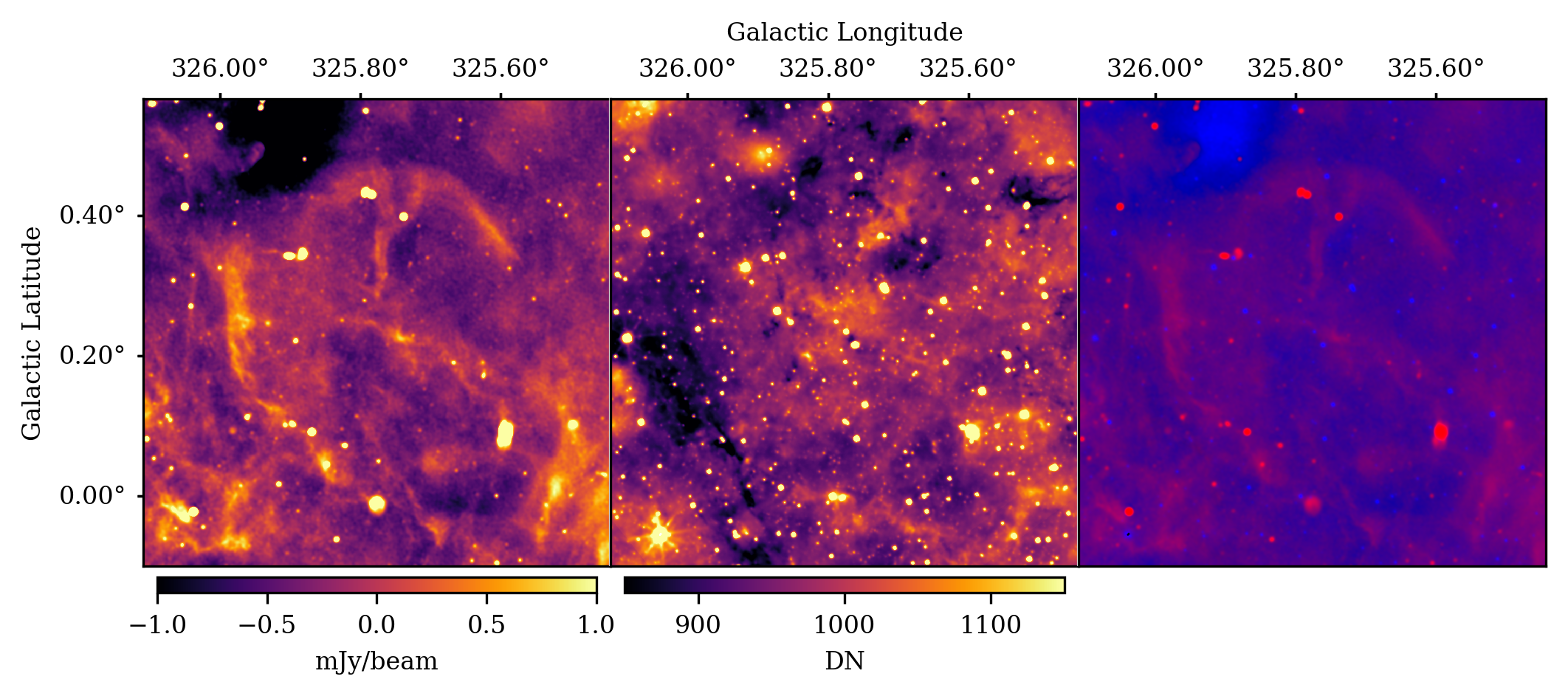}
    \caption{G325.8$+$0.3}
    \label{fig:24}
\end{figure*}

\begin{figure*}
    \centering
    \includegraphics[width=0.85\textwidth]{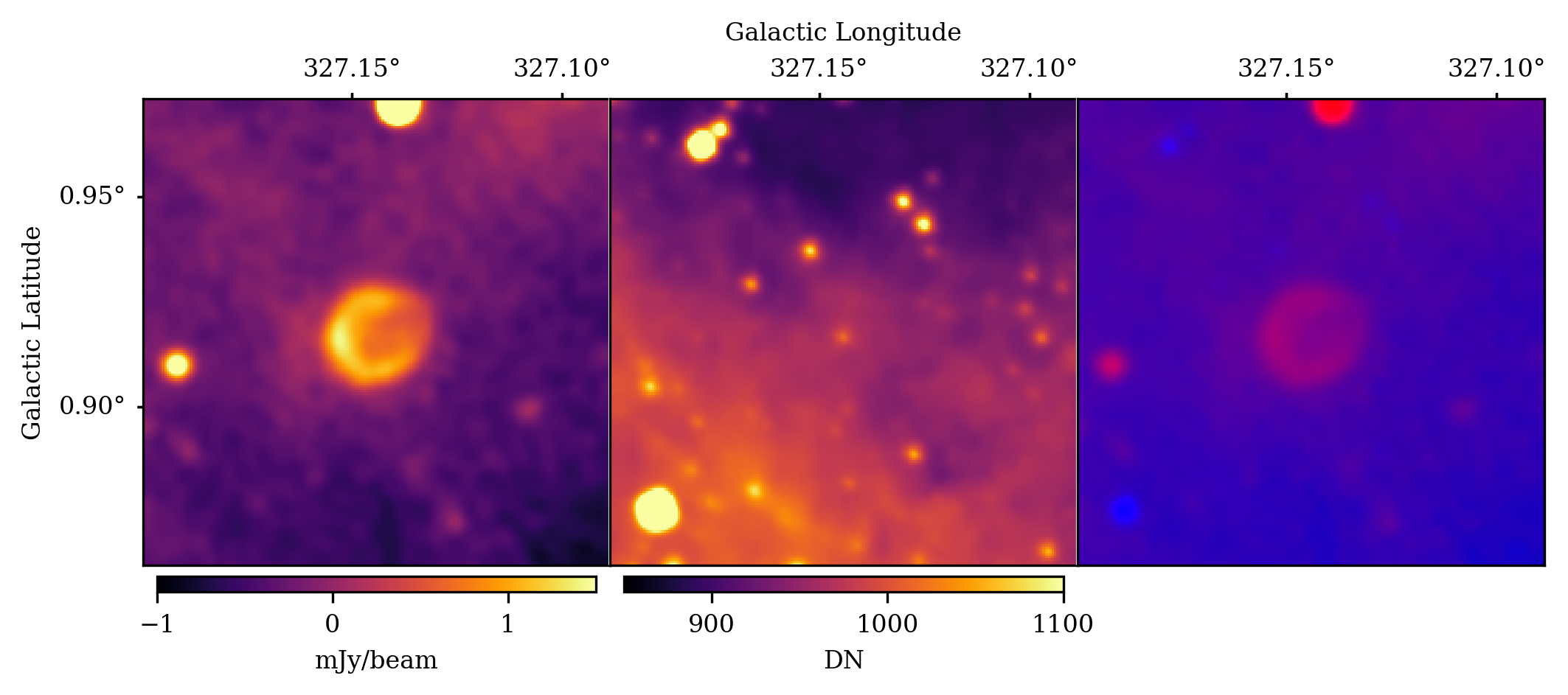}
    \caption{G327.1$+$0.9}
    \label{fig:16}
\end{figure*}

\begin{figure*}
    \centering
    \includegraphics[width=0.85\textwidth]{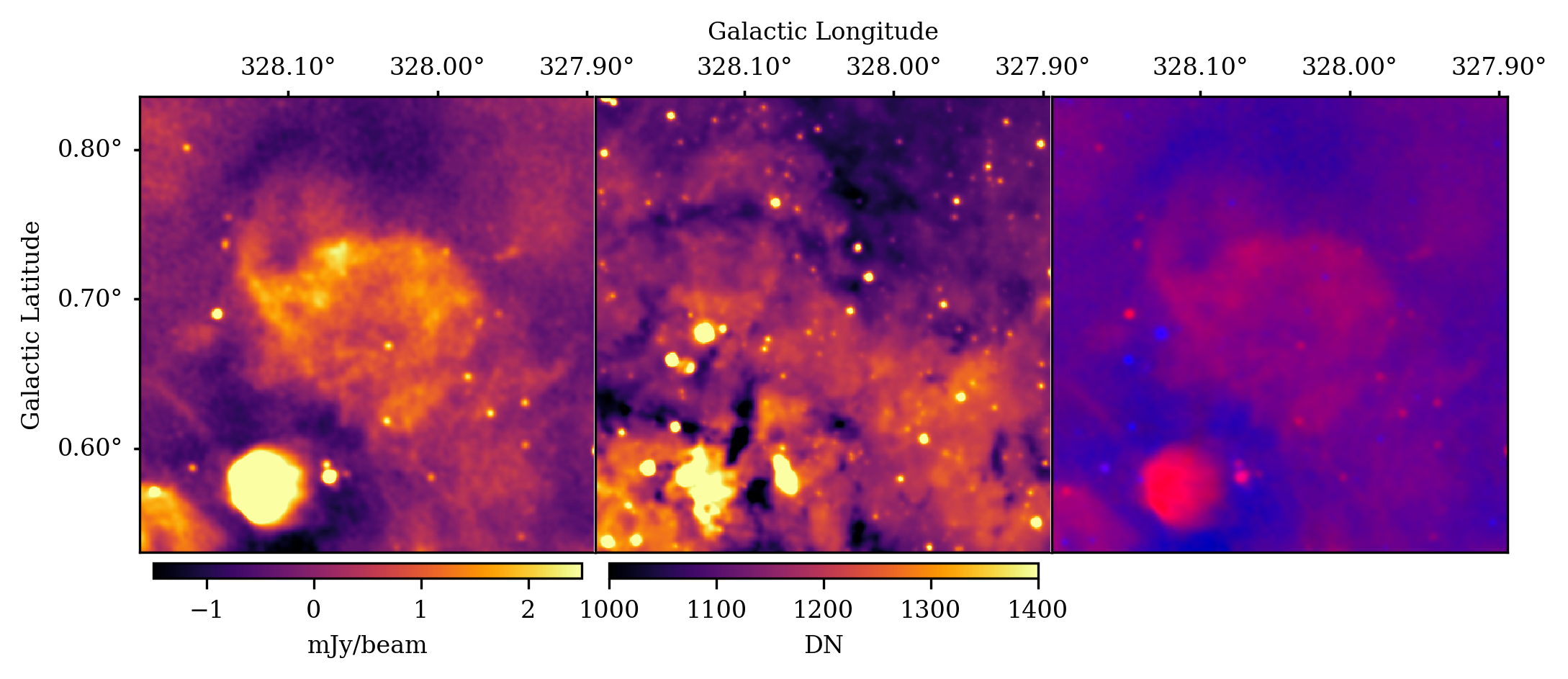}
    \caption{G328.0$+$0.7}
    \label{fig:17}
\end{figure*}

\begin{figure*}
    \centering
    \includegraphics[width=0.85\textwidth]{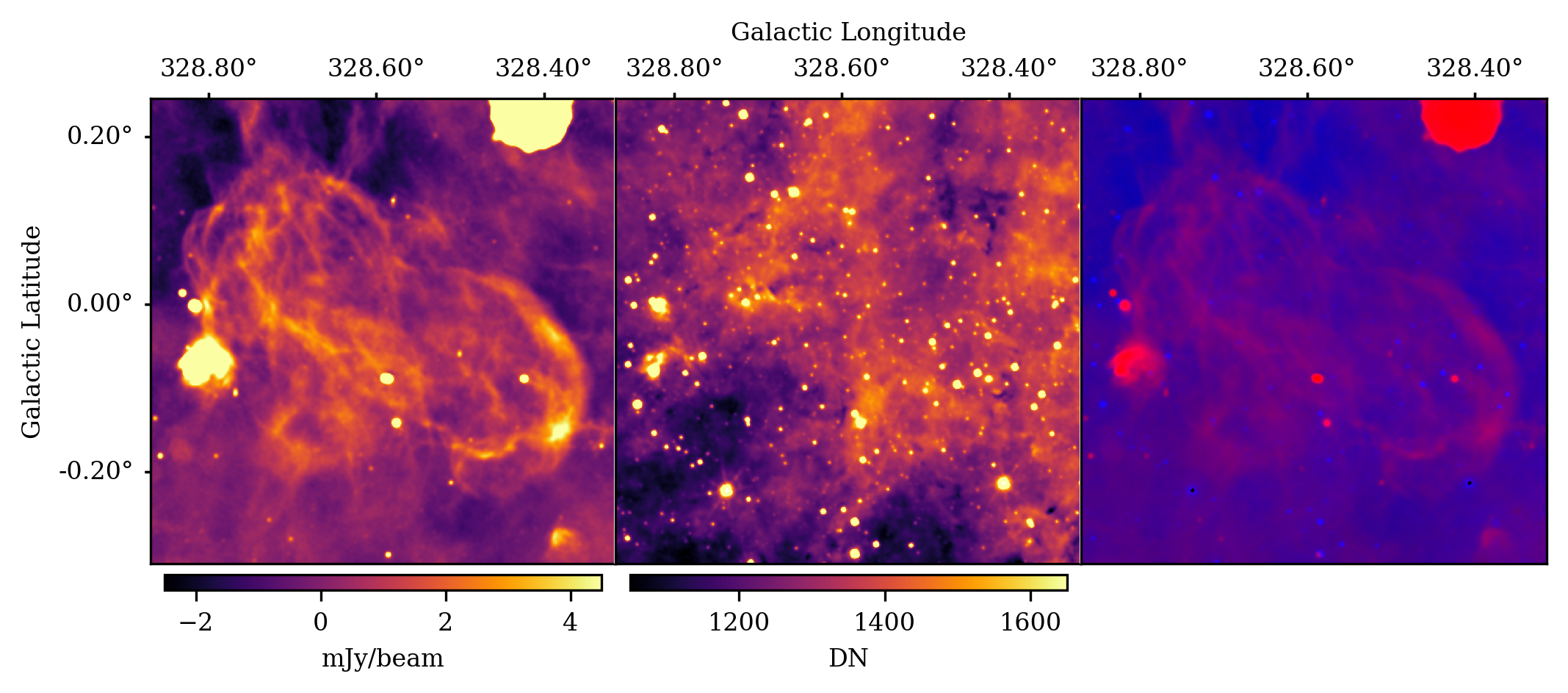}
    \caption{G328.6$+$0.0}
    \label{fig:11}
\end{figure*}

\begin{figure*}
\centering
    \includegraphics[width=0.85\textwidth]{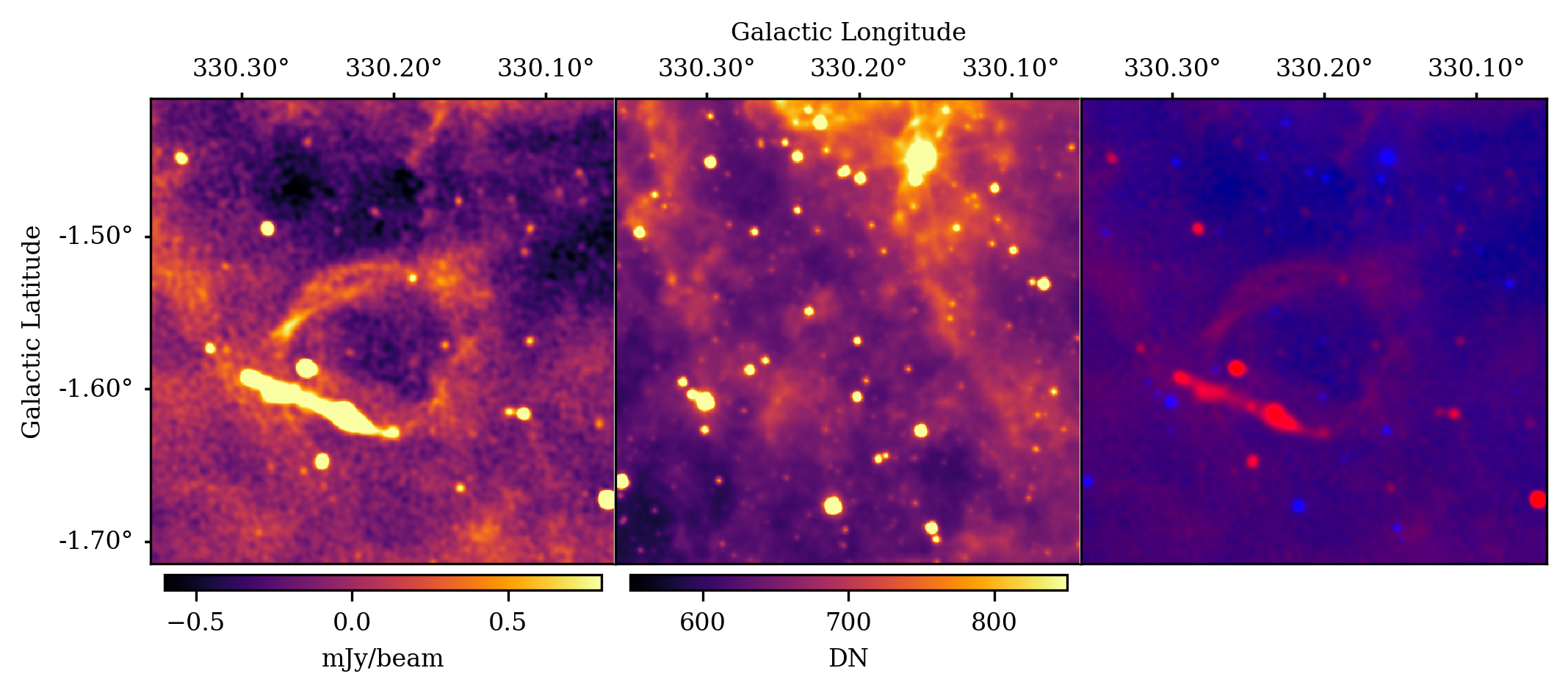}
    \caption{G330.2$-$1.6}
    \label{fig:25}
\end{figure*}

In Figures~\ref{fig:13} to \ref{fig:25} we present images of the SNR candidates in the EMU/POSSUM Galactic pilot II field using the same format described in Appendix \ref{sec:appendixA}. These are sources that do not appear in the \cite{Green2022} radio SNR catalogue.

\bsp	
\label{lastpage}

\end{document}